\documentclass[printer]{aa}
\usepackage{natbib}
\bibpunct{(}{)}{;}{a}{}{,} % to follow the A&A style\usepackage{mathcomp}%\usepackage{floatflt}
\usepackage{verbatim}
\usepackage{graphicx}%,amsmath,multicol,epsfig,floatflt}
\usepackage{tabularx}%, hyperref}%
\usepackage{txfonts}
\usepackage{amsmath}
\usepackage{amssymb}
\usepackage{caption}
\usepackage{booktabs}
\usepackage{subfig}
\usepackage{mathtools}
\usepackage{afterpage}

\newcommand{\kms}{\hbox{km~s$^{-1}$}}
\newcommand{\obj}{HD~206893B }
\newcommand{\host}{HD~206893 }
\newcommand{\Mjup}{M$_{\rm Jup}$ }
\newcommand{\Rjup}{R$_{\rm Jup}$ }

\newcommand{\teff}{T$_{e\!f\!f}$\xspace}
\newcommand{\logg}{log~\emph{g}\xspace}

\newcommand{\mic}{$\mu$m\xspace}
\newcommand{\as}{\hbox{$^{\prime\prime}$}\xspace}

\begin{document}

\title{In-depth study of moderately young but extremely red, very dusty substellar companion HD206893B\thanks{}} 
 
%\subtitle{II. An example text with infinitesimal
%  scientific value\\
%  whose title and subtitle may also be split}

\author{
  P. Delorme\inst{1},  T. Schmidt\inst{2} , M. Bonnefoy \inst{1}, S. Desidera \inst{3}, C. Ginski\inst{4,26}, B.Charnay\inst{2}, C. Lazzoni\inst{3}, V. Christiaens\inst{5,23}, S. Messina\inst{6} , V. D'Orazi\inst{3}, J. Milli\inst{7}, J.E. Schlieder\inst{8,9}, R. Gratton\inst{3}, L. Rodet\inst{1}, A-M. Lagrange\inst{1}, O. Absil\inst{5}\fnmsep\thanks{F.R.S.-FNRS Research Associate}, A. Vigan\inst{10}, R. Galicher\inst{2}, J. Hagelberg\inst{1}, M. Bonavita\inst{11}, B. Lavie\inst{22,12}, A. Zurlo\inst{10,13}, J. Olofsson\inst{8,14}, A.Boccaletti\inst{2}, F. Cantalloube\inst{8}, D. Mouillet\inst{1}, G.Chauvin\inst{1}, F.-J. Hambsch\inst{15}, M. Langlois\inst{25,10}, S. Udry\inst{22}, T. Henning\inst{8}, J-L. Beuzit\inst{1}, C. Mordasini\inst{12}, P.Lucas\inst{16}, F. Marocco\inst{16}, B. Biller\inst{24}, J. Carson\inst{8,17}, A. Cheetham\inst{22}, E. Covino\inst{18}, V. De Caprio \inst{18}, A. Delboulbe \inst{1}, M. Feldt\inst{8}, J. Girard\inst{7}, N. Hubin\inst{19}, A-L. Maire\inst{8}, A. Pavlov\inst{8}, C. Petit\inst{20}, D. Rouan\inst{2},  R. Roelfsema \inst{21}, F. Wildi\inst{22}
 }

% ,  ...
\offprints{P. Delorme, \email{Philippe.Delorme@univ-grenoble-alpes.fr} Based on observations made with ESO Telescopes at the Paranal Observatory under programs ID 097.C-0865(D) (SPHERE GTO, SHINE program) and Program ID: 082.A-9007(A) (FEROS) 098.C-0739(A), 192.C-0224(C) (HARPS). This work has made use of the SPHERE Data Centre.
}

\institute{ Univ. Grenoble Alpes, CNRS, IPAG, F-38000 Grenoble, France 
\and LESIA, Observatoire de Paris, PSL Research University, CNRS, Sorbonne Universités, UPMC Univ. Paris 06, Univ. Paris Diderot, Sorbonne Paris Cité, 5 place Jules Janssen, 92195 Meudon, France
\and INAF-Osservatorio Astronomico di Padova, Vicolo dell'Osservatorio 5, 35122 Padova, Italy
\and  Leiden Observatory,Niels Bohrweg 2, NL-2333 CA Leiden, The Netherlands
\and  Space sciences, Technologies and Astrophysics Research (STAR) Institute, Universit\'e de Li\`ege, 19c All\'ee du Six Ao\^ut, B-4000 Li\`ege, Belgium
\and  INAF-Catania Astrophysical Observatory, via S. Sofia 78, I-95123 Catania, Italy)
\and European Southern Observatory (ESO), Alonso de Córdova 3107, Vitacura, Casilla 19001, Santiago, Chile
\and Max-Planck-Institute for Astronomy, Koenigsthul 17, D 69117 Heidelberg, Germany%JOSH
\and Exoplanets and Stellar Astrophysics Laboratory, Code 667, NASA-Goddard Space Flight Center, Greenbelt, MD 20771, USA%JOSH
\and Aix Marseille Univ, CNRS, LAM, Laboratoire d'Astrophysique de Marseille, Marseille, France %arthur
\and Institute for Astronomy, The University of Edinburgh, Royal Observatory, Blackford Hill, Edinburgh, EH9 3HJ, UK
\and Physikalisches Institut, University of Bern, Sidlerstrasse 5, 3012 Bern, Switzerland
\and N\'ucleo de Astronom\'ia, Facultad de Ingenier\'ia, Universidad Diego Portales, Av. Ejercito 441, Santiago, Chile %alice
\and Instituto de Física y Astronomía, Facultad de Ciencias, Universidad de Valparaíso, Av. Gran Bretaña 1111, Playa Ancha, Valparaíso
\and Remote Observatory Atacama Desert (ROAD), Vereniging Voor Sterrenkunde (VVS), Oude Bleken 12, B-2400 Mol, Belgium
\and Centre for Astrophysics Research, Science and Technology Research Institute, University of Hertfordshire, Hatfield AL10 9AB, UK
\and Department of Physics \& Astronomy, College of Charleston, 58 Coming Street, Charleston, SC 29424, USA
\and INAF Napoli: INAF, Astrophysical Observatory of Capodimonte, Salita Moiariello 16, 80131 Napoli, Italy
\and European Southern Observatory (ESO), Karl-Schwarzschild-Straße 2, 85748 Garching bei München, Germany
\and ONERA: ONERA, 29 avenue de la Division Leclerc, 92322 Châtillon Cedex, France
\and NOVA: NOVA Optical-Infrared Instrumentation Group at ASTRON, Oude Hoogeveensedijk 4, 7991 PD Dwingeloo, The Netherlands
\and Geneva: Geneva Observatory, University of Geneva, Chemin des Maillettes 51, 1290 Versoix, Switzerland 
\and Departamento de Astronom\'ia, Universidad de Chile, Casilla 36-D, Santiago, Chile
\and Institute  for  Astronomy,  University  of  Edinburgh,  Blackford  Hill View, Edinburgh EH9 3HJ, UK
\and CRAL,  UMR  5574,  CNRS,  Université  Lyon  1,  9  avenue  Charles André, 69561 Saint Genis Laval Cedex, France
\and Anton Pannekoek Institute for Astronomy, University of Amsterdam, Science Park 904, 1098 XH Amsterdam, The Netherlands 
} 
%  \and The University of Texas at Austin, Department of Astronomy, 2515 Speedway, Stop C1400, Austin, Texas 78712-1205, USA
%    \and Carnegie Institution of Washington DTM, 5241 Broad Branch Road NW, Washington, DC 20015, USA
%\and NASA Sagan fellow
%   \and Institut UTINAM, CNRS UMR 6213, Observatoire des Sciences de l’Univers THETA Franche-Comt\'e Bourgogne, Univ. Bourgogne Franche-Comt\'e, 41 bis avenue de l’Observatoire
% \and IfA, University of Hawai'i, 2680 Woodlawn Drive, Honolulu, HI 96822, USA
%\and Institut de Recherche sur les Exoplan\`etes,
 % Universit\'e de Montr\'eal, C.P. 6128, Succursale Centre-Ville,
 % Montr\'eal, QC H3C 3J7, Canada
%  \and Univ Lyon, Ens de Lyon, Univ Lyon1, CNRS, Centre de Recherche Astrophysique de Lyon UMR5574, F-69007, Lyon, France
%  Cedex 07, France
 %  \and Canada-France-Hawaii Telescope Corporation, 65-1238 Mamalahoa 
 %  Highway, Kamuela, HI96743, USA
 %\and  UC Santa Cruz, ISB 159 / 1156 High St, Santa Cruz, CA 95060 

\abstract{The substellar companion HD206893b has recently been discovered by direct imaging of its disc-bearing host star with the Spectro-Polarimetric High-contrast
Exoplanet REsearch (SPHERE) instrument.}
{We investigate the atypical properties of the companion, which has the reddest near-infrared colours among all known substellar objects, either orbiting a star or isolated, and we provide a comprehensive characterisation of the host star-disc-companion system.}
{ We conducted a follow-up of the companion with adaptive optics imaging and spectro-imaging with SPHERE, and a multi-instrument follow-up of its host star. We obtain a R=30 spectrum from 0.95 to 1.64~$\mu$m of the companion and additional photometry at 2.11 and 2.25~$\mu$m.  We carried out extensive atmosphere model fitting for the companions and the host star in order to derive their age, mass, and metallicity.}
 {We found no additional companion in the system in spite of exquisite observing conditions resulting in sensitivity to 6\Mjup (2\Mjup) at 0.5\as for an age of 300~Myr (50~Myr).  We detect orbital motion over more than one year and characterise the possible Keplerian orbits. We constrain the age of the system to a minimum of 50~Myr and a maximum of 700~Myr, and determine that the host-star metallicity is nearly solar. The comparison of the companion spectrum and photometry to model atmospheres indicates that the companion is an extremely dusty late L dwarf, with an intermediate gravity (log $g \sim$4.5$--$5.0) which is compatible with the independent age estimate of the system.}
{ Though our best fit corresponds to a brown dwarf of 15--30 \Mjup aged 100-300~Myr, our analysis is also compatible with a range of masses and ages going from a 50~Myr 12\Mjup planetary-mass object to a 50 \Mjup Hyades-age brown dwarf. Even though this companion is extremely red, we note that it is more probable that it has an intermediate gravity rather than the very low gravity that is often associated with very red L dwarfs. 
%An empirical comparison with known objects shows that the best fit is obtained with atypically red L3-L5 dwarfs, with additionnal reddening modelled by extra-photospheric absorption by fosterite dust. 
 We also find that the detected companion cannot shape the observed outer debris disc, hinting that one or several additional planetary mass objects in the system might be necessary to explain the position of the disc inner edge. }
% insister sur l'opposition pas si jeune et très rouge très dustry
% 

\date{}


\authorrunning{P. Delorme et al.}
\titlerunning{}
\maketitle

\section{Introduction}
 The discovery of young extrasolar giant planets found with high-contrast imaging techniques \citep{Chauvin.2004,Marois.2008,Lagrange.2010,Rameau.2013,Delorme.2013,Bailey.2014,Macintosh.2015,Gauza.2015} offers the opportunity to directly probe the properties of their photosphere. The improved contrast and spectroscopic capabilities of the new generation of adaptive optics (AO) instruments such as Spectro-Polarimetric
High-contrast
Exoplanet REsearch (SPHERE) \citep{Beuzit.2008} and GPI \citep{Macintosh.2012}  have made it possible to study the molecular composition and physical processes taking place in the atmospheres of extrasolar giant planets, \citep{Zurlo.2016,Bonnefoy.2016,Vigan.2016,Derosa.2016,Chilcote.2017}.\\

 These previous studies have shown that while young exoplanets have a spectral signature quite distinct from field brown dwarfs of equivalent effective temperature, they have many atmospheric properties in common with isolated brown dwarfs recently identified in young moving groups \citep{Liu.2013,Gagne.2015c,Aller.2016,Faherty.2016}. They notably share a very red spectral energy distribution (SED)  in the near-infrared (NIR) that can be attributed to the presence of very thick dust clouds in their photosphere. This trend was qualitatively expected by atmosphere models because the lower surface gravity of these planetary mass objects inhibits dust settling and naturally increases the dust content within the photosphere. However, all atmosphere models fail to quantitatively match the very red NIR colours of young planetary mass objects via a self-consistent physical model,and have to resort to parametrising the sedimentation efficiency of the dust to match these observations, as done for instance in the {\it Dusty} models where there is  no dust settling \citep[][]{Allard.2001} or the parametrised cloud models \citep{Saumon.2008,Madhusudhan.2011,Morley.2014,Baudino.2015}. The recent work by \citet{Liu.2016} focuses on a sample of substellar objects with known ages and distances and has confirmed that low-gravity objects populate an area of NIR colour-magnitude diagram that is distinct from the area occupied by field brown dwarfs of the same spectral type. This study also claims  that  low-mass companions might have properties distinct from those of isolated low-mass objects, while they should have similar gravity, hinting that other parameters might also shape the NIR SED of substellar objects.\\

 In this light the recent discovery of the very red close-in companion \obj (or HIP107412~b) identified by \citet{Milli.2017}, offers the opportunity to explore the physical properties of a relatively young mid-L dwarf substellar companion and to try to understand its atypical properties. We carried out a systematic study of the debris disc-bearing host star, including new observations, to determine its age and metallicity (Section 2). Through the  SpHere INfrared survey for Exoplanets (SHINE) GTO large programme with SPHERE, we obtained additional resolved observations of the companion under very stable observing conditions, including the low-resolution spectroscopy that we present in Section 3. These observations confirm that \obj is the reddest known substellar object, even redder in NIR than 2M1207~b \citep[admittedly by a slight margin of approximately 1$\sigma$, ][]{Chauvin.2004} or the recently discovered 2MASS~J223624+4751 \citep{Bowler.2017}. Section 4  focuses on the comparison of \obj with known red L dwarfs. Section 5  is dedicated to the characterisation of the atmospheric properties of this atypical young substellar companion, through atmosphere  and substellar evolution models. In Section 6 we detail the orbital constraints that can be derived from the observed Keplerian motion of the companion and determine that this detected companion cannot shape the observed disc alone.

\section{ Host star properties}
The determination of the properties of the substellar object around the star
depends sensitively on the adopted stellar parameters, and especially
on stellar age. Unfortunately, the age of mid-F stars like \host is particularly
difficult to determine. On the one hand, their evolution is slow, making isochrone fitting
not very constraining when close to the main sequence. On the other hand, the
indicators based on rotation and activity are more uncertain than 
for colder stars due to the very thin convective envelope.
\cite{Milli.2017} adopted an extended age range 0.2-2.1 Gyr,  while \cite{galicher16}
adopted $200^{+200}_{-100}$ Myr from X-ray and UV luminosity.
Our approach is based on a comprehensive evaluation of several age indicators
following the methods described in \citet{desidera15}.
Spectroscopic and photometric datasets were acquired to gather the
required information.

\subsection{Kinematic analysis and young groups membership}
\label{s:kin}

HD~206893 is included in Gaia DR1 \citet{gaia_dr1} and we adopted 
Gaia DR1 trigonometric parallax and proper motion (using a corresponding distance of 40.6$^{+0.5}_{-0.4}$pc; see Table~\ref{t:param}),
unlike Milli et al. who adopted the Hipparcos parallax \citep[38.3$^{+0.8}_{-0.7}$pc;][]{Vanleeuwen.2007}. 
%These values are  within the errorbars of previous determination
%but of better accuracy.
For the radial velocity we adopt RV=$-11.92\pm0.32$ km\,s$^{-1}$, as derived from
the analysis of the FEROS spectrum.
This is similar to the values of $-12.9\pm1.4$ km\,s$^{-1}$ measured by \citet{nordstrom04}, and also consistent with our measurements from HARPS data of $-12.2\pm0.1$\kms. 
The presence of the brown dwarf companion would imply a maximum RV semi-amplitude  only of the order of 0.15\kms\ (from the family of orbits fitting the available astrometric data, see Section.\ref{s:astrom}), which is smaller than  our current uncertainties on the RV of the system and is hence neglected in the kinematic analysis. These spectroscopic data also show that \host is not a spectroscopic binary.

The resulting space velocities of the system are U,V,W =-20.15$\pm$0.23, -7.40$\pm$0.15, -3.40$\pm$0.26 ~\kms. We  used these kinematics to investigate  the  
possible  membership  of HD~206893 to a young moving group using the BANYAN2 online tool 
\citep{Gagne.2014a}. Assuming that HD~206893 is younger than
1 Gyr (see below), we obtain a 13.5\% probability for membership
in the Argus association and 86.5\% for the young field.

%with the following setups:
%- "regular" which uses prior accounting for the fact that YMG
%stars are much less common than field stars in the sky. 
%It particularly affects Argus because Jonathan attributes only 11 stars to
%Argus,  which  for  a  similar  match  in  kinematics,  would  lower
%the probability of Argus by roughly a factor 5 compared to AB-
%Doradus or Tucana which are attributed 50 stars. 
%This yelds a 9\% probability that it is a member of Argus, 60\% of the young
%field (which is also compatible with our data), and 31\% to the
%old field
%- "without prior", which drops assumptions on numbers, and
%makes BANYAN2 equivalent to banyan 1 with updated kinematics distributions: 
%These yields a 97\% membership probability to Argus
%- With "young prior": since we know HD~206893 is less than
%1Gyr, it makes no sense to include the old field hypothesis in the
%study. SInce the old field kinematics distribution is quite wide,
%and since most of the stars of the milky way are old field, it has
%a strong prior in the "regular hypothesis", that is not relevant,
%i believe. In this case we get 13.5\% probability for Argus and
%86.5\% for young field, which I believe is a decent guess.
%
%In summary, kinematic provide some evidence for a possible membership
%to Argus to be considered when addressing the various age indicators 
%but a young disc kinematic is also fully viable.

\subsection{Spectroscopic parameters}
\label{s:spec}

The star was observed with the FEROS spectrograph on 2008 November 13 as part of the preparatory program for the SPHERE GTO survey. 
%performed at La Silla 2.2m telescope as part of Max Planck Institut
%f\"ur Astronomie (MPIA) observing time.
Details on the instrument set-up and reduction procedure are 
given in \citet{desidera15}.

As done in our previous investigations  (see e.g. \citealp{Vigan.2016} for GJ 758~B) 
we  derived spectroscopic parameters using the code MOOG by Sneden 
(\citealt{sneden73}, 2014) and the Kurucz grid (\citeyear{kur1993}) of model 
atmospheres, with the overshooting option switched on. Effective temperature 
($T_{\rm eff}$) was obtained by zeroing the slope between abundances from 
Fe~{\sc i} lines and the excitation potential of the lines, 
while microturbulence velocity ($\xi$) comes from removing spurious trends 
between the reduced equivalent widths and abundances from Fe~{\sc i} lines. 
We note that due to the relatively high rotational velocity ($ v\ sin i \sim$ 32 km\,s$^{-1}$, see Section. \ref{s:liba}) we were able to exploit  only 27 Fe~{\sc i} lines.
Surface gravity (log$g$) was derived imposing the ionisation equilibrium condition that  abundances from Fe{\sc i} and Fe{\sc ii} lines are in agreement within 0.05 dex.
We found T$_{\rm eff}$ = 6500$\pm$100 K, log$g$=4.45$\pm$0.15 dex, and $\xi$=1.35$\pm$0.2 km\,s$^{-1}$ 
(uncertainties were derived following \citealt{desilva13} and \citealt{Vigan.2016}; we refer the 
reader to those papers for details on error computation). 
Our analysis results in A(Fe~{\sc i})=7.54 $\pm$ 0.02 dex, which implies 
[Fe/H]=+0.04$\pm$0.02 having adopted A(Fe~{\sc i})=7.50 for the Sun.

The spectroscopic temperature based on excitation equilibrium is fully 
consistent with those derived from the fitting of the wings of the $H_{\alpha}$ line, 
which results in $6550\pm100$~K, and from B$-$V, V$-$I, and 
V$-$K colours using the calibrations determined by \citet{casagrande10}, which yield 6519$\pm$49 K.

\subsection{Lithium and barium}
\label{s:liba}
The lithium abundance was derived by synthesising the Li~{\sc i} doublet at 
6708~\AA, using the driver {\it synth} in MOOG, as done in all previous 
works (e.g. \citealt{dr09}; \citealt{desidera11}; \citealt{desilva13}).
As a by-product of the spectral synthesis calculations we  also obtained the projected rotational velocity ($v\sin{i}$), finding 32$\pm$2 km\,s$^{-1}$.%, which is similar to the value reported by \citet{nordstrom04}.%29.0 km\,s$^{-1}$ no error bar in article

From our analysis, the Li abundance for HD~206893  is A(Li~{\sc i})=2.38$\pm$0.10 dex. 
For comparison with other works, we have also measured the EW of Li~{\sc i} doublet at 
6708 \AA, which results in 28.5$\pm$7.0 m\AA, which is lower than what is expected 
for an Argus member ($\sim$100\,m\AA\, at V$-$I= 0.51\,mag) and/or for stars of young ages ($\ga$100\,m\AA\, at ages $\lesssim$ 100 Myr; \citealt{Torres08}). 
The observed Li EW is intermediate between those measured for Hyades 
and Pleiades members, suggesting an intermediate age.

Finally, we have investigated the barium (Ba) content for our target star, since Ba abundances 
have been shown to be enhanced in young open cluster and field stars 
\citep[e.g.][]{dorazi09,desidera11,desilva13}.
In particular, stars younger than $\sim$ 100 Myr has been found to 
exhibit an extreme Ba enhancement, up to $\sim$0.65 dex, more than a factor 
of four above the solar abundance.
For consistency, we derived the Ba abundance employing the same line list, code, 
and procedure as in previous papers (e.g. \citealt{desilva13}). From the synthesis of 
the strong Ba~{\sc ii} line at 5853 \AA~ (including hyperfine structure and isotopic 
splitting with solar mix composition) we obtained [Ba/Fe]=0.2$\pm$0.2 dex. 
The relatively high uncertainties are related to the saturated behaviour of the Ba~{\sc ii} line. 
This result seems to confirm that HD~206893 is unlikely to be a member of the Argus association 
since the average Ba abundance for Argus is [Ba/Fe]=0.53$\pm$0.03 (from \citealt{desilva13}).

\begin{center}
\begin{figure*}
\includegraphics[width=18cm]{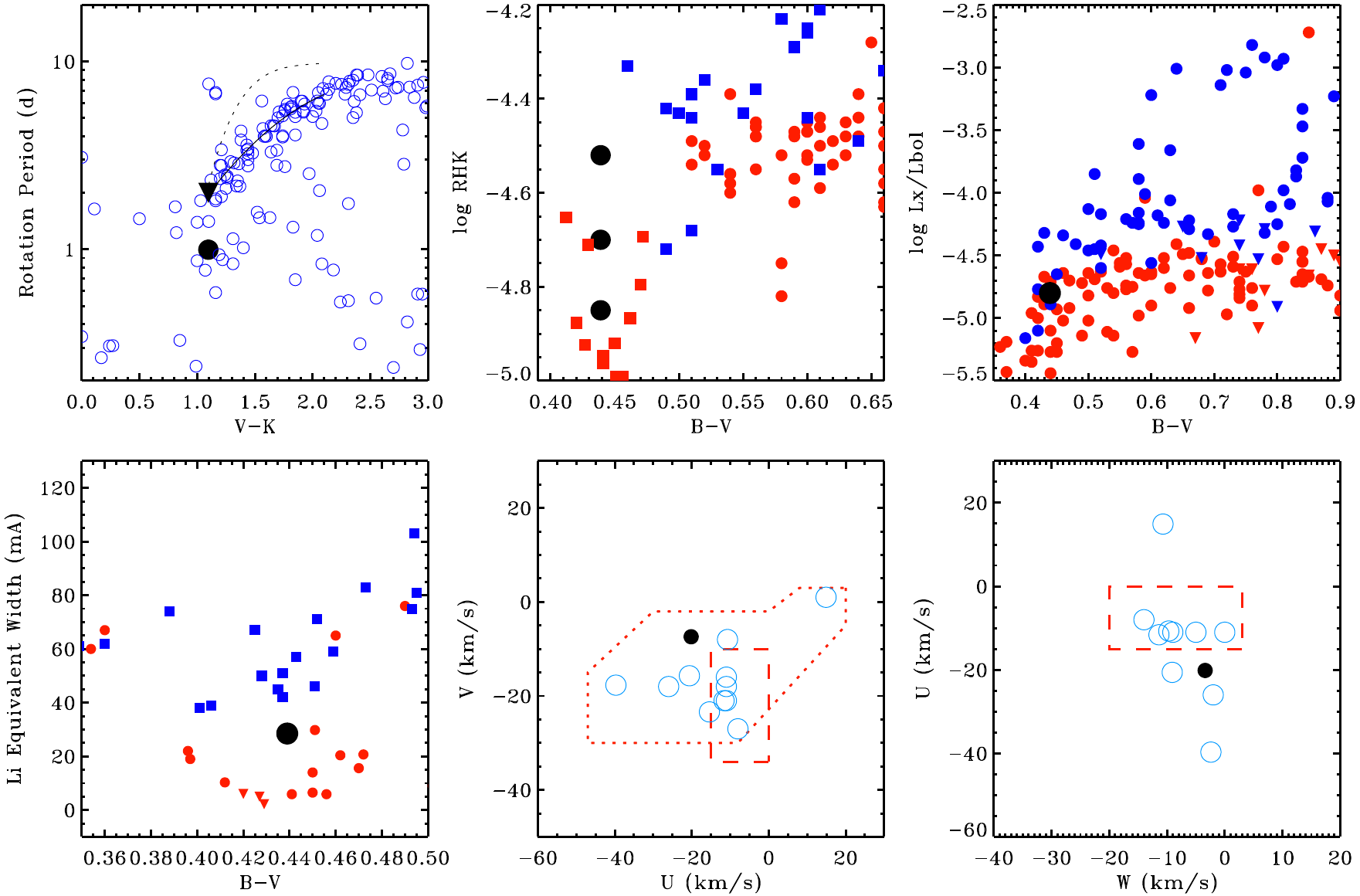}
\caption{Comparison of rotation period, chromospheric emission $\log_{RHK}$, coronal emission
$\log L_{X}/L_{bol}$, and Li EW of HD~206893 to those
of the Hyades (red symbols) and Pleiades (blue symbols) open clusters. 
In the first panel (rotation period), the filled circle shows the photometric period and the triangle the upper limit derived from the projected rotational velocity.
In the same panel the continuous and dashed lines are the sequences of slow rotators
for Pleiades and Praesepe, respectively \citep{Stauffer16}. They overlap at 
the colour of \host. The last two panels show 
the kinematic parameters of \host and those of several young moving groups. In these panels, the dashed  box represent the locus of the 
Nearby Young Population as defined in \citet{Zuckerman.2004}. 
The dotted contours in the U vs V plot mark the locus of kinematically 
young stars, as proposed by \citet{Montes.2001}.
} \label{f:param}
\end{figure*}
\end{center}

\subsection{Rotation}
\subsubsection{Photometric periodicity}

Photometric observations of HD~206893 were gathered between  October 29 and December 5, 2016, with a 40 cm (f6.3) telescope located at the Remote Observatory Atacama Desert.
%(ROAD) in the Atacama Desert close to the town of San Pedro de Atacama, Chile  (2450\,m a.s.l.)
%and equipped with a 4K$\times$4K pixel FLI ML16803 CCD camera (9\,$\mu$m pixel size) with  a  40$^{\prime}\times$40$^{\prime}$ field of view and BVI and Clear filters.
The star was observed for  34 almost consecutive nights spanning a time interval of 38 days.
We collected a total of 452 frames in B filter and 462 frames in V filter (generally
three telescope pointings per night, separated by $\sim$1.5\,hr 
with five consecutive frames in each filter on each pointing, which were averaged
to get three average magnitudes per night and corresponding uncertainties). We used the tasks 
within IRAF\footnote{IRAF is distributed by the National Optical Astronomy Observatory, which is operated by the Association of the Universities for Research in Astronomy (AURA), Inc., under cooperative agreement with the National Science Foundation.}  for bias correction, flat fielding, and aperture photometry to extract the magnitudes of HD~206893 and of two nearby stars: HD207006 (V = 7.55\,mag, B = 8.55\,mag), which served as comparison ($c$), and HD206926 (V = 9.03\,mag; B = 10.15\,mag), which served as a reference star ($ck$). Differential magnitudes of HD~206893 ({\it v\rm}) were computed with respect to HD207006.
The photometric precision is 0.015\,mag in B filter and 0.014\,mag in V filter. Differential magnitudes of comparison minus check turned out to be constant at the level $\sigma_{B(c-ck)}$ = 0.016\,mag.
\begin{center}
\begin{figure}
\includegraphics[width=6cm,angle=90]{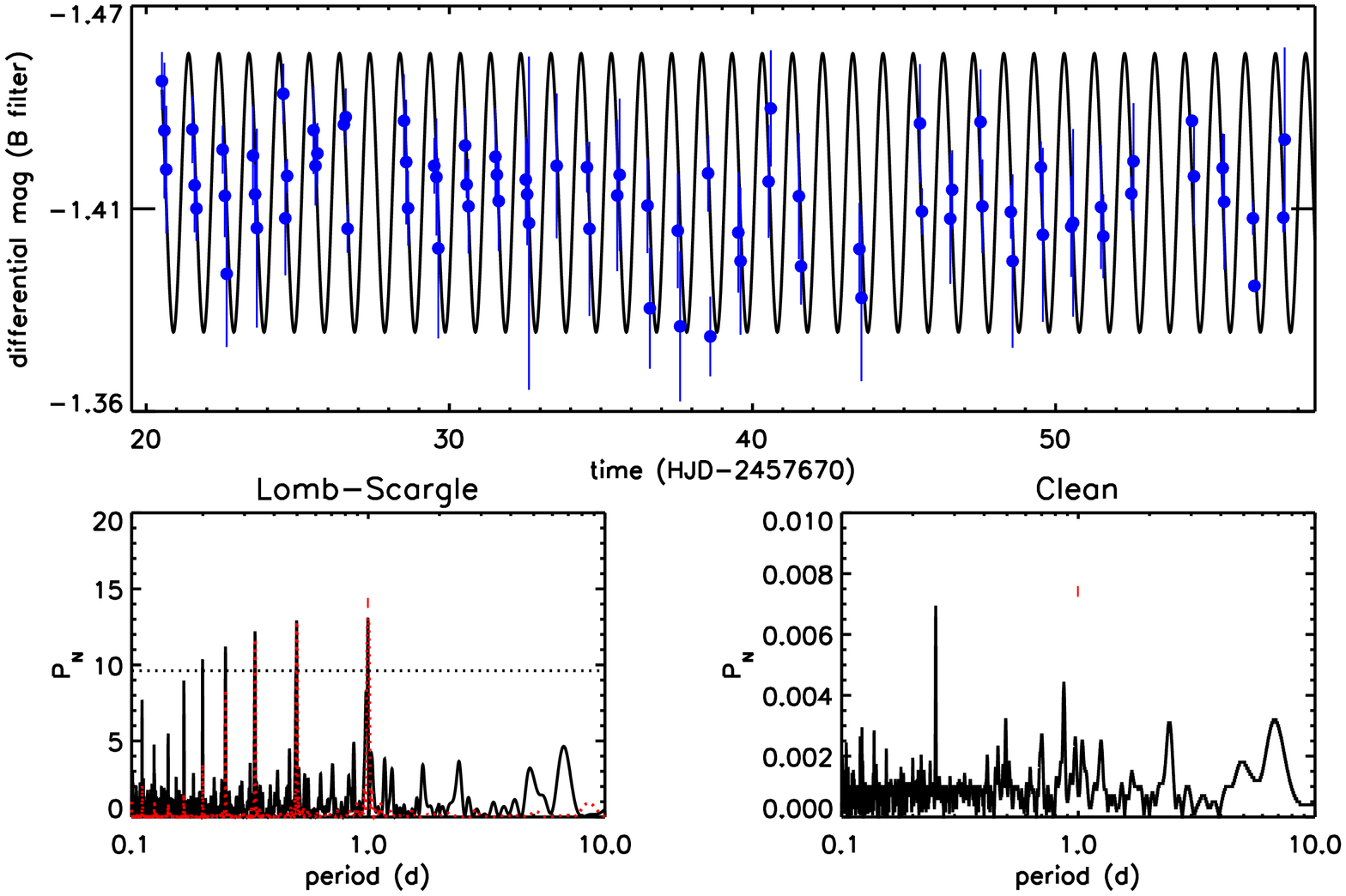}

\includegraphics[width=6cm,angle=90]{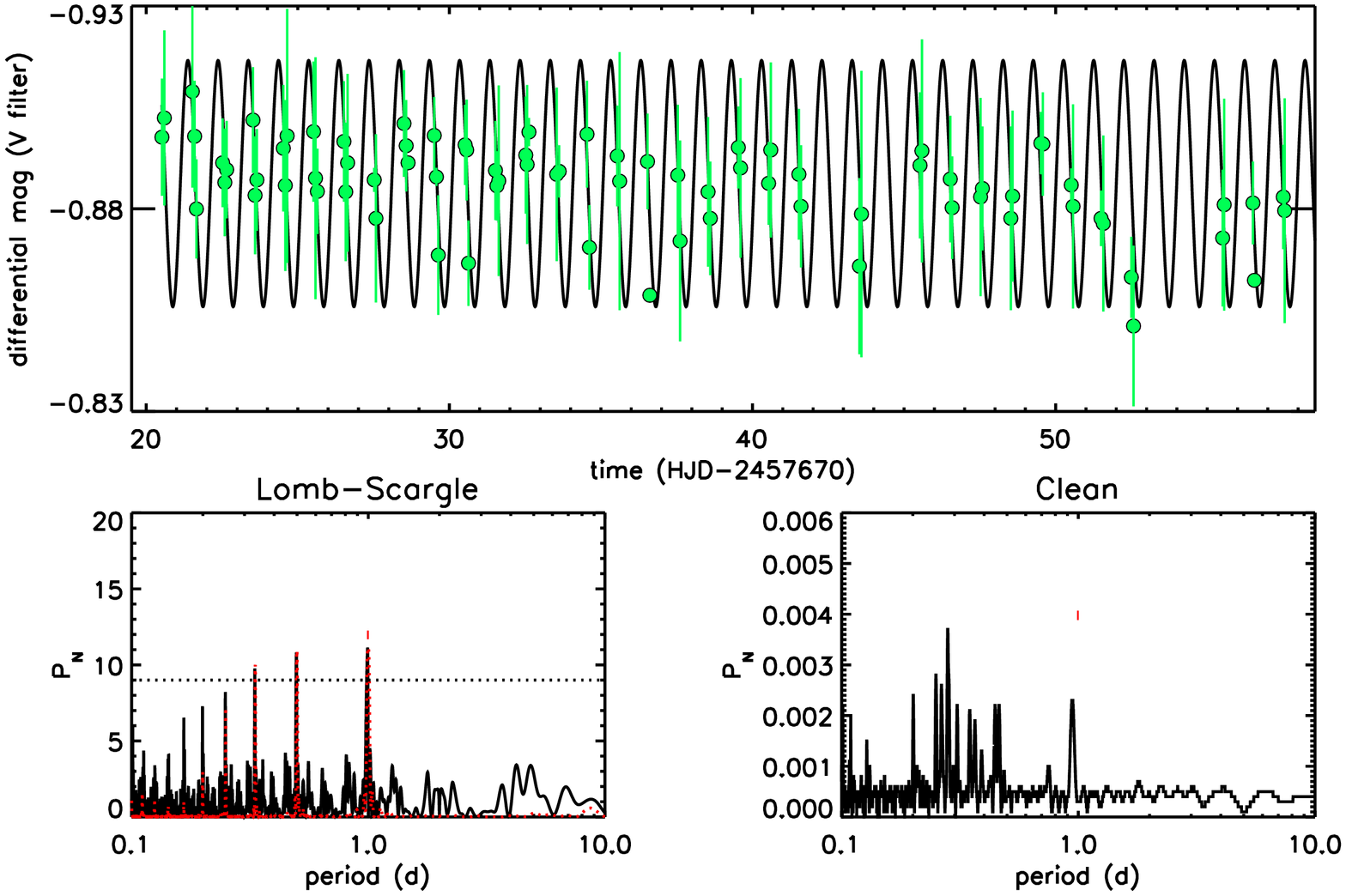}
\caption{Results of photometric analysis of HD~206893 for B (top) and V filter (bottom) data. We plot the differential magnitude time series, the Lomb-Scargle periodogram (red line is the window function, dotted line the power at 99\% confidence level), the Clean periodogram, and the light curve with the sinusoid fit overplotted.} \label{lightcurve}
\end{figure}
\end{center}
We performed Lomb-Scargle \citep{Scargle82} and CLEAN \citep{Roberts87} periodogram analysis on the
average data for the B time series and the V time series.
The results are summarised in Fig.\,\ref{lightcurve}. We find the same period
P = 0.996$\pm$0.003\,d
in the B and V filters, and light curve amplitudes $\Delta$B = 0.07\,mag and $\Delta$V = 0.05\,mag.
The period is highly significant (confidence level $>$ 99\%) in both  filters.
The detection of a rotation period very close to one day is very challenging
since it is very close to the observation sampling. The spectral window significantly
powers periods at one day and its harmonics.

However, as explained in detail in Appendix \ref{app:rot}, we can reasonably infer that the P=0.996\,d signal has a stellar origin and is most likely linked to the rotation of the star.

\subsubsection{Implications for system age and orientation}

The measured rotation period of HD~206893 is clearly below the `interface' \citep{barnes07}
sequence of the Pleaides as obtained by K2 observations \citep{Stauffer16}, suggesting
an age younger than about 130\,Myr. Our RV measurements do not show indications of the presence of close companions altering the star's angular momentum, meaning than the angular momentum evolution of \obj should be similar to what is expected for single stars.
However, fast rotators (P$\la$2\,d) belonging
to the `convective' sequence are also observed at older ages.
Stars with similar rotation period and  colour to HD~206893
can be seen in clusters older than the Pleiades, for example, in M35 (\citealt{meibom09})
and NGC 6811 (\citealt{meibom11}). %ages 220 and 1 Gyr respectively.
Therefore, we conclude that HD~206893 is too close to the boundary
at which rotation is not a sensitive age indicator (see overlap
of Pleiades and Praesepe sequences of slow rotators at the colour of our target;
Fig.~\ref{f:param}).
Nevertheless, the fast rotation most likely suggests a moderately young age,
considering the observed tendency toward slower rotation
of stars with debris discs with respect to coeval disc-less stars in some associations (Messina et al. 2017; submitted).

By coupling the stellar radius of 1.26$\pm$0.02\,$R_{\odot}$ (see Sect. \ref{s:age}) with the measured
$v \sin i$ and $P_{rot}$, we infer a system inclination of 30$\pm$5$^{\circ}$, not far
from the estimated inclination of the disc \citep[40$\pm$10$^{\circ}$]{Milli.2017}.

\subsection{Coronal and chromospheric emission}

The X-ray flux from the star was detected by ROSAT \citep{rosat-bright} 
yielding values of % $L_{X}$ = xxx and  
$\log L_{X} / L_{bol} = -4.80$.
This is compatible with both Pleiades and Hyades stars, whose distributions slightly 
overlap in this colour range.

The chromospheric flux $\log R_{HK}$ was measured from FEROS and HARPS spectra 
yielding a value of $-4.85$ and $-4.70$, respectively, using the calibration 
by \citet{desidera15} for FEROS and a similar one for 
HARPS\footnote{A much larger value of $\log R_{HK}$, $-4.466$, was derived 
by \citet{gray06}. 
While temporal variability cannot be ruled out, the most likely explanation is
a measurement effect due to the low spectral resolution of the \citet{gray06} data
(3\AA~~vs the 1\AA~~window of the M. Wilson system). 
Indeed, a $\log R_{HK}$ discrepancy between  Gray et al.  and other studies was
already noticed for F-type stars, characterised by shallow H \& K absorption features
and thus more vulnerable to a lower spectral resolution \citep{desidera15}.}.

Due to the paucity of comparison stars with known age at the colour of our target, we retrieved from the 
ESO archive the spectra of a dozen of Hyades F stars observed with HARPS (B$-$V between 
0.41 to 0.47).
%{\bf if needed we can have a dedicated appendix with the details of 
%measurements of Hyades stars}
A median value of $\log R_{HK}$ = $-$4.92 was obtained for these stars with a median colour B$-$V = 0.45.
Therefore, HD~206893 appears to be slightly more active than the average of Hyades of the same colour,
although within the scatter of individual objects.

%\subsection{Isochrone age}
%\label{s:isoc}
%
%Using the adopted parameters (see Table \ref{t:param}), the resulting
%isochrone age, derived using the PARAM web interface \citep{param} and \citet{bressan12}
%stellar models, is 877$\pm$779 Myr.
%This range is mostly consistent with previous investigations \citep[e.g., ][]{casagrande11} but broader  
%than the results of other methods.
%The resulting stellar mass and radius from the same models are 1.288$\pm$0.031 $M_{\odot}$
%and 1.289$\pm$0.031 $R_{\odot}$.
%When limiting the allowed ages to the range derived by lithium and other indicators,
%the stellar mass and radius from the same models result
% 1.32$\pm$0.02 $M_{\odot}$ and 1.26$\pm$0.02 $R_{\odot}$, respectively.

\subsection{Age summary}
\label{s:age}

The stellar properties of HD~206893 confirm the difficulty in obtaining
high-precision ages for mid-F stars.
The placement on the colour-magnitude diagram is close to the zero age main sequence  and compatible with a
broad range of ages, up to well above 1 Gyr \citep[][ and references therein]{Milli.2017}.

A non-negligible probability of membership in the Argus MG is obtained
but the lithium and barium  abundances point to an older age,
making membership unlikely. 
%It should also be considered that the Argus MG is not as well-defined as%other ones; \cite{bell15} argue that some fraction of the stars are not actually members 
%of a coeval population.

The tightest age constraint is obtained by lithium, which falls
between the observed values for Pleiades and Hyades.
Coronal and chromospheric activity support this estimate, but are also compatible with
a broader range of ages.
The observed rotation period (both the tentative detection and the firm upper limit
from $v\sin{i}$) are also compatible with these values, once considering the
presence of F-type fast rotators at both young and moderately old age.
We  adopt in the following an age of $250^{+450}_{-200}$ Myr, where the error bars
should be considered as very conservative (max allowed range), including the age of
Argus MG as a lower limit.
Adopting this age range,
the stellar mass and radius from \citet{bressan12}  and the PARAM web interface 
\citep{param} result in 1.32$\pm$0.02 $M_{\odot}$ and 1.26$\pm$0.02 $R_{\odot}$, respectively.

\begin{center}
\begin{table}
\caption{Stellar parameters of HD~206893.}\label{t:param}
\begin{tabular}{lcl}
\hline\hline
Parameter      & Value  & Ref \\
\hline
V (mag)                &  6.69       & Hipparcos \\
B$-$V (mag)                 &  0.439      & Hipparcos \\
V$-$I (mag)                 &  0.51       & Hipparcos \\
J (mag)                   &  5.869$\pm$0.023  & 2MASS \\
H (mag)                   &  5.687$\pm$0.034  & 2MASS \\
K (mag)                   &  5.593$\pm$0.021  & 2MASS \\
Parallax (mas)       &   24.59$\pm$0.26  & GaiaDR1 \\
$\mu_{\alpha}$ (mas\,yr$^{-1}$)  & 94.236$\pm$0.044  & GaiaDR1 \\
$\mu_{\delta}$ (mas\,yr$^{-1}$)  &  0.164$\pm$0.031  & GaiaDR1 \\
RV   (km\,s$^{-1}$)                  &  -11.92$\pm$0.32  & this paper \\
U (km\,s$^{-1}$)                   &  -20.15$\pm$0.23 & this paper \\
V   (km\,s$^{-1}$)                   &  -7.40$\pm$0.15, & this paper \\
W   (km\,s$^{-1}$)                   & -3.40$\pm$0.26 & this paper \\
$T_{\rm eff}$ (K)      &  6500$\pm$100 & this paper \\
$\log g$             &   4.45$\pm$0.15 & this paper \\
$\rm [Fe/H]$               &   +0.04$\pm$0.02 & this paper \\
EW Li (m\AA)         &   28.5$\pm$7.0 & this paper \\
A(Li)                &   2.38$\pm$0.10 & this paper \\
$\rm [Ba/Fe]$              &   0.20$\pm$0.20 & this paper \\
$v \sin i $  (km\,s$^{-1}$)          &   32$\pm$2 & this paper \\
$\log L_{X}/L_{bol} $  &   -4.80$\pm$ & this paper \\
$\log R_{HK}$         &   -4.77$\pm$ & this paper \\
$P_{rot}$ (d)             &   0.996$\pm$0.003 & this paper \\
Age (Myr)            &   $250^{+450}_{-200}$ & this paper \\
$M_{star} (M_{\odot})$   &  1.32$\pm$0.02 & this paper \\
$R_{star} (R_{\odot})$   &  1.26$\pm$0.02 & this paper \\
$i$ (deg)   &     30$\pm$5     & this paper \\
\hline\hline
\end{tabular}
\end{table}
\end{center}

\section{New SPHERE observations of \obj }\label{observations}  
  We observed \obj on September 16, 2016, with VLT/SPHERE as part of the Guaranteed
Time Observations dedicated to exoplanet search
(SpHere INfrared survey for Exoplanets or SHINE, Chauvin
et al. in prep.). These observations took advantage of the extreme adaptive optics system SAXO \citep{Fusco.2006,Sauvage.2016} and used the Apodised Lyot Coronograph \citep{Carbillet.2011,Guerri.2011}. We selected the IRDIFS\_EXT observing mode, with which the Integral Field Spectrograph \citep[IFS;][]{Claudi.2008} provides a spectral resolution of R=30 in
the wavelength range between 0.95\mic and 1.63\mic on
a field of view of 1.7\as x 1.7\as, and  Infra-Red Dual-beam
Imager and Spectrograph \citep[IRDIS;][]{Dohlen.2008}
simultaneously operating in the Dual-Band Imaging \citep[DBI;][]{Vigan.2010} mode in the $K$ band with the $K12$ filter
pair. The $K1$ filter is centred at  2.110\mic and the $K2$ filter at 2.251\mic. The observations lasted from 02:08:05 to 03:49:14 UTC under stable conditions and very good seeing varying between 0.4 and 0.5\as in the NIR and an atmospheric coherence time of 4 to 5~ms. During this sequence 80 IFS and IRDIS science frames with 64s exposure time were acquired. The total parallactic angle variation during the science observations was 76\,deg. The observing sequence also included two observations using the waffle mode active (a small fixed waffle pattern was applied to the deformable mirror to create four echoes of the PSF) for an accurate determination of the stellar position behind the coronagraphic mask. We also obtained unsaturated observations without coronograph, using short exposure times and a neutral density filter to provide an accurate flux calibration.

  \subsection{Data reduction \label{reduc}}
 Our data was reduced by the SPHERE Data Centre (hereafter SPHERE-DC) \footnote{http://sphere.osug.fr/spip.php?rubrique16\&lang=en}. For IRDIS data, the first reductions steps (dark, flat, and bad pixel correction) rely on the SPHERE Data Reduction and Handling (DRH) pipeline \cite{Pavlov.2008} provided by ESO. For IFS data the SPHERE-DC complements the DRH pipeline with additional steps that improve the wavelength calibration and the cross-talk correction \citep{Mesa.2015}. For both IRDIS and IFS data, the SPHERE-DC also provides a frame-by-frame accurate  determination of the parallactic angle for our pupil tracking observations.
%An automatic sorting In the case of our observations the conditions were very good and stable, and no frame was removed by the automated frame sorting. 

To perform the astrometric calibration of the IRDIS and IFS dataset on sky, we used the tools described by \citet{Maire.2016astrom}. The astrometric field 47\,Tuc was observed on the same night as the science observations, and the data aligned to the true north by applying a rotation of -1.762$^\circ$. The uncertainty on the true north calibration is 0.048$^\circ$.  \\

The resulting reduced master cubes were then used as input of several distinct Angular DIfferential \citep[ADI][]{Marois.2006,Lafreniere.2007} algorithms. We detail here the one we retained for the spectral analysis, a dedicated algorithm based on the TLOCI-ADI \citep{Marois.2010} routines given in \citet{Galicher.2011}. This TLOCI-ADI routine has been tested by the SHINE consortium among many other reduction algorithms and  is shown to  provide extracted spectra with the best agreement with the known injected spectra of fake planets, within its estimated errors bars (Galicher et al., in prep). The resulting spectra is shown in Fig. \ref{spectra}.
The contrast within the first eight channels of the IFS data with respect to the host star is too high to indisputably identify the companion. We thus include only upper limits for our final TLOCI reduction flux up to 1.07 $\mu$m in wavelength. Our final error estimation includes fitting errors, PSF, and atmospheric variation errors as well as an estimation of the speckle noise within a corona of 1 FWHM width located at the separation of the companion. At the low separation of the companion, this type of noise dominates the error statistics in the blue part of the spectrum up to about 1.25 $\mu$m.

 The extracted spectrum of  \obj after the TLOCI-ADI analysis is presented in Fig. \ref{spectra}. We also applied other independent reduction and analysis pipelines, and checked that the spectrum of \obj extracted with these other methods was consistent, within error bars, with our TLOCI-ADI reduction. This comparison is detailed in Appendix.\ref{app:otherred}

\begin{center}
\begin{figure*}
\includegraphics[width=18cm]{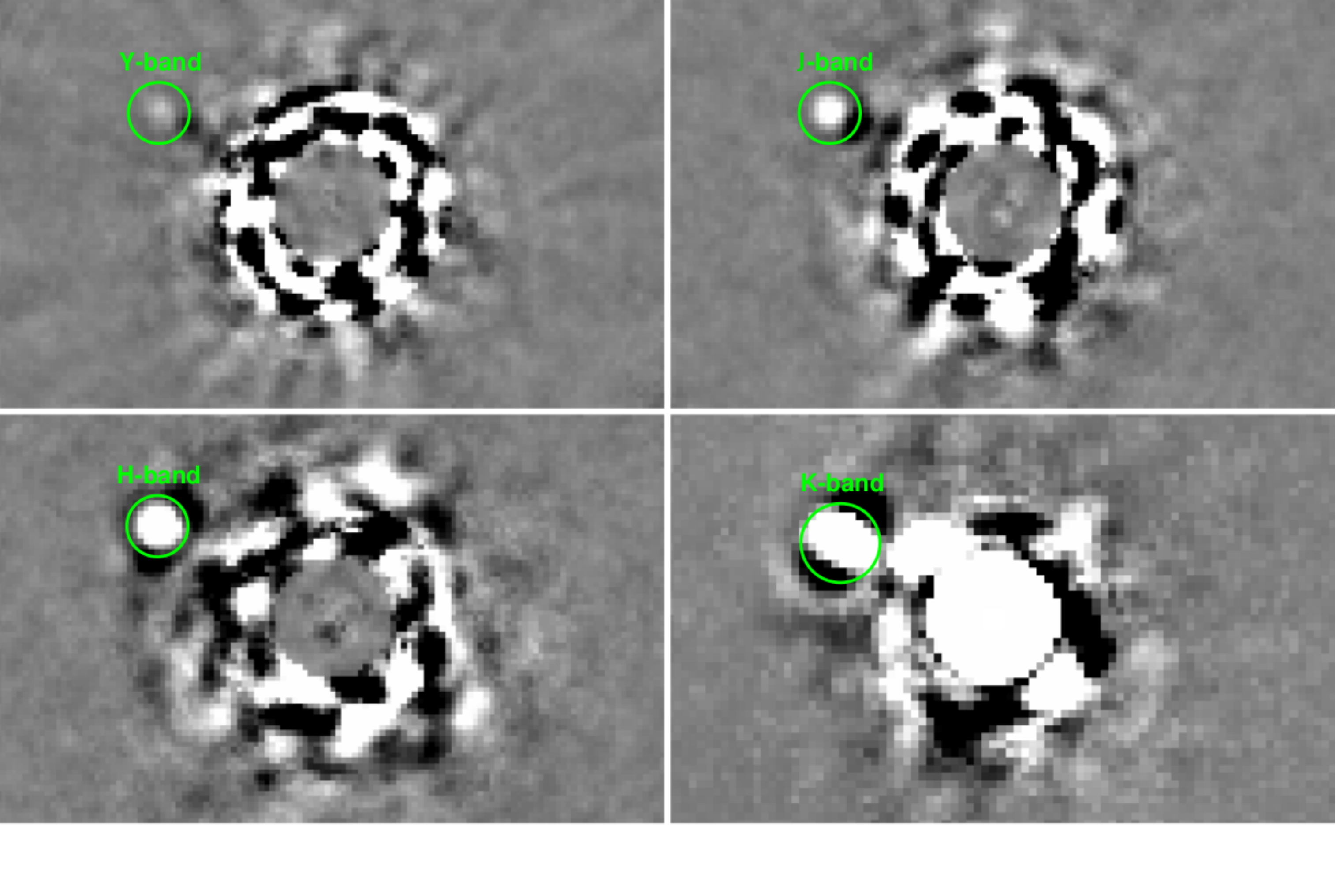}
 \caption{Residual images of \obj in $Y$, $J$, $H$, and $K$ bands in linear scale.  The scale is identical in $Y$, $J$, and $H$, but is 4 times wider in $K$ to accommodate  the brighter speckle residuals and brighter companion in the IRDIS $K$-band image .\label{resimages}}
\end{figure*}
\end{center}

\begin{figure}
\includegraphics[width=8.4cm]{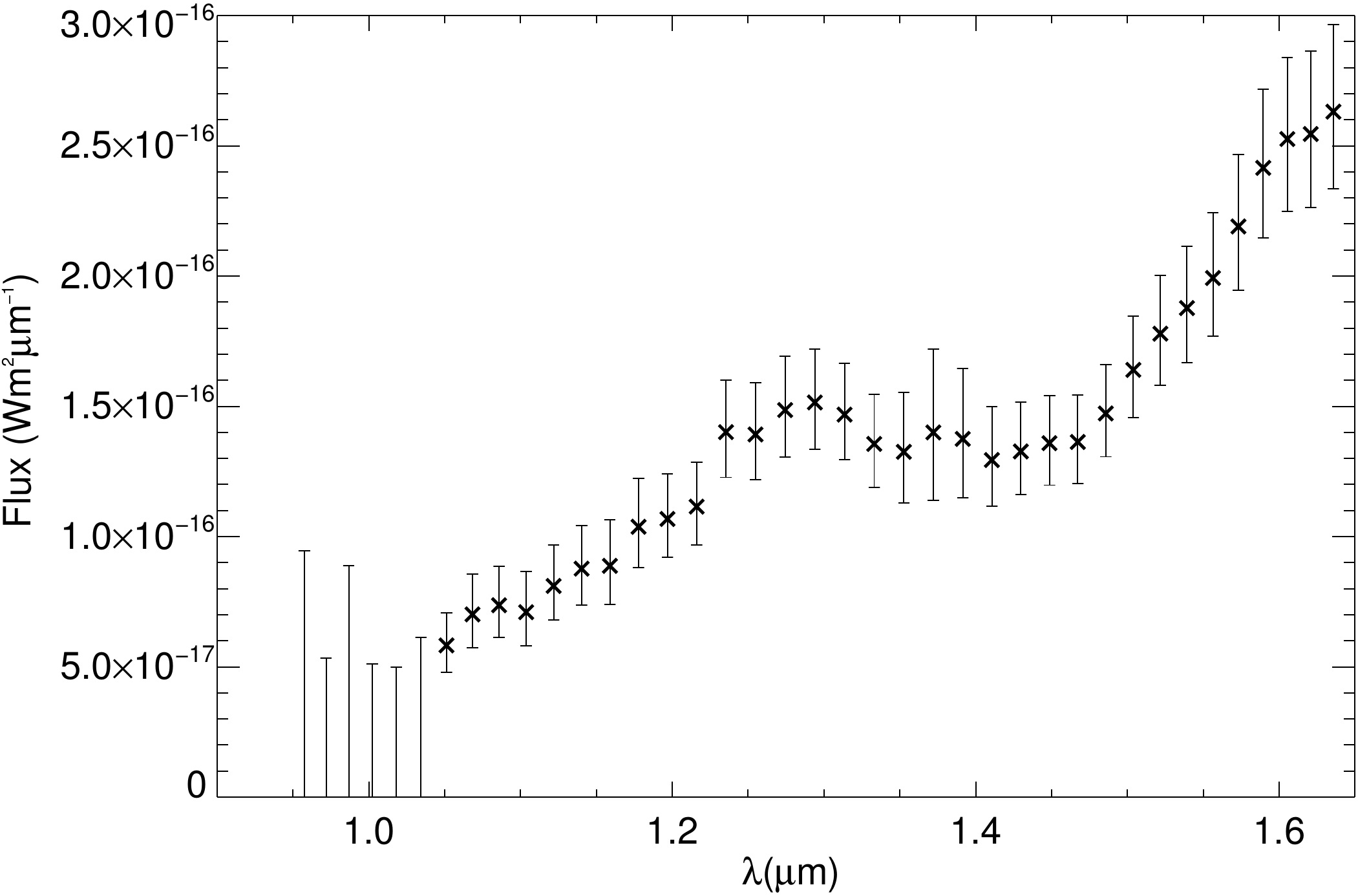}
 \caption{Extracted spectrum of \obj through TLOCI-ADI\label{spectra}}
\end{figure}

  \subsection{Constraints on additional companions in the system}
\begin{figure}
\begin{center}
\includegraphics[width=8.4cm]{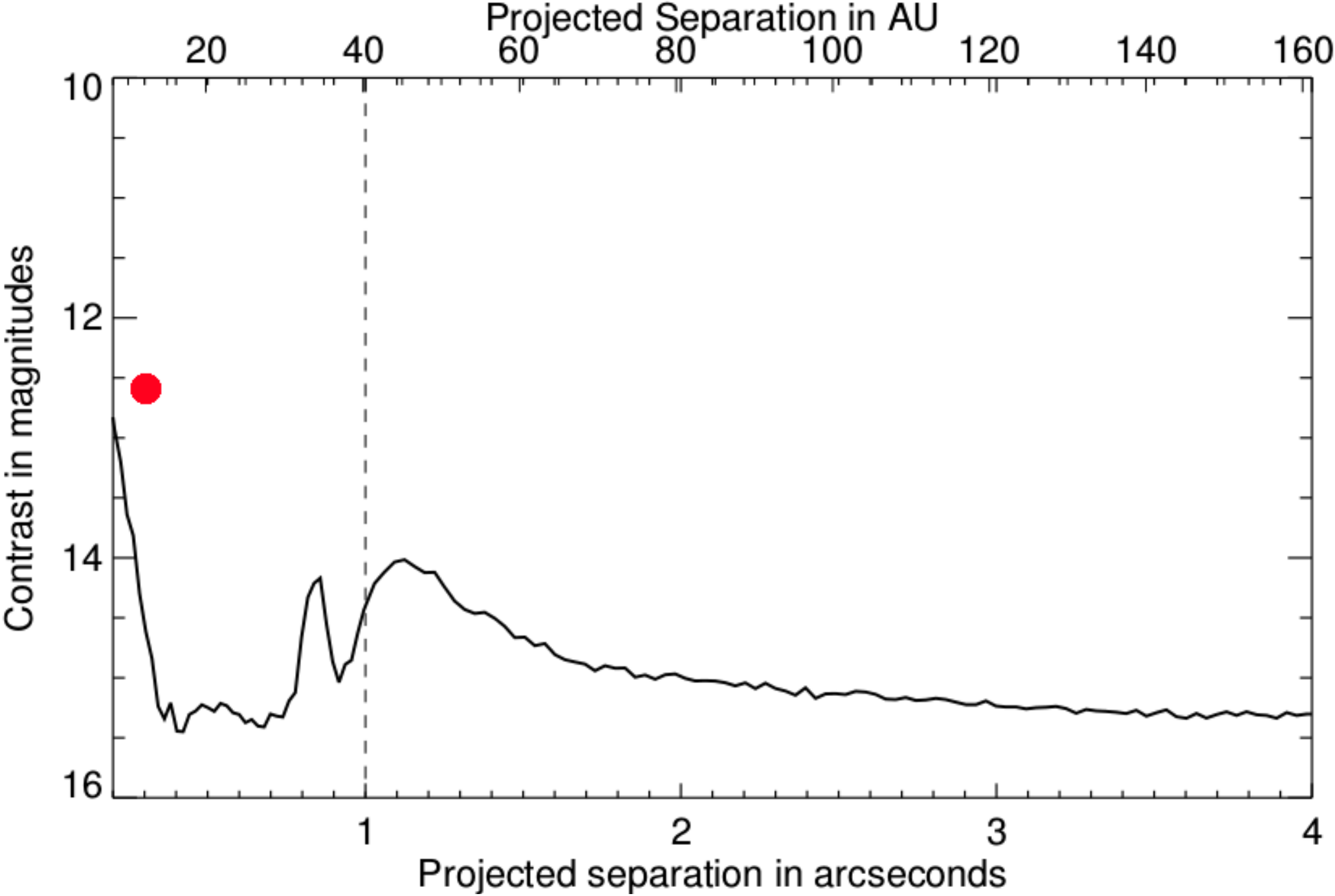}\caption{Five $\sigma$ detection limits combining IFS and IRDIS information. Left of the dashed line, IFS sensitivity is best and we show IFS TLOCI-ASDI sensitivity using the mean of the channels covering $Y$, $J$, and $H$ band. Right of the dashed line, we show IRDIS sensitivity after TLOCI-ADI reduction, using the mean of the $K1$ and $K2$ images. The red dot indicates \obj. }
\label{detlim}
\end{center}
\end{figure}

  Our TLOCI-ADI analysis was dedicated to extract the spectrum of the companion and in order to limit the biases we used only the angular information (ADI analysis) contained in the observation to remove the speckle pattern. As shown by e.g. \citet{Rameau.2015}, using the spectral information (SDI) in addition to the angular information makes it very difficult to extract an unbiased spectrum of a detected companion. However, the same study points out that the speckle pattern is very efficiently removed when using the spectral information. Our own tests within the SHINE Consortium also show that the best spectral fidelity, i.e. the ability to retrieve, within the error bars, the spectrum of an injected fake companion is best when using only ADI, while the best point-source sensitivity is achieved by combining ADI and SDI. When we probed the possible presence of additional companions in the system we used TLOCI and PCA using this combined `ASDI' approach with IFS data. Figure \ref{resimages} shows the resulting residuals maps in the $Y$, $J$, $H$, and $K$ bands. The $Y$, $J$, and $H$ are a stack of the IFS channels after TLOCI-ASDI reduction that have a wavelength corresponding to each photometric band, while the $K$-band image is a stack of the IRDIS $K1$ and $K2$ filters after TLOCI-ADI reduction. No companion was detected even though our detection limits reach respectively 14.5 and 15.2 magnitudes of contrast for separations of 0.3 and 0.5\arcsec\ (see Fig. \ref{detlim}). These detection limits are computed for a companion with a uniform contrast and are hence conservative because the deep absorption bands in the spectra of real substellar companions slightly improve the sensitivity of spectral differential imaging methods \cite[see e.g.][]{Rameau.2015}. For an age of 300~Myr (50~Myr) these limits correspond to detection limits in mass, assuming \citet{Baraffe.2003} \textit{hot-start} evolution models, of respectively 6 and 8~\Mjup (2 and 3~\Mjup). 

\subsection{No disc re-detection in SHINE DATA}
   We initially believed that our $K$-band IRDIS SHINE data showed a re-detection of the faint structure highlighted as a probable disc detection in the $H$ broad-band IRDIS images analysed by \citet{Milli.2017}. However, given the faintness of this signal, we checked whether the observed extended structure was indeed an astrophysical signal or an observational bias. Instead of de-rotating our sequence of images according to the actual parallactic angle of the observations, we applied a de-rotation in the opposite direction, which should effectively kill any  astrophysical signal. Since the same fuzzy extended structure was still visible, it meant it was caused by an observational bias, probably the wind extending the coronographic halo along a privileged direction. As a sanity check, we also applied this procedure to the $H$-band data analysed by \citet{Milli.2017}, and we show that as expected for an astrophysical disc signal, applying an incorrect  de-rotation angle destroys the signal in this case.  We therefore confirm that \citet{Milli.2017}  have probably detected a faint disc structure at large separation in their SPHERE broad-band data, but we have no independent confirmation of this signal from our new observations. We note that this is not unexpected because the higher exposure time and larger bandwidth of the \citet{Milli.2017} observations makes them more sensitive to disc signal at large separations.

  \subsection{Conversion to fluxes}
\label{sec:convf}
\begin{figure}
\begin{center}
\includegraphics[width=8.4cm]{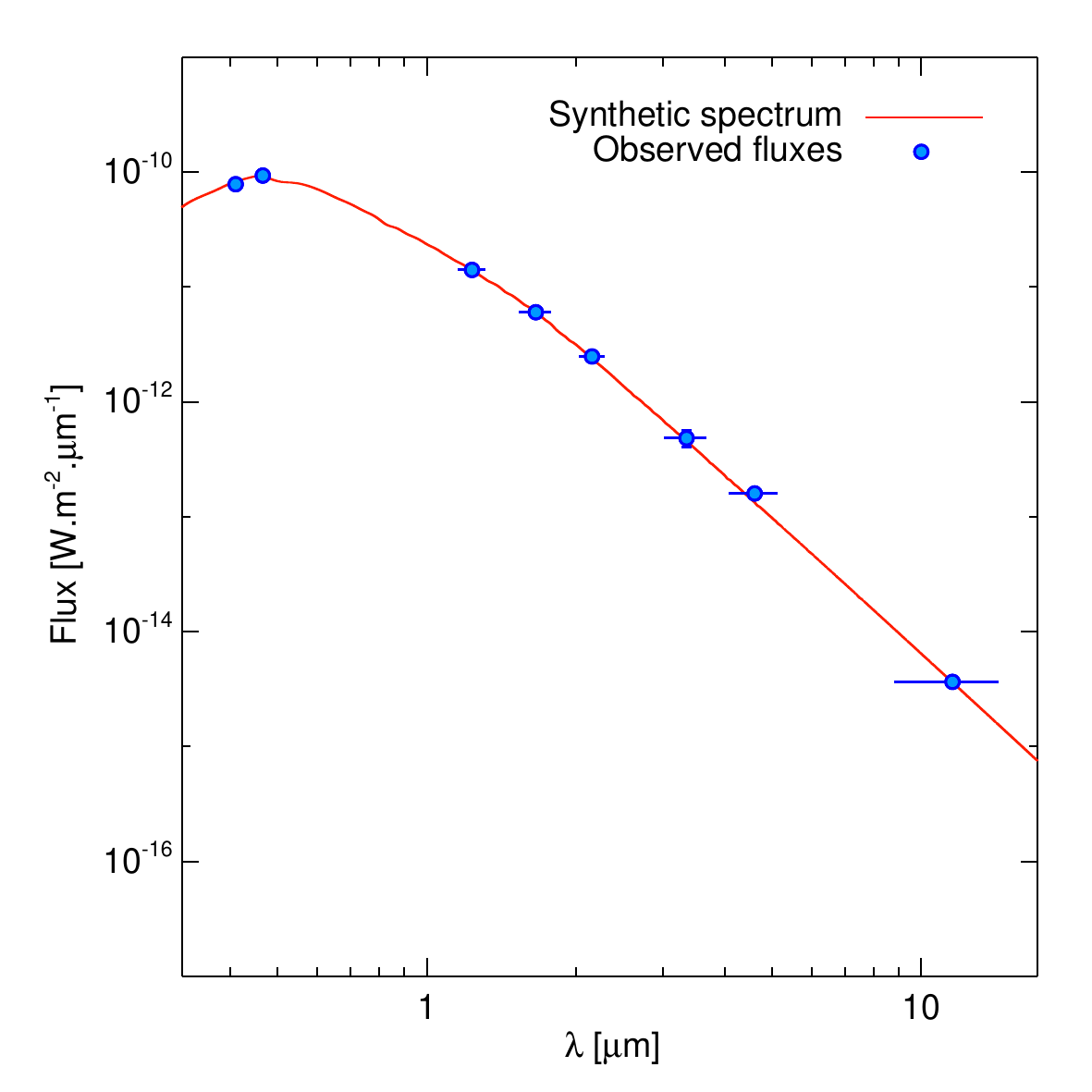}
\caption{BT-NEXTGEN synthetic spectrum (red curve) adjusted onto the star photometry (blue dots).}
\label{Fig:SEDstar}
\end{center}
\end{figure}

A prior knowledge of the star spectrum in the NIR  is needed to convert the companion contrast extracted from the SPHERE data to fluxes. To get a simple estimate of the stellar spectrum fit for our calibration purposes, we  compiled the  Stromgren b and v, 2MASS JHK$\mathrm{_{s}}$, and WISE W1, W2,  and W3-band photometry \citep{2015A&A...580A..23P, 2003tmc..book.....C, 2012yCat.2311....0C}.  We converted this photometry to apparent fluxes with the VOSA tool \citep{2008A&A...492..277B}. We then smoothed a BT-NEXTGEN synthetic spectrum \citep{2012RSPTA.370.2765A} with $\mathrm{T_{eff}=6600}$K,  log g=4.5 dex, and M/H=0.0 dex to the IFS spectral resolution (R$\sim$30), and ajusted its flux onto the star fluxes (Fig. \ref{Fig:SEDstar}).  The $\mathrm{T_{eff}}$, \logg, and metallicity are very close to the values found in the previous section.

We used this spectrum to  compute  the average stellar flux at the IFS and IRDIS wavelengths. To do so, we took  the mean of the stellar flux weighted  by the response curves of the IFS channels (Gaussian transmission), and of the K1 and K2 passbands.  The weighting also accounts for the atmospheric transmission using a model of the telluric absorptions generated with the Cerro Paranal Sky Model \citep{2012A&A...543A..92N, 2013A&A...560A..91J}. We repeated this procedure for the L' band photometry of HD~206893b gathered by \cite{Milli.2017}. The resulting photometry is shown in Table \ref{photom}

\begin{table}
\caption{Photometry of \obj. \label{photom}}
\begin{tabular}{l|c} \hline 
Filter& Magnitude  \\ \hline \hline
$Js^1$&  18.33 $\pm$ 0.17  \\
$H^2$ &  16.79 $\pm$ 0.06   \\
$K1^3$ &  15.2 $\pm$ 0.10 \\
$K2^3$ &  14.88 $\pm$ 0.09\\
$L'^2$ &  13.43$^{+0.17}_{-0.15}$  \\ \hline \hline
\end{tabular}
\tablefoot{ $^1$ Synthesised using IFS TLOCI-ADI spectroscopy
 $^2$ From \citet{Milli.2017}
$^3$ This article, from TLOCI-ADI analysis of IRDIS data.
}
\end{table}

\section{Atmospheric properties of \obj: a spectral analysis by comparison with reference objects}
\label{sec:refobj}

\subsection{Colour-magnitude diagrams}
\begin{figure*}
\begin{center}
\begin{tabular}{cc}
\includegraphics[width=8.4cm]{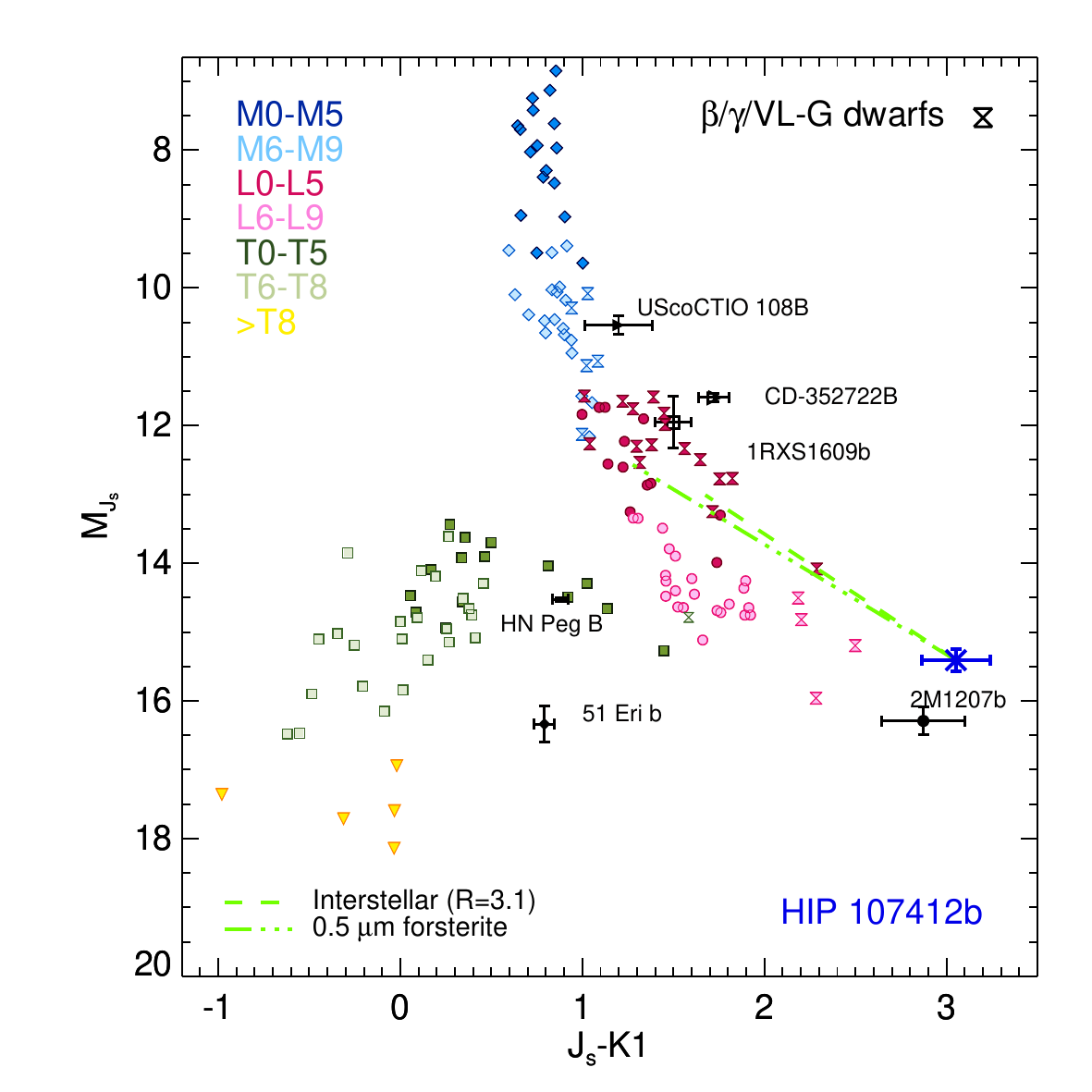} &
\includegraphics[width=8.4cm]{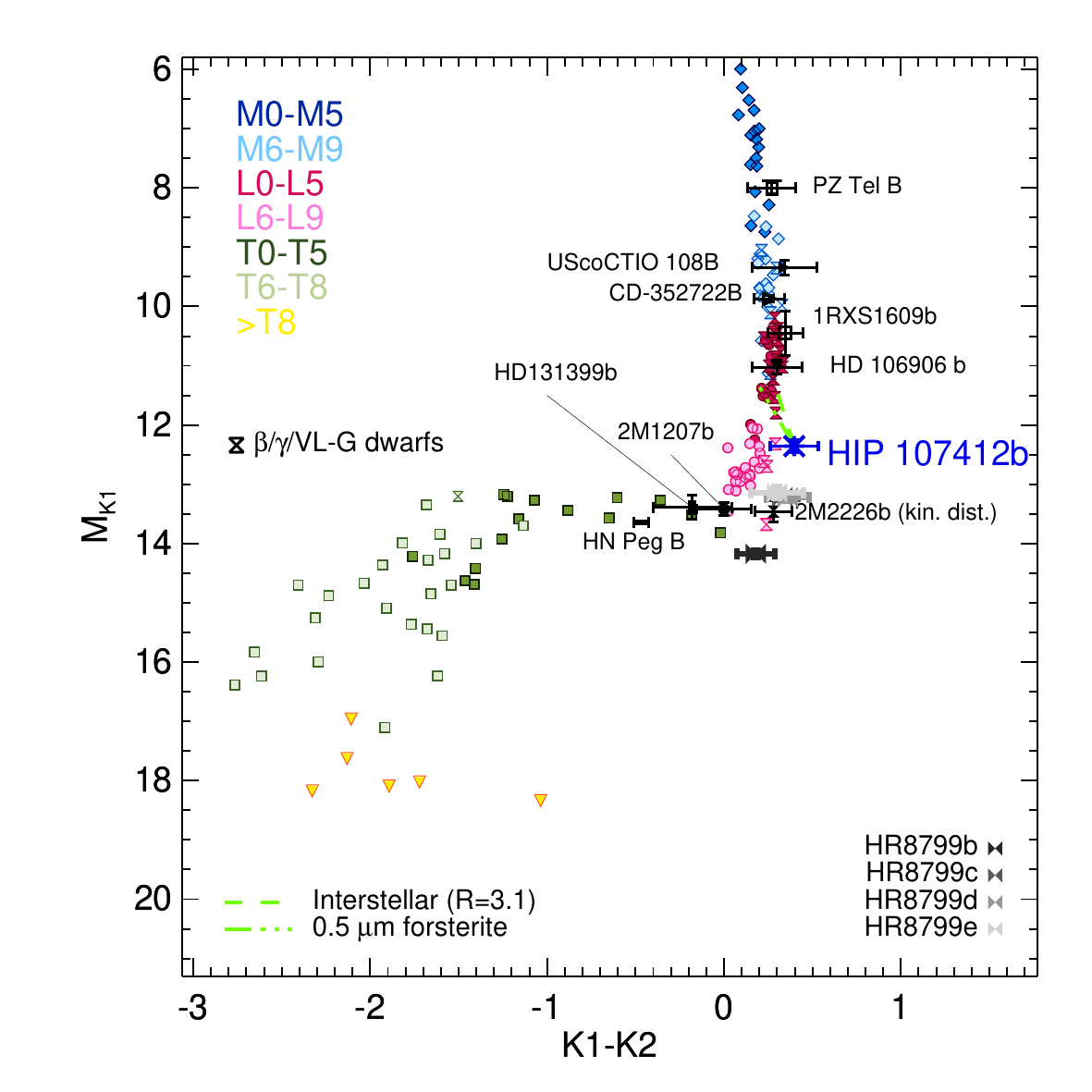} 
\end{tabular}
\caption{Placement of the IRDIS photometry of HD~206893~B (HIP107412~B) in  colour-magnitude diagrams. The dash-dotted line corresponds to the reddening vector caused by forsterite grains with mean size 0.5$\mu$m. The dashed line correspond to the interstellar reddening (reddening parameter R=3.1).}
\label{fig:CMD}
\end{center}
\end{figure*}

HD~206893~b has  extremely red H-L' colours \citep{Milli.2017}. The K1K2 photometry confirms that HD~206893b is indeed a remarkably red object. We show in Fig. \ref{fig:CMD} the location of the companion in a K-band colour-magnitude diagram  \citep[see][for details]{2016A&A...593A.119M}. The companion has a K1-K2 colour (0.40$\pm0.13$ mag) in the same range as the HR8799d planet \citep[$0.36\pm0.13$ mag,][]{Zurlo.2016}. Nonetheless, its absolute flux is consistent with those of mid-L dwarfs (both young and old). It has a slightly redder K1-K2 colour  than that of the already red L/T transition companion 2MASS J22362452+4751425 ($0.28\pm0.10$ mag\footnote{Synthetic photometry computed from the flux-calibrated K-band spectrum of the source.}) identified by \cite{Bowler.2017}. We synthesised the J-band photometry  $J_{s}=18.33\pm0.17$ mag of HD~206893b corresponding to a virtual filter with a 100\% transmission between 1.2 and 1.3 $\mu$m, and report the corresponding colour-magnitude diagram in Fig. \ref{fig:CMD}.  The companion has a J-band luminosity that is more consistent with those of late L dwarfs. It is the reddest object in that diagram, even slightly redder than the extremely red companion 2M1207b, which  is located  nearby in the magnitude-colour space. 

%The comparison of the two diagrams shows that the young, and/or low-gravity, and/or dusty objects follow a distinct sequence in the Js-K1 diagrams (in agreement with the conclusions of \cite{2016ApJ...833...96L} for the MKO photometry), while this class of object falls closer to the sequence of regular field dwarfs in the K-band diagram. This shows that young and peculiar objects  which fall inside the field of view of the IFS can already be distinguished from background stars with colors more typical of GKM dwarfs using the informations coming from the two instruments of SPHERE, while it is more difficult to make that distinction if only the IRDIS K-band data are used. 

\subsection{Spectral templates}
We compared the new SPHERE NIR spectrum and K1K2 photometry to reference low-resolution (R$\sim75-120$) spectra acquired with the SpeX instrument. The spectra were taken for the most part from the SpeXPrism library \citep[][532 objects]{2014ASInC..11....7B} and from  \cite{2015ApJ...814..118B} (122 objects). Those two libraries include spectra of peculiar dusty  L dwarfs. We used additional 1-2.5$\mu$m continuous NIR spectra of L and T dwarfs from the \cite{2013ApJS..205....6M} library (72 objects; contains some spectra of peculiar L/T dwarfs),  young M and L dwarfs of \cite{Allers.2013} (17 spectra), and of red/dusty L dwarfs later than L4 taken from the literature (see Appendix \ref{App:A}).  All spectra were smoothed to  a resolution R$\sim$30, corresponding to those of our data. We then took the weighted mean of the flux within the K1 and K2 filter passbands, and the IFS channels. We considered  $G$, the goodness-of-fit indicator  defined in \cite{2008ApJ...678.1372C}, which  accounts for the filter and spectral channel widths ( see Appendix \ref{app:goodness} for  details). 

\begin{figure*}
\begin{center}
\begin{tabular}{cc}
\includegraphics[width=8.4cm]{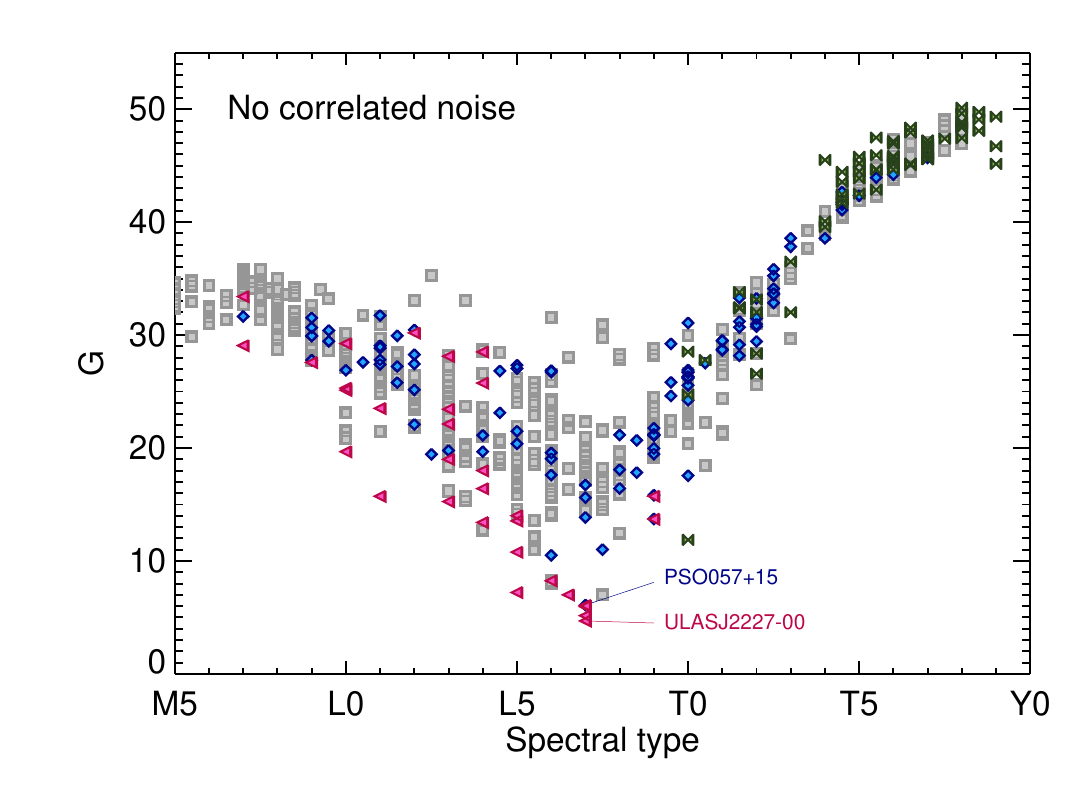} &
\includegraphics[width=8.4cm]{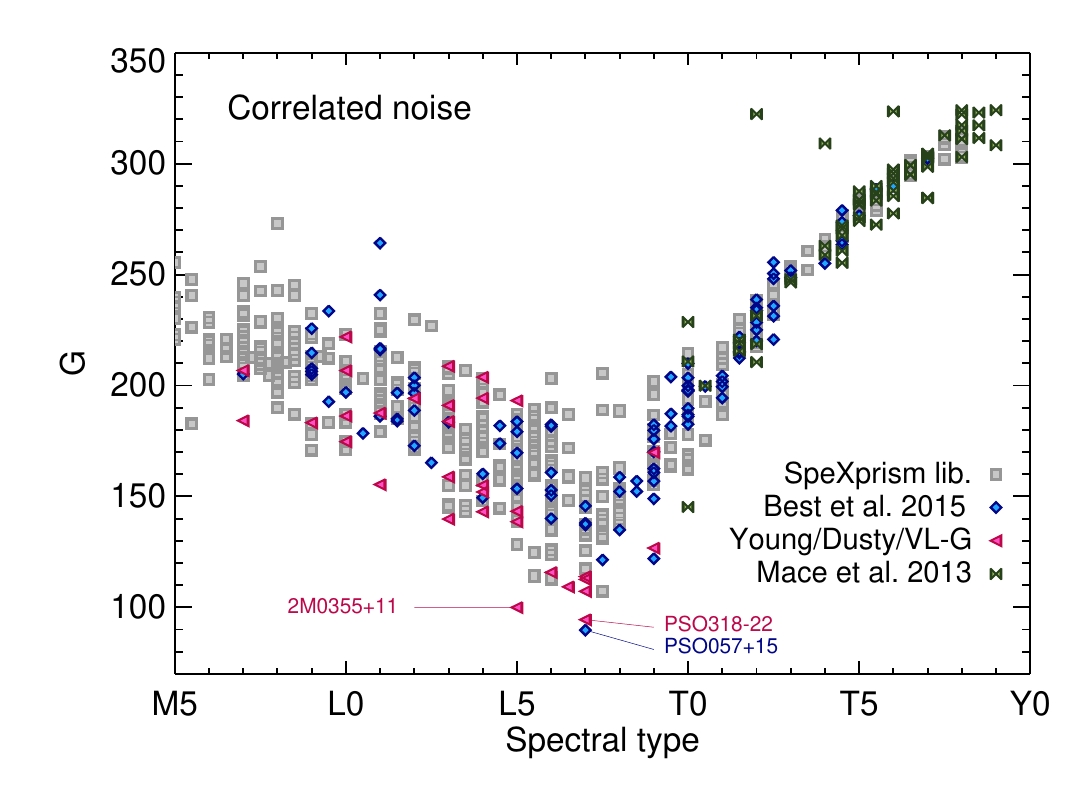}\\
\end{tabular}
\caption{Comparison of the NIR spectrum and photometry of HD~206893b to those of reference spectra. On the left the fit does not account for the correlation noise of the IFS spectrum; 
on the right, it does.}
\label{fig:GSpeX}
\end{center}
\end{figure*}

The  goodness of fit  indicates that the HD~206893b  spectral slope is better represented by those of L5 to L7 dwarfs (Fig. \ref{fig:GSpeX}). This is in agreement with the placement of the object in the Js-K1 colour-magnitude diagram (Fig. \ref{fig:CMD}).  Among all the considered objects, those of PSO J318.5338-22.8603  and PSO\_J057.2893+15.2433  \citep[][]{Liu.2013} represent  the companion spectrophotometry most closely.  These two objects are proposed members of the $\beta$ Pictoris moving group and have estimated masses in the planetary mass range ($\sim 8 M_{Jup}$). A visual inspection of the fit reveals nonetheless that the companion slope is still too red compared to that of these two objects. In addition, the water band from 1.35 to 1.5 $\mu$m is less deep in the HD~206893b spectrum.  In general, the red,  young, and/or dusty L dwarfs tend to produce lower G values than regular dwarfs for a given spectral type because of their redder spectral slope. Nevertheless,  these dusty objects also fail to reproduce the extreme red slope of the companion. The  goodness of fit  follows the same trend whether or not the correlation of noise between spectral channels is accounted for (see Appendix \ref{app:correl} for  details). This suggests that despite its important contribution to the G value,  the noise correlation is not sufficient to influence the fit.

\begin{figure}
\begin{center}
\includegraphics[width=8.4cm]{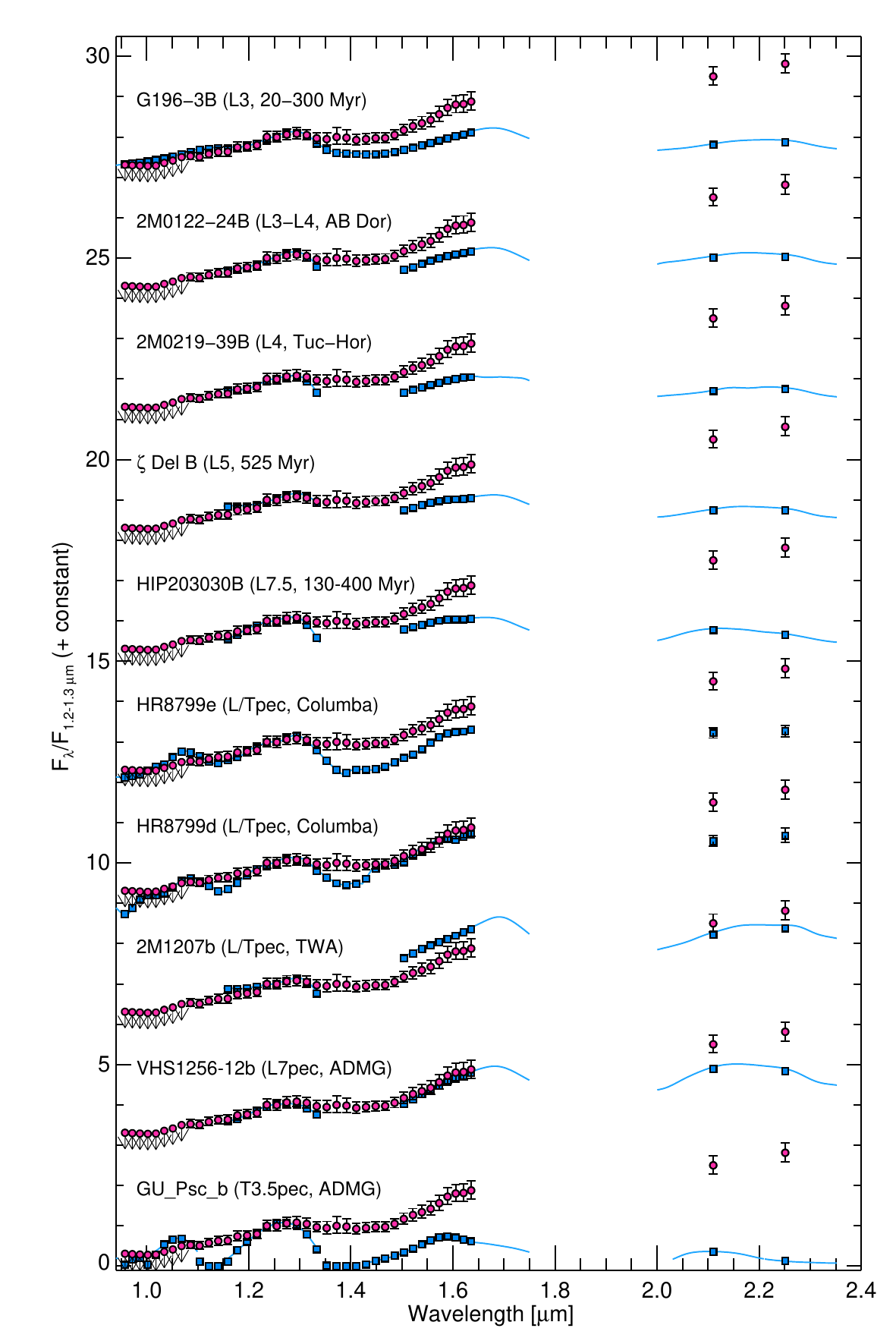}
\caption{Comparison of the spectrum  and photometry of HD~206893b (pink dots) to those of L-type companions (blue).}
\label{fig:Youngcomp}
\end{center}
\end{figure}

We visually compared the  HD~206893b spectrum to those of mid- to late-L companions (Fig. \ref{fig:Youngcomp}) taken from the literature (GU Psc b, \cite{2014ApJ...787....5N}; VHS J125601.92-125723.9 ABb, \cite{Gauza.2015};  2M1207b, \cite{2010A&A...517A..76P}; HR8799d and e, \cite{Zurlo.2016}; HIP203030B, Bonnefoy et al.  in prep; $\zeta$ Del B, \cite{2014MNRAS.445.3694D}; 2M0219-39b, \cite{2015ApJ...806..254A}; 2M0122-24B, \cite{2015ApJ...805L..10H}; G196-3B, \cite{1998Sci...282.1309R}). Unlike the dusty dwarfs, the primary stars enable more robust estimates of the age of the companion. The slope is best reproduced by those of young companions (age $\leq$ 150 Myr)   at the L/T transition, although still redder than the templates. The fact that the best fits for \obj are obtained with young companions hints that its age might be in the lower part of the age range estimated for \host. The best fit on the full NIR wavelength range is provided by 2M1207b, a late L dwarf planetary-mass companion much younger than the estimated age range for \host. When excluding the $K$ band, a very good fit is obtained with the spectrum of VHS1256-12b, a member of AB-Doradus, with an age compatible with the young part of the age range estimated for \host.  However we note that in both cases the 1.35-1.5 $\mu$m water-band absorption is slightly weaker in the HD~206893b spectrum.  The spectra of the older companions (130-525 Myr) HIP203030B and  $\zeta$ Del B do not represent the pseudo-continuum of  HD~206893b  well.  Similarly, an  inspection of  Fig. 3 in \cite{2016ApJ...829L...4K} indicates that the companion HR 2562B (L7, 300-900 Myr) has a bluer 1.1-2.4 $\mu$m spectral slope than VHS J125601.92-125723.9 ABb, which already has a bluer slope than  HD~206893b. 

\subsection{Strength of the 1.4 $\mu$m band}

The discrepancy between the slope and main absorptions of the companion and those of benchmark objects can be quantified through spectral indices. These indices have to be defined to work on the SPHERE IFS wavelength grid.  We therefore propose two indices based on the flux ratio of two spectral channels chosen to be correlated with the object's spectral type and which are less affected by systematics: 

\begin{equation}
H2O_{S}-A=\dfrac{ F_{1.4107\mu m}}{ F_{1.2744\mu m} } \\
\end{equation}

\begin{equation}
H2O_{S}-B=\dfrac{ F_{1.4107\mu m}}{ F_{1.5731\mu m}}\\
\end{equation}

\begin{figure}
\begin{center}
\includegraphics[width=8.4cm]{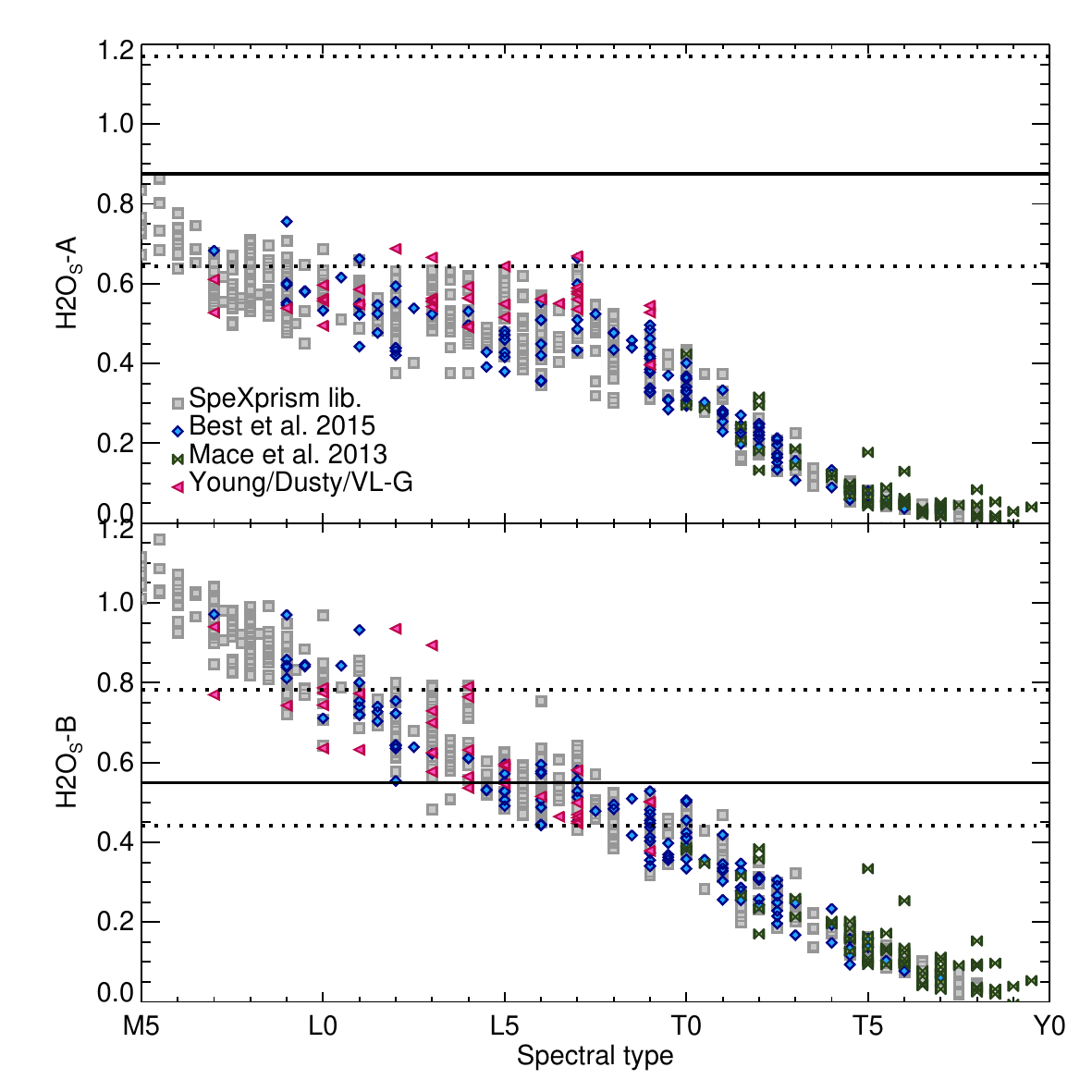}
\caption{Evolution of the spectral indices $H2O_{S}-A$ and $H2O_{S}-B$. We overlaid the value of the indices (thick line) and the associated error (dotted line) computed on the HD~206893b spectrum.}
\label{fig:indices}
\end{center}
\end{figure}

 The $H2O_{S}-A$  and  $H2O_{S}-B$ indices measure the strength of the 1.4 $\mu$m band compared to the J and H-band pseudo-continua, respectively.  Those indices are close to the H2O1 and H2O-B indices defined by \cite{2004ApJ...610.1045S} and \cite{2003ApJ...596..561M}, respectively. Their evolution with spectral type is shown in Fig. \ref{fig:indices}.

The  $H2O_{S}-B$ index is strongly correlated with the spectral type. The dusty, low-gravity, and/or young  M and L dwarfs have values which are on average in the same range as the field dwarf counterparts. HD~206893b  has an $H2O_{S}-B$ index value ($0.55^{+0.23}_{-0.11}$) compatible with those of M9-T1 dwarfs. This is consistent with the fit of the 1-2.5$\mu$m  spectrophotometry (Fig. \ref{fig:GSpeX}). Conversely, the dusty,  low-gravity, and/or young  M and L dwarfs tend to have lower $H2O_{S}-A$ values than the mean population of field objects at comparable spectral types.  The companion HD~206893b has a $H2O_{S}-A$ index ($0.88^{+0.29}_{-0.23}$) compatible with those of early L dwarfs. This confirms the unusual weak 1.4$\mu$m absorption of the companion spectrum and suggests that it corresponds to one extreme of the trend shown by dusty and/or young objects. The dusty L7 object  PSO\_J057.2893+15.2433 is the only source for which the two indices are compatible within the error bars with those of HD~206893b.

In summary, HD~206893b has a very red spectral slope coupled with a shallow 1.4 $\mu$m absorption which cannot be completely reproduced by any of the  comparison objects considered here, some of which  are young objects.  We investigate possible origins of these properties in the following section.

\subsection{Exploring extinction/reddening by dust}
        \cite{2014MNRAS.439..372M}, and \cite{2016ApJ...830...96H}  found that the spectra of young and/or dusty dwarfs in the L0-L7 regime could be modelled by spectra of standard L objects  reddened by an extra layer of submicron-sized  dust grains. We extended these conclusions to the domain of young companions (HR8799) in \cite{Bonnefoy.2016}. This extra-opacity may correspond to a haze of grains high in the atmosphere, which is also now proposed to explain the spectrophotometric variability of  mid- and late-L dwarfs \citep{2015ApJ...798L..13Y, 2016ApJ...829L..32L}. %While this empirical approach enables to explore the impact of the dust opacity on the spectrophotometry, we caution that  such a haze should in principle modify the temperate-pressure profile of the object  and redistribute the flux to other wavelengths. 
                
\begin{figure}
\begin{center}
\begin{tabular}{c}
\includegraphics[width=8.4cm]{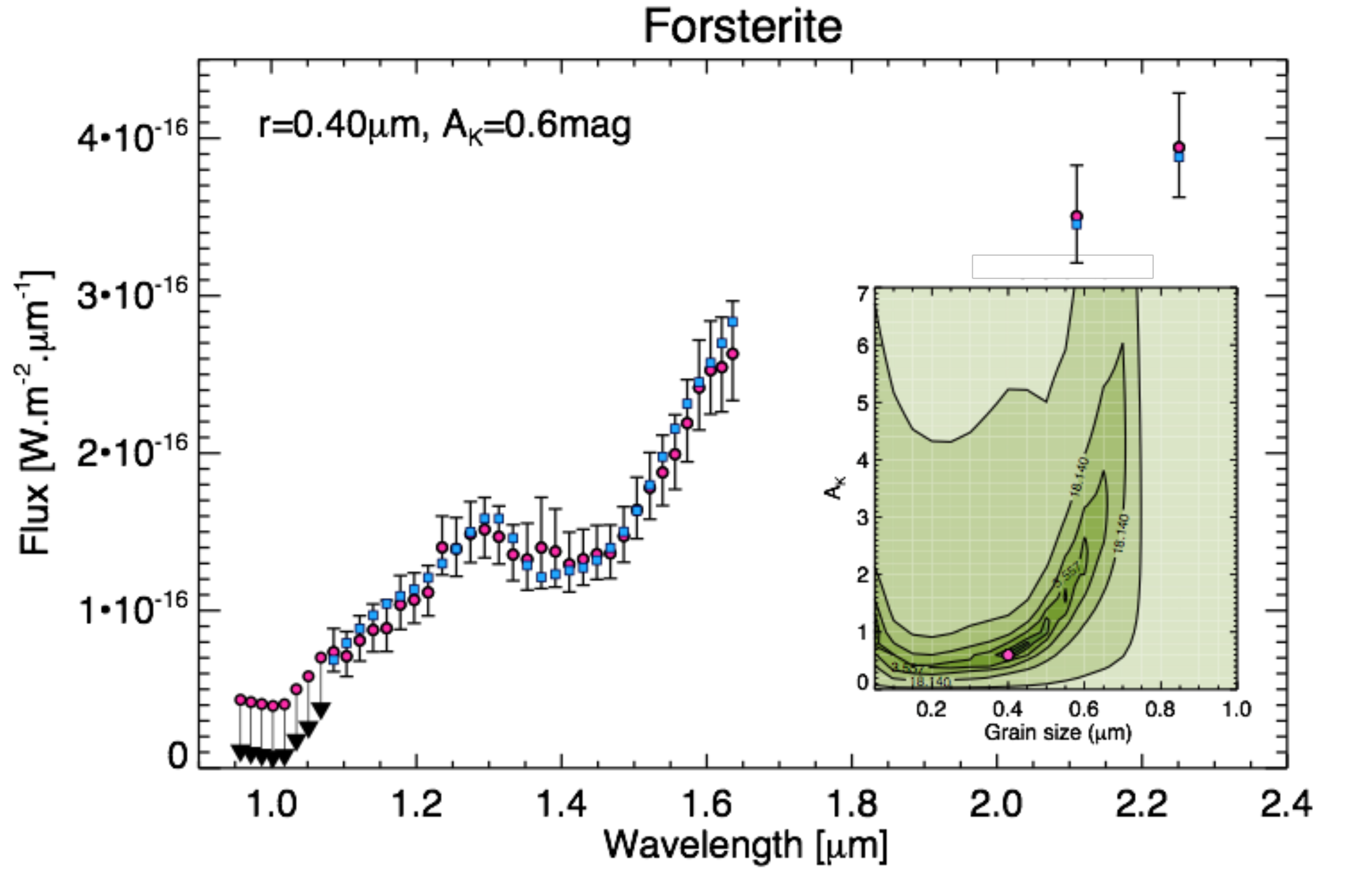}\\
\includegraphics[width=8.4cm]{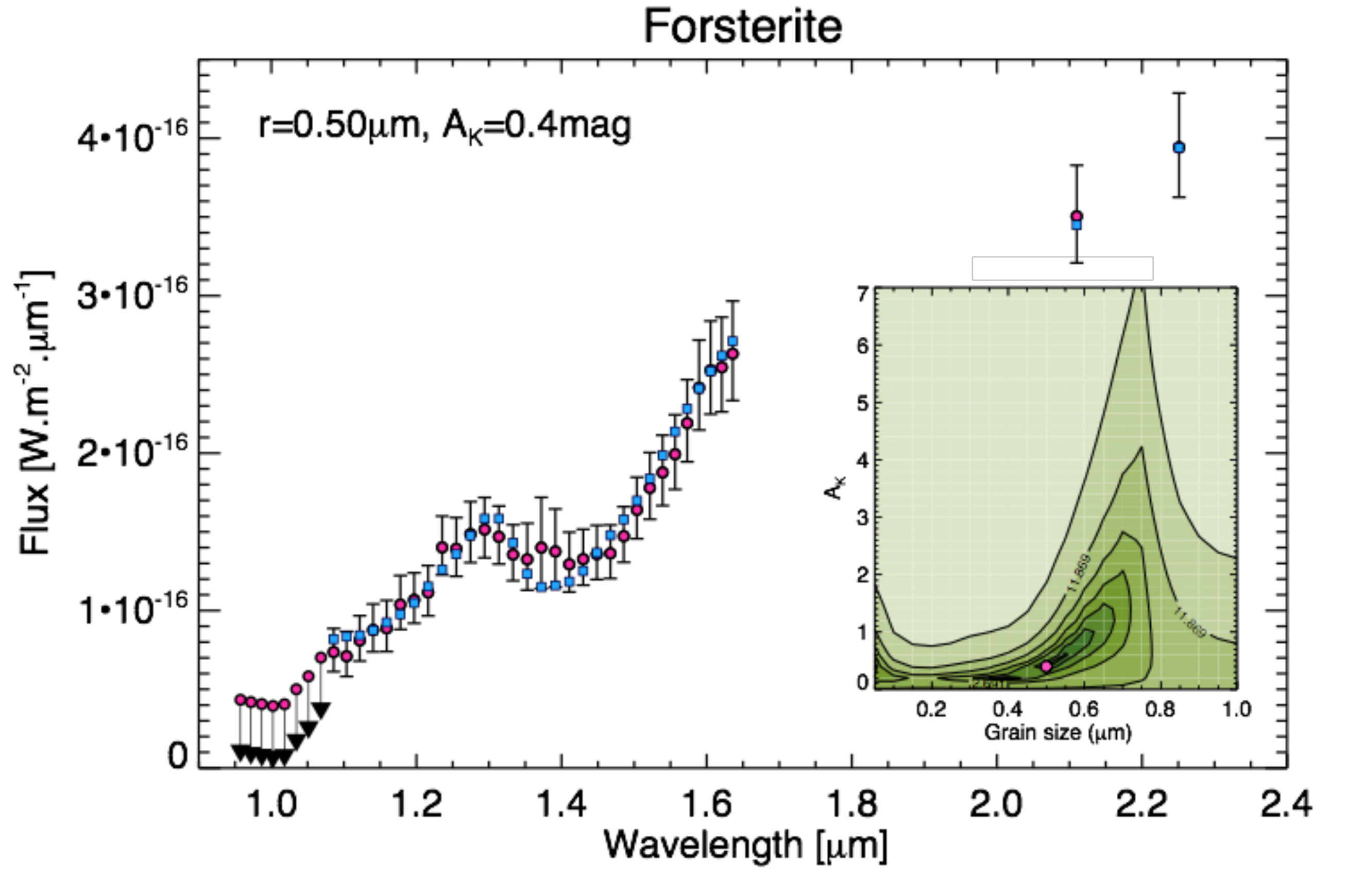}
\end{tabular}
\caption{Comparison of the spectrum  and photometry of HD~206893b (pink dots) to that of the L7 red dwarf PSO\_J057.2893+15.2433 \citep[bottom,][]{Liu.2013} and the L0 optical standard dwarf 2MASP J0345432+254023 \citep[top,][]{Kirkpatrick.1997} reddened by a layer of forsterite grains with mean sizes of 0.50 and 0.40$\mu$m, respectively. The mean flux of the reddened template spectrum is re-normalised to fit that of the companion. The G map is overlaid in the lower right portion of the plots.}
\label{Fig:redL0}
\end{center}
\end{figure}

                We tested the hypothesis applying the extinction curves of forsterite,  enstatite, corundum, and iron grains presented in  \cite{2014MNRAS.439..372M} and \cite{Bonnefoy.2016} on the spectra of optical standard L dwarfs and on the spectrum of the red L7 dwarf PSO\_J057.2893+15.2433. We considered extinctions ($A_{k}$)  up to 10 mag in steps of 0.1 mag and computed the G parameter for each combination of $A_{k}$ and mean grain size (0.05 to 1 $\mu$m in steps of 0.05 $\mu$m). The test was repeated for each considered grain species. The G parameter is minimised for grains with a mean size around 0.5$\mu$m.  We adopted a size distribution of width $\sqrt{2\sigma} = 0.1 \times r$ with r corresponding to the characteristic grain radii. A full description is given in \citet{Marocco.2014,Bonnefoy.2016}. The silicate and corundum grains reproduce better the shape of the continuum from 1.1 to 1.3 $\mu$m. The reddened spectrum of PSO\_J057.2893+15.2433 matches the slope of HD~206893b and reproduces the depth of the water band at 1.4 $\mu$m of  HD~206893b within the error bars. The reddened L0 to L3 dwarfs can also reproduce both the absorption and the slope. The parameters of the best fit are reported in Table \ref{tab:Red}. The reddened spectrum of the L0 dwarf and of PSO\_J057.2893+15.2433 are shown  in Fig. \ref{Fig:redL0}. 
                
\begin{table}
\caption{Parameters found when fitting  the SPHERE spectrophotometry of HD~206893b with spectra of L dwarfs reddened by an extra layer of dust grains.}
\label{tab:Red}
\begin{center}
\begin{tabular}{lllll}
\hline\hline            
Template        &       Species          &       Size ($\mu$m)   &       $A_{K}$ (mag)   &       G       \\
\hline
PSO057.28       &       Forsterite      &       0.45            &       0.3             &       0.1092  \\
                                &       Enstatite       &       0.65            &       0.6             &       0.1303  \\              
                                &       Corundum        &       0.45            &       0.3             &       0.1332  \\
                                &       Iron            &       0.20            &       0.3             &       0.1793  \\
%ULAS227                &       Forsterite      &       0.50            &       0.3             &       0.3510  \\
%                               &       Enstatite       &       0.50            &       0.3             &       0.3377  \\              
%                               &       Corundum        &       0.50            &       0.3             &       0.2795  \\
%                               &       Iron            &       0.20            &       0.2             &       0.4636  \\
\hline          
L3 opt std      &       Forsterite      &       0.50            &       0.9             &       0.2769  \\
                &       Enstatite       &       0.50            &       0.6             &       0.1950  \\              
                &       Corundum        &       0.50            &       0.8             &       0.1619  \\
                &       Iron            &       0.20            &       0.6             &       0.4917  \\
L2 opt std      &       Forsterite      &       0.50            &       0.6             &       0.2061  \\
                &       Enstatite       &       0.65            &       0.5             &       0.5150  \\              
                &       Corundum        &       0.50            &       0.8             &       0.1972  \\
                &       Iron            &       0.20            &       0.6             &       0.5635  \\
L1 opt std      &       Forsterite      &       0.05            &       1.0             &       0.3138  \\
                &       Enstatite       &       0.50            &       0.8             &       0.2488  \\              
                &       Corundum        &       0.45            &       0.8             &       0.3072  \\
L0 opt std      &       Forsterite      &       0.40            &       0.6             &       0.1340  \\
        &       Corundum        &       0.45            &       0.8             &       0.2154  \\
        &       Enstatite       &       0.50            &       0.8             &       0.2867  \\
        &       Iron            &       0.20            &       0.8             &       0.7697  \\
\hline          
\end{tabular}
\end{center}
\end{table}

\begin{figure}
\begin{center}
\includegraphics[width=8.4cm]{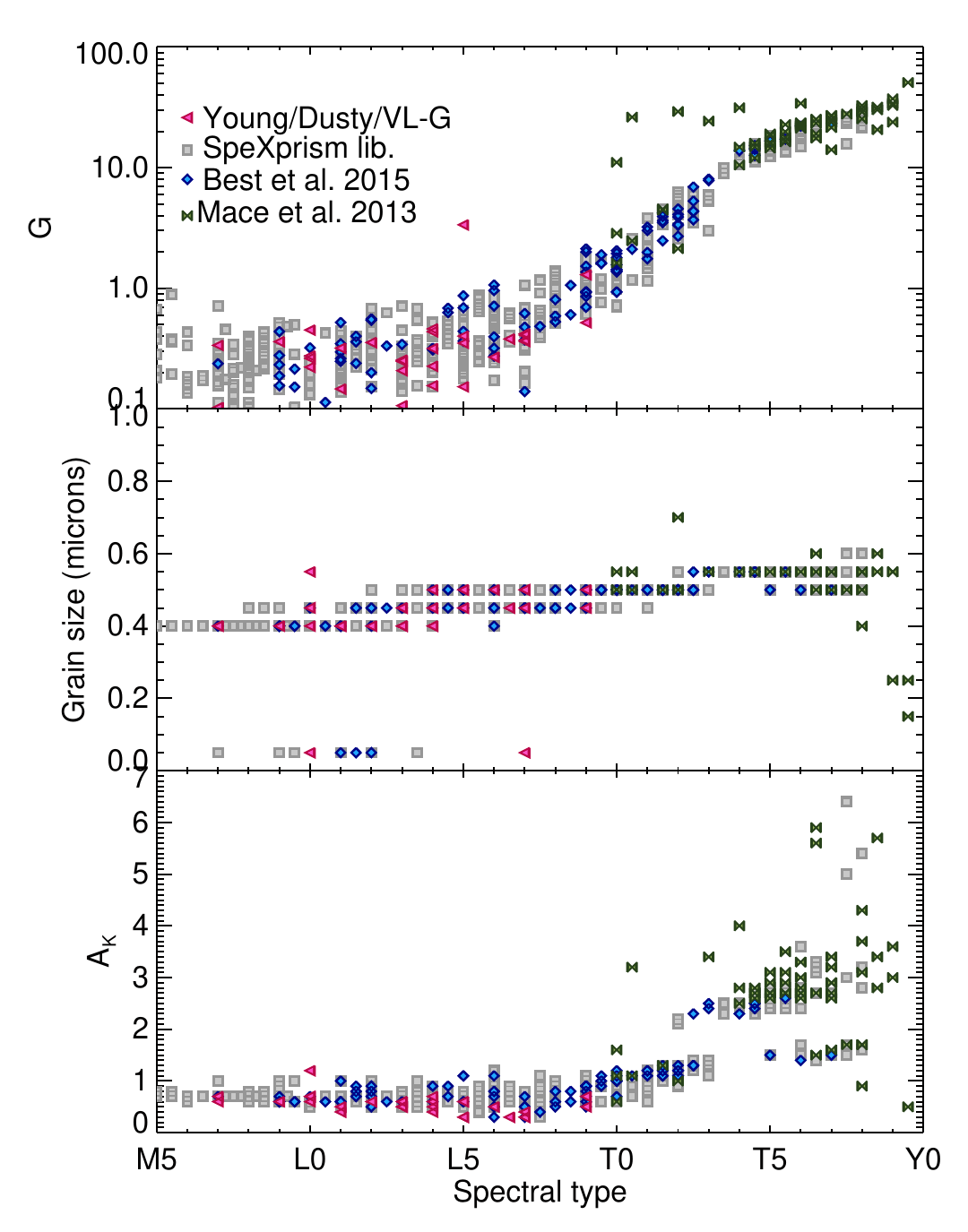}
\caption{Best fit parameters for each dwarf spectra considered in Fig. \ref{fig:GSpeX} reddened by a layer of forsterite grains.}
\label{Fig:redglobal}
\end{center}
\end{figure}

We generalised this analysis to all the spectra considered in Fig. \ref{fig:GSpeX} for the case of the forsterite grains (which are expected to be more abundant than  corundum and iron at high altitude). The reddening parameters found for  each object are reported in Fig. \ref{Fig:redglobal}. The majority of spectra are de-reddened by 0.4-0.6 $\mu$m grains, e.g.  a size consistent with that found by \cite{2014MNRAS.439..372M}. L5-L7 dwarfs require the lower extinction ($A_{K}$). 
                 
\begin{figure*}
\begin{center}
\begin{tabular}{cc}
\includegraphics[width=8.4cm]{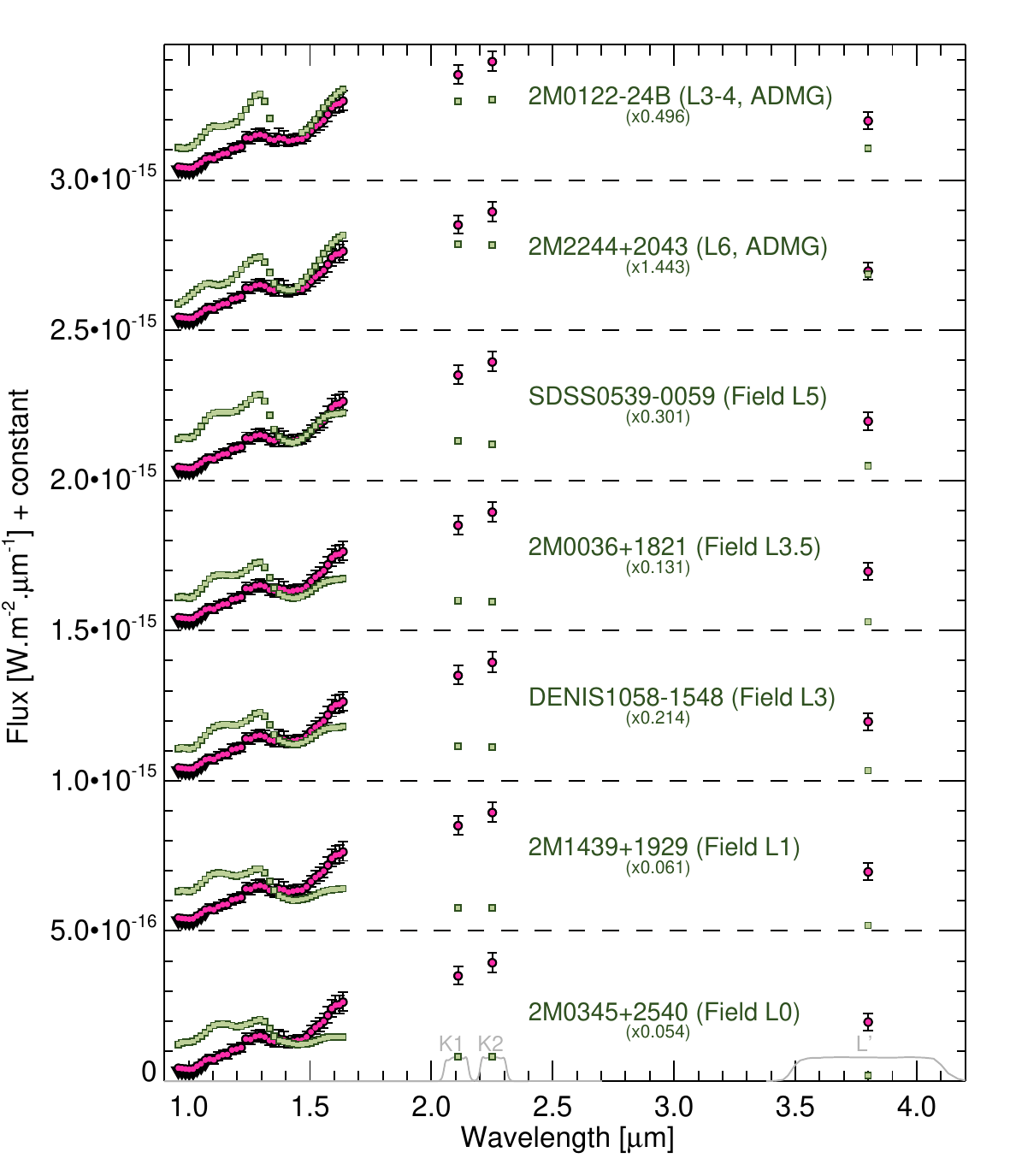} &
\includegraphics[width=8.4cm]{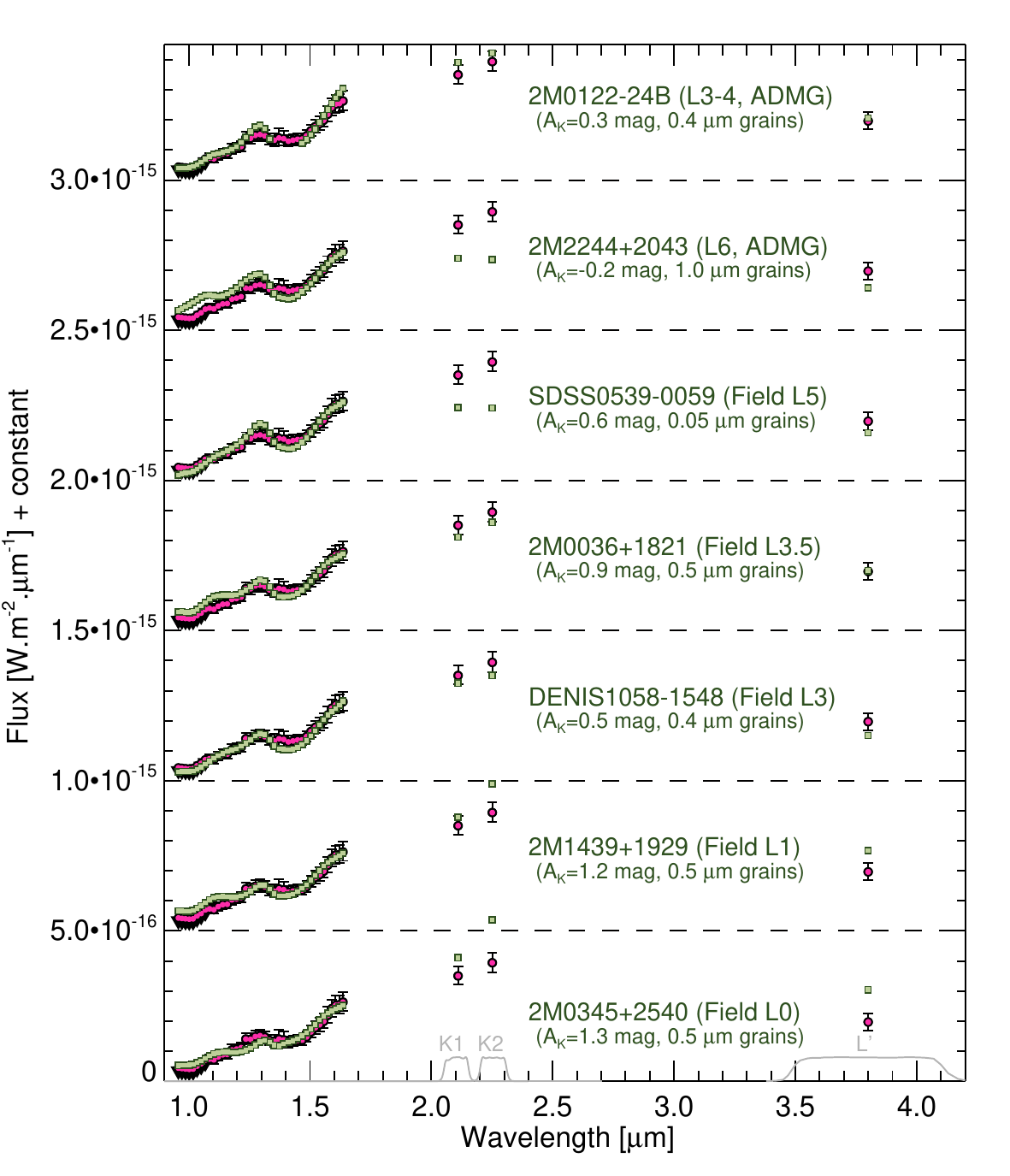} \\
\end{tabular}
\caption{Comparison of spectral energy distribution of HD~206893b to those of flux-calibrated spectra of benchmark objects re-scaled to the distance of HD~206893b. On the left, the template spectra are re-normalised by an additional factor to match the flux of HD~206893.  On the right, they are reddened by forsterite grains, but an additional normalisation factor is not applied.}
\label{Fig:redSEDf}
\end{center}
\end{figure*}

                 The reddening vectors in  Fig. \ref{fig:CMD} intersect the field dwarf sequence at the location of mid-L dwarfs.  This suggests that an  extinction caused by forsterite grains could explain consistently both  the deviation of the 1-2.25 $\mu$m spectral slope and the absolute flux of the companion with respect to the  sequence of field dwarfs.  We investigated whether this hypothesis still hold  up to 3.8$\mu$m adding the L' band photometry. We calibrated in flux the NIR spectra of 2MASPJ0345432+254023 (L0), 2MASSWJ1439284+192915 (L1), DENIS-PJ1058.7-1548 (L3), 2MASSWJ0036159+182110 (L3.5), SDSSpJ053951.99-005902.0 (L5),  2MASSWJ2244316+204343 (L7pec, see Table \ref{tab:AppA}), and 2M0122-24B (L3-4) using the 2MASS or MKO H-band magnitude of the objects and the corresponding zero points reported in VOSA. The L-band magnitudes were taken from \cite{2004AJ....127.3516G} and \cite{2013ApJ...774...55B}, and the zero point (MKO) was estimated using a flux-calibrated spectrum of Vega, and the MKO-L' filter pass band. The parallaxes were taken from \cite{2012ApJ...752...56F} and \cite{Liu.2016}, apart for 2M0122-24B where we used the photometric distance of the star \cite{2013ApJ...774...55B}. The central wavelength and width of the MKO L-band filter was considered to be similar to that of NaCo here.  We compare in Fig. \ref{Fig:redSEDf} the SED of HD~206893b to those templates. The companion is redder than all the objects considered here. The object is more luminous than the young L7 dwarf and less luminous than the L5 field dwarf template.  When an extra reddening by forstertite grains is applied, L3.5 dwarfs from the field or from the AB Dor moving group can reproduce both the shape and absolute flux of  the companion over 1 to 4~$\mu$m. While the reddened L0 template reproduces the best the shape of normalised spectrum and K-band photometry (Table \ref{tab:Red} and Fig. \ref{Fig:redL0}), it cannot  reproduce  the whole SED when the L' band flux is accounted for and it is too bright compared to the companion. 
                 
                 We therefore conclude that HD~206893~b spectrophotometry (continuum and $H_{2}O$ band)  is most consistent with that of a L3-L5 object affected by an extra extinction. Two dusty L dwarfs, ULAS J2227 and 2MASS J21481628+4003593, are not related yet to a young moving group but share some of the properties of HD~206893 b. This suggests that other parameters beyond gravity,  such as the enrichments at formation \citep[e.g.][]{2016arXiv161003216L}, the variability \citep[along the reddening vector, see][]{2016ApJ...829L..32L}, or the presence of a circumplanetary disc \citep{2017MNRAS.464.1108Z} may also be responsible for the peculiar properties of the companion.
        
         %\cite{2017MNRAS.464.1108Z} find that the spectral energy distribution (SED) of the moderately young (20-300 Myr) L3 companion G196-3B may be explained by the presence of a  vertically extended ring of optically thick debris. HD~206893b may lie in the same are range as G196-3B. Although both companions have close K-band luminosities, the near-infrared spectrum of  HD~206893b is much redder than the one of G196-3B (Fig.\ref{fig:Youngcomp}). Discs around younger targets can produce shallower water-band absorptions and red slopes \citep{2007ApJ...666.1219L}. But  circumstellar or circumsecondary  discs in the HD~206893 system  are unlikely to be optically thick given the estimated age of the star.  

\section{Spectral synthesis using brown dwarf model atmospheres}
\label{secspectro}
Model  fitting is a standard approach used to study substellar atmospheres, but it should be noted that we only have  43 data points, so our model fitting  has  few degrees of freedom and should not be overinterpreted. However the SED of our object is extremely peculiar, and pushes models to their limits in two ways. First, \obj has an an extremely red $J-K$ colour and second, the water absorption bands that dominate the spectra of L dwarfs are extremely dampened, notably between the $J$ and $H$ bands. Though several model atmospheres provide a fit with a reduced $\chi ^2$ close to 1 (or even below) because of the few data points that we have, most of the models fail to reproduce the two above-mentioned extreme spectral features.
\subsection{Ensuring that best fit atmospheres can be related to the physically plausible substellar objects}
\label{absflux}

 Atmosphere models provide the set of effective temperature, gravity, and sometimes other parameters such as metallicity or cloud sedimentation efficiency that best match the observed spectrum. However, atmosphere models  do not necessarily guarantee by design that every atmosphere within their grid can actually be related to a physically plausible substellar object. Some substellar atmosphere models provide grids going to \logg =6.0, for which Newton's law and the physics of electron degeneracy \citep{Kumar.1963} predict that no substellar object can exist. Evolutionary models such as \citet{Baraffe.2002,Baraffe.2003} include this well-established physics, as well as more complex cooling mechanisms, and show that the highest gravity that a substellar object can reach is around \logg=5.6. Many atmosphere models also cover gravities as low as \logg=3.5  or even 3.0 (i.e. Earth's gravity), which can only be compatible with extremely young ($<1$~Myr)or even accreting giant planets and are inconsistent when studying exoplanets around older host stars, such as young moving group members. Even though extending the atmosphere model grids to such extremes  is useful to allow the interpolation of extremal boundary conditions and might help determine whether an object has, qualitatively, `very low' or `very high' gravity, it can be misleading when trying to quantitatively determine the physical parameters of a given object.
  
\begin{figure*}
\begin{center}
\includegraphics[width=17cm]{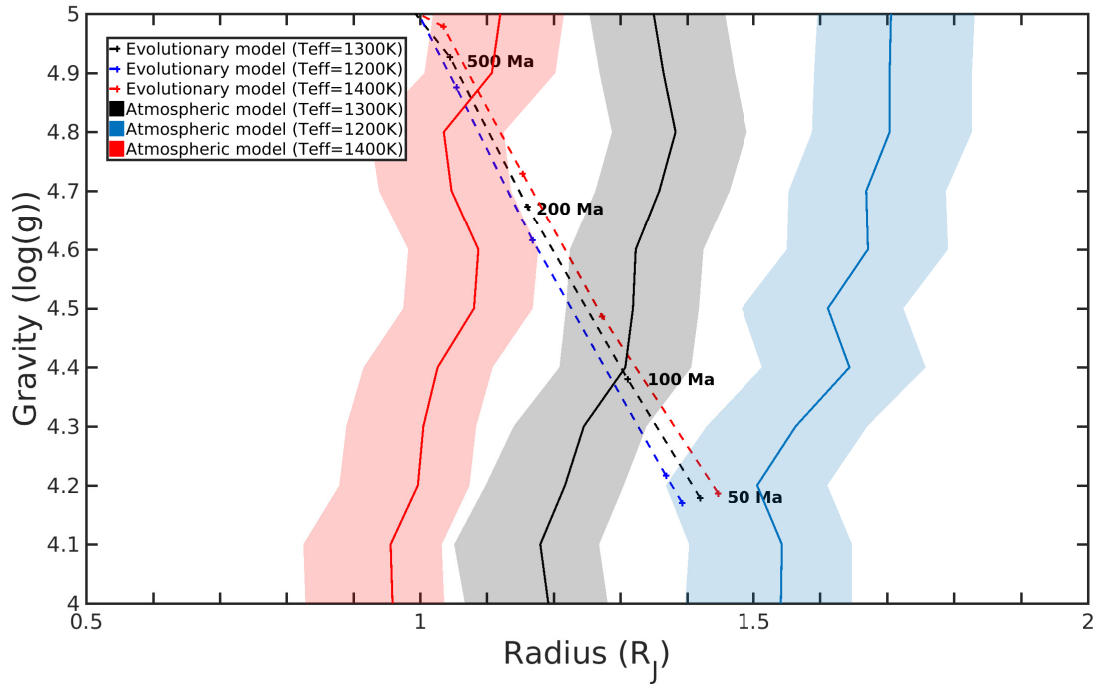} 
\caption{ Radius and gravity associated with atmosphere models (full line) and evolutionary models (dashed line). The shaded area corresponds to the uncertainty on the radius of the atmosphere models to match the absolute flux of \obj. We used the ExoREM atmosphere models, which provide the best fit to \obj (see Sect. \ref{exorem}), and the \citet{Baraffe.2003} evolutionary models.}
\label{evol}
\end{center}
\end{figure*}

%determines a radius of 0.46\Rjup for a mass of 80\Mjup. SUch a radius is well below the miminum radius allowed by electron degenrac
In the following we systematically relate the best fit atmospheres to physical objects and ensure they are plausible.
%We do so through our observational constraints on the absolute magnitude of \obj that, for a given effective temperature, is compatible with only a given range of radius.
 %As described in more details in Section \ref{absflux}, since our observational coverage, from 1$\mu$m to 4$\mu$m, include most of the energy emitted by such a cool object, our overall constraints on the radius only weakly depends upon the atmosphere modelisation. Given the uncertainties that also affect evolutionary models, we don't use this as an quantitatively accurate validation, but we ensure the best fit atmosphere can be related to a plausible substellar object or reject the corresponding unphysical solution.
Using our measurement of the absolute flux of the companion, we explored which parts of the parameter space that are allowed by atmosphere models fits are also consistent with the associated evolutionary model solution. From the knowledge of the distance, we directly estimated the physical radius of the object  associated with each atmosphere model from the dilution factor necessary  to match the observed absolute flux of \obj. Given the gravity of the model atmosphere, this radius can be converted into a mass using Newton's law. The resulting combinations of mass, radius, and effective temperature are then compared with substellar evolution models.  In order to visualise the area of the parameter space that are associated with a self-consistent set of effective temperature and gravity/radius, we plotted in Fig. \ref{evol} the radius derived from the atmospheric models for several effective temperatures together with the radius from evolutionary models for a range of age and the same effective temperatures. Error bars on the radius were derived by considering the maximum and minimum scaling factor, and hence associated radius, for each of the photometric bands for which we have significant flux ($J$, $H$, $K$, and $L'$).
% We stacked the IFS channels in order to obtain higher S/N absolute photometry in $J$ and $H$ bands, see Table \ref{photom}. 
This difference in scaling factor depending on the photometric band considered is responsible for most of the error on radius  and dominates over photometric error. Given the systematic uncertainties within both atmosphere and evolutionary models, we also chose to consider the extreme plausible radii according to our observational errors. We therefore considered the extreme radius obtained when matching the photometric band providing the smallest radius with a negative flux noise excursion (including photometric error and parallax errors) and the photometric band providing the largest radius with a positive flux noise excursion. Therefore, if a model atmosphere gravity/radius estimation  in Fig. \ref{evol} does not match an evolutionary model track, it means that the observed absolute of \obj is inconsistent in all photometric bands with evolutionary model prediction, even taking into account favourable  photometric noise excursion at plus or minus 1$\sigma$. We also note that if we had the bolometric flux of the object, we would have no systematics at all when deriving the radius associated with a given effective temperature and gravity. Given our large spectral coverage ($1--4 \mu$ m), which encompasses most of the energy emitted by \obj, the systematic uncertainties of the models are already significantly limited by the hard physics of black-body law.As can be seen in Table \ref{recapmodels}, this is confirmed by the narrow range of bolometric luminosity ($\log (Lbol/ L\odot)$ = -4.3 to -4.5) spanned by our best fitting unextincted models. In the following, we therefore consider that model atmospheres that cannot match evolutionary model tracks within our very conservative error bars cannot be related to physically plausible substellar objects.

\subsection{Lyon's group atmosphere models}
\begin{figure*}
\begin{tabular}{cc}
\includegraphics[width=8.4cm]{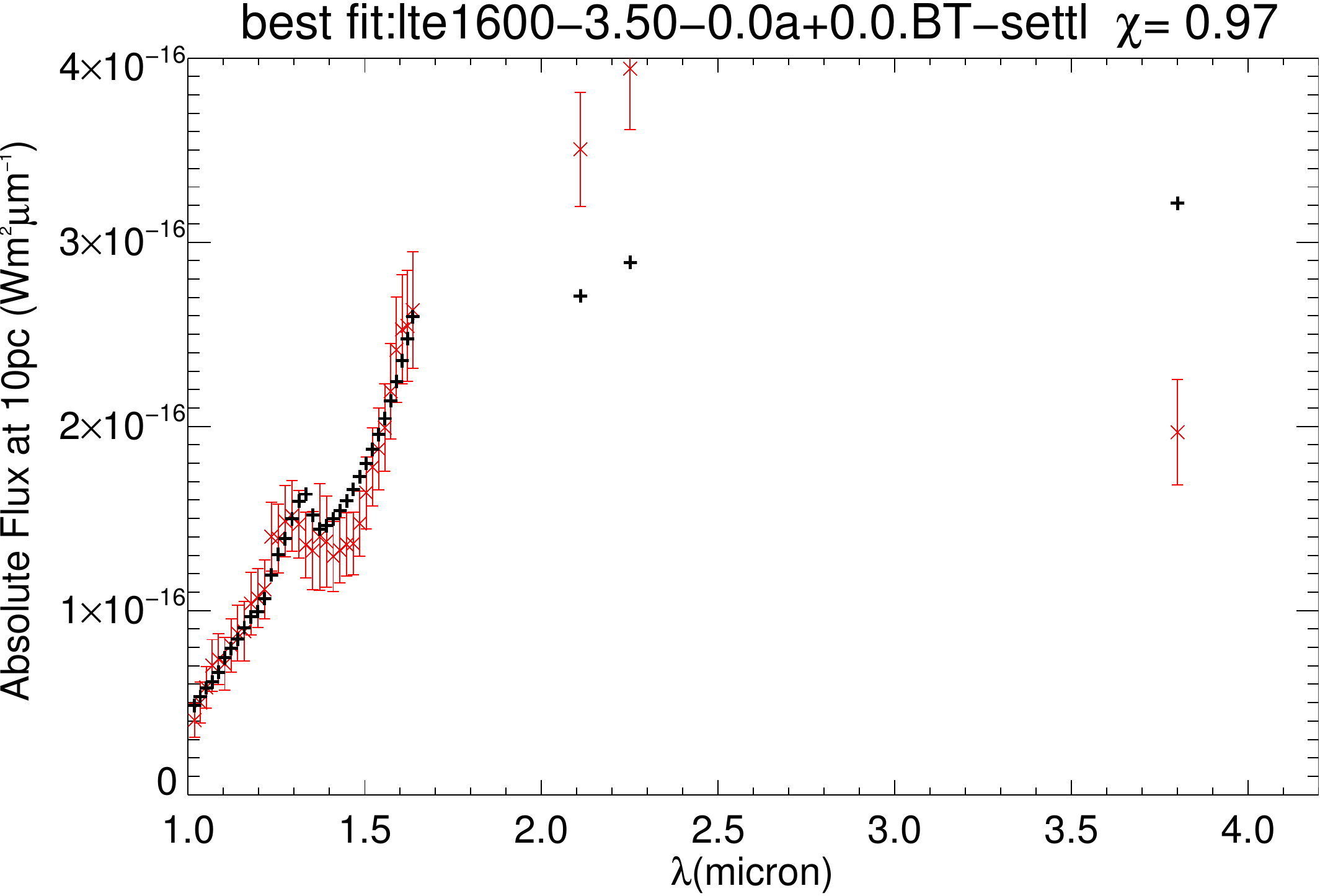}&\includegraphics[width=8.4cm]{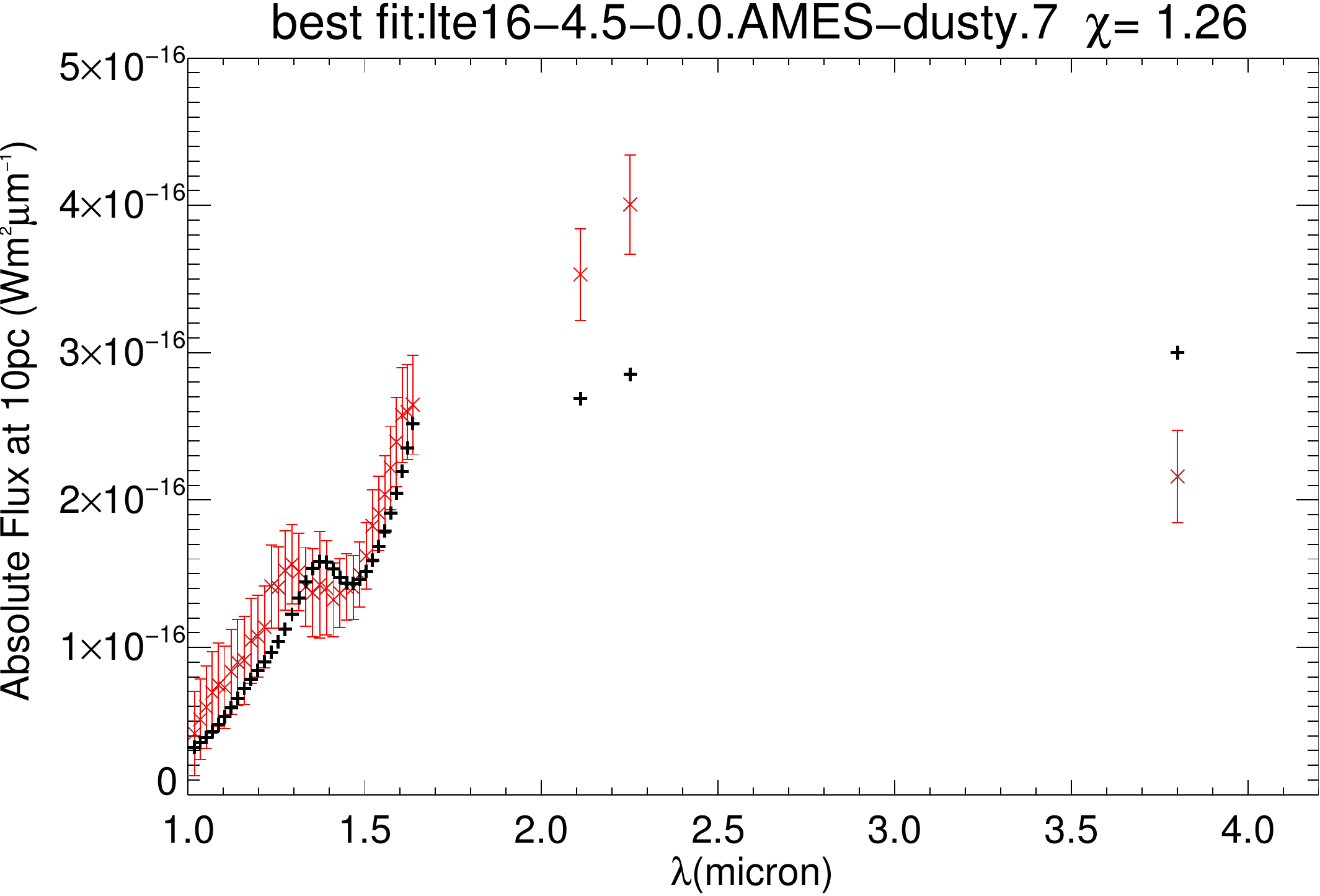} \\
\end{tabular}
 \caption{{\bf Left:} \textit{BT-Settl} model atmosphere best fit (black) for the spectra of HD~206893b (red) {\bf Right:} \textit{Dusty} model atmosphere best fit (black) for the spectra of HD~206893b (red) 
\label{dustyfit}}
\end{figure*}

  We first compared the observed spectrometric and photometric data of \obj to the state of the art  \textit{BT-Settl} model \citep{Allard.2014proc}.
 Our model grid covered effective temperatures from 1100\,K to 1950\,K with steps of 50\,K and gravities from log $g$ =3.5~dex to 5.5~dex (cgs) with steps of 0.5~dex. We included both solar and super-solar metallicity models (+0.3~dex, i.e. twice the solar abundance). We also calculated the correlated $\chi ^2$ that takes into account the spectral correlation of our data \citep[see][]{Derosa.2016,Samland.2017}, see Appendix \ref{app:correl} for more details. By construction the correlated $\chi ^2$ values are higher than the regular reduced $\chi ^2$, and are only equal if all the data is perfectly uncorrelated. However, taking into account the spectral correlation did not change the best fits, so in the following we refer only to regular reduced $\chi ^2$ values, which are easier to interpret than the correlated $\chi ^2$ values. 
 % Keep this in in case the referee asks for it.
%Also since the spectral resolution of the IFS is 30, our resulting IFS spectra has three data points per resolution elements, meaning that if we assign a weight of one for each IFS channel in the $\chi ^2$ calculation, we overestimate the information content of IFS data points compared to that of the IRDIS and NACO ones. On one hand we have 3 data points per resolution element, but on the other hand, the S/N of each resolution element is sqrt(3) higher than for individual IFS channels, meaning that we overestimate the information content by  3/sqrt(3). We therefore applied a weight of sqrt(3)/3 to individual IFS channels in $\chi^2$ calculation, and in the calculation of the degree of freedom for the reduced $\chi^2$, but we highlight that this weighting did not actually change the best fit models. 
We fitted the observed spectra to the models after normalising both observation and models to their observed $J$-band peak, between 1.255--1.29~$\mu$m, which is a common approach in substellar atmosphere studies where parallaxes and hence absolute fluxes are not always available. 
 The best fitting model with a reduced  $\chi ^2$ of 0.98 corresponds to a 1600~K atmosphere at low gravity (\logg=3.5 dex ($cgs$)) and solar metallicity, see Fig. \ref{dustyfit}.  The radius that would match the observed absolute flux of \obj is 0.92\Rjup and the associated mass is only 1\Mjup. This combination of physical parameters is  in contradiction with substellar evolution models that predict that a 1\Mjup isolated object never reaches 1600~K, with \citet{Baraffe.2003} evolutionary models predicting an effective temperature of less than 1000~K at 1~Myr, naturally cooling down at older ages.  This is confirmed by Fig. \ref{evol}, which shows that atmospheres with temperatures higher than 1400~K can be consistently associated only with very small radius and very high gravity.
  We tried the opposite approach, and instead of normalising observation and models to their $J$-band peak, we directly used the absolute flux derived from a physically consistent set of radii and masses in the input of each  \textit{BT-Settl} atmosphere model to fit the absolute flux of \obj. The best fit was extremely poor,  that of a 1000~K object with \logg=5.0, which would correspond to a mid- or late-type field T dwarf, whose deep methane and water absorption bands and rather blue NIR spectra have nothing to do with the very red and smooth spectra of \obj. This means that this model set can match either the absolute flux of our target or its SED, but struggle to match both together. The same problem was already encountered by \citet{Ducourant.2008, Barman.2011,Currie.2011,Marley.2012,Skemer.2012} during early analysis of young planetary companions such as 2M1207b or HR8799bcde, for which the atmosphere models which matched the SED would lead to unphysically small radii when accounting for absolute flux.

% In fact \citep{Allard.2014proc} model atmospheres are derived from a self consistent locus in effective temperature and gravity, that is taken from evolutionary model. The header of our best fit  \textit{BT-Settl} atmosphere indicates that this 1600~K, \logg=3.5 atmosphere is that of 2.7\Mjup object with a radius of 1.55\Rjup, confirming that though this atmosphere is a good fit to the overall shape of the SED, it cannot match its absolute flux with a physically consistent combination of mass and radius. 
%  We tried the opposite approach, and directly used the physically consistent set of radius and masses in input of each  \textit{BT-Settl} atmosphere model to directly fit the absolute flux of \obj. The best fit was extremely poor,  that of 1000~K object with \logg=5.0, which would correspond a mid/late field T dwarf, whose deep methane and water absorption bands, and rather blue NIR spectra have nothing to do with the very red and smooth spectra of \obj. This means that this model set can match either the absolute flux of our target or its SED, but struggle to match both together.
   Given that the best empirical fits for the object were very dusty L dwarfs, the unphysical fit obtained with the  \textit{BT-Settl} model might be caused by its self-consistent cloud model not managing to produce enough clouds to match the SED of this extremely atypical object. We therefore also tested our data against the simpler \textit{Dusty} atmosphere models of the same team \citep{Allard.2001}. This model set has no self-consistent cloud model and simply assumes there is no settling of dust in the photosphere of the objects, naturally leading to very dusty atmosphere.
  \textit{Dusty} models also produce very good fits to our data (see Fig. \ref{dustyfit}), with reduced $\chi ^2$ as low as 1.26. Again, introducing the correlated noise approach did not change the best fit model, that of a 1600~K atmosphere at moderately low gravity (\logg=4.5 dex ($cgs$)) and solar metallicity. However, the corresponding radius (0.75\Rjup) and masses (6\Mjup) are also unphysical.
We also explored other publicly available atmosphere models, notably by \citet{Madhusudhan.2011} and \citet{Morley.2012}, but the match to our data were similar or slightly worse than those found with \textit{BT-Settl}, and were also associated with unphysical radii, so we do not describe them here.
We therefore decided to test another set of models with more free parameters to explore whether some combination of gravity, effective temperature, dust settling efficiency, or metallicity could better match the observed properties of \obj.

\subsection{ExoREM atmosphere models}
\label{exorem}
In a second step, we compared our data to ExoREM, a 1-D radiative convective model developed for young giant exoplanets \citep{Baudino.2015}.
This model solves the temperature profile for radiative-convective equilibrium iteratively. We computed a dedicated grid of ExoREM model atmospheres for the study of \obj.
The atmospheric grid consists of 64 logarithmically equally spaced pressure levels. Radiative transfer is performed using the correlated-k method from 0.67 to 500~$\mu$m. Scattering is not taken into account, but non-equilibrium chemistry is included by comparing a chemical timescale with a vertical mixing timescale, parametrised by an eddy mixing coefficient ($K_{zz}$) profile from \citet{Ackerman.2001}. In this version of ExoREM, we also implemented a cloud scheme similar to the one described in  \citet{Ackerman.2001}. We use iron, silicate and sulfide clouds with saturated pressure from \citet{Visscher.2010,Morley.2012}. The vertical profiles of cloud mixing ratio and particle size are computed by fixing parameter fsed, 
which is the ratio of sedimentation velocity to the typical vertical mixing velocity. We use the same $K_{zz}$ for clouds as for non-equilibrium chemistry. 
The basic input parameters of the model are the effective temperature, the surface gravity at 1 bar, the elemental abundances, and fsed.
Our model grids cover effective temperatures from 400K to 2000K with steps of 100K, and gravities from \logg = 3 to 5 (g expressed in cm.s$^{-2}$) with steps of 0.1. 
The grids include cases with 0.3x, 1x, and 3x solar metallicity. We computed these grids with fsed=1, 2, and 3. 
While fsed=2 generally provides the best match for field L dwarfs, only cases with fsed=1 were cloudy enough to reproduce the red colour of \obj.

\subsubsection{Best fit ExoREM model atmospheres}
\begin{center}
\begin{figure*}
\begin{tabular}{c}
\includegraphics[width=18cm]{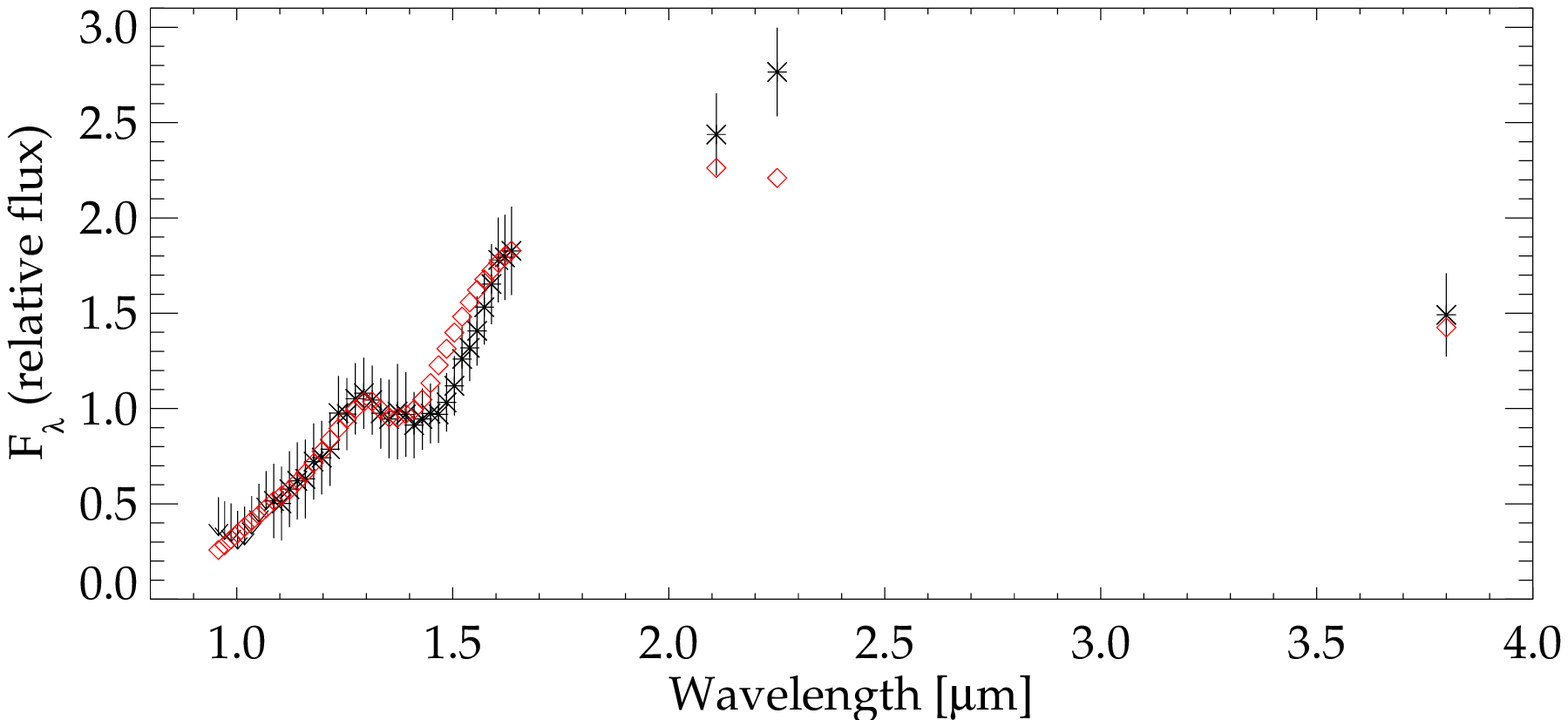}\\
\includegraphics[width=18cm]{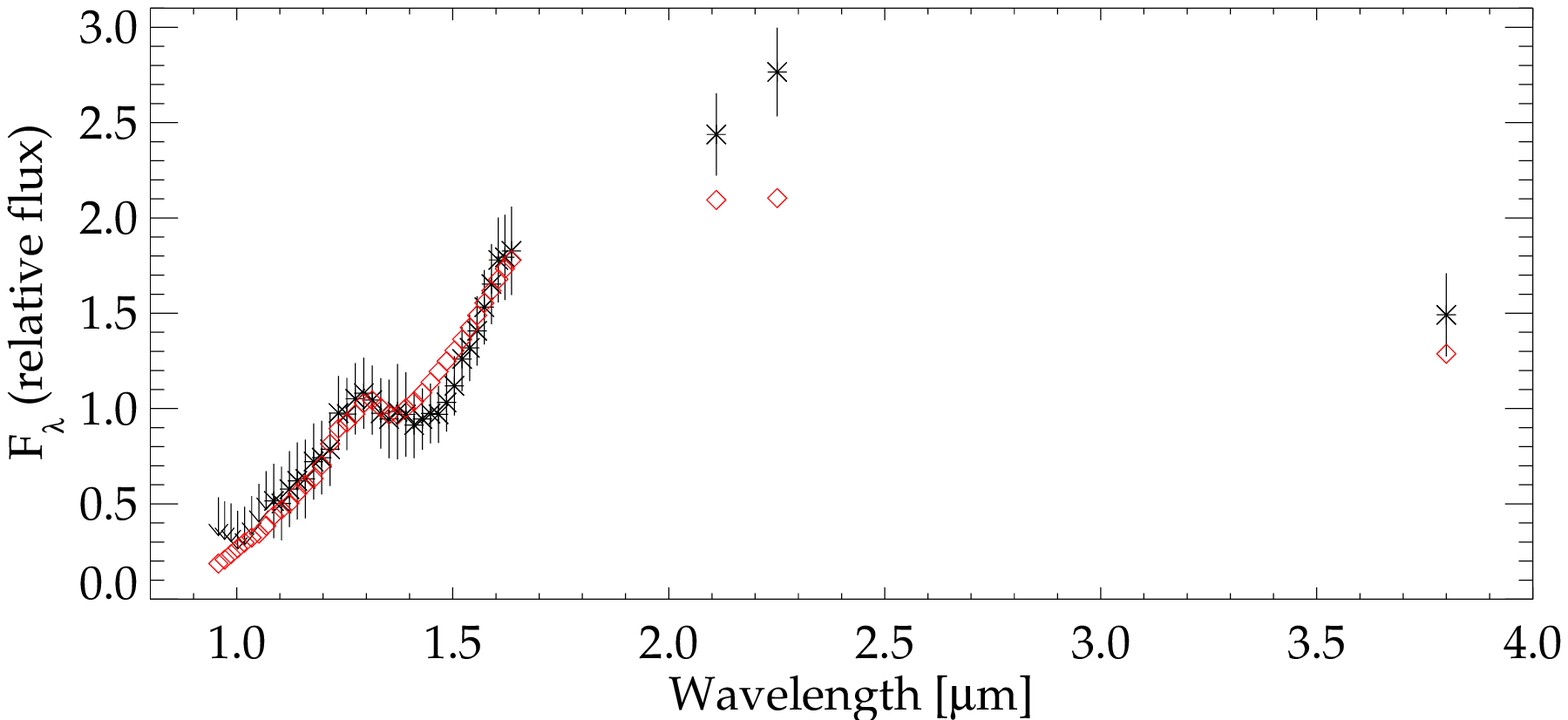}
\end{tabular}
\caption{{\bf Upper panel:} Best fit Exorem atmosphere model (1300~K, [M/H]=0.0, \logg=4.4, in red), compared to the \obj spectrum (in black). {\bf Lower panel:} Best fit Exorem atmosphere model when super-solar metallicity is forced (1500~K, [M/H]=0.5, \logg=4.5, in red), compared to the \obj spectrum (in black). }
\label{exoremfit}
\end{figure*}
\end{center}

From the $\chi^2$ fitting of these models with the additional free parameter of optical extinction we achieve a best fit at \teff= 1300 K, \logg= 4.4 for solar metallicity and no extinction with a reduced $\chi^2$ of 0.98, given in Fig.~\ref{exoremfit}. We note that fits with several magnitudes of extinction give similarly good results, meaning that our observations are compatible with additional extinction, as in the scenario discussed in Section \ref{sec:refobj}. However, our reduced $\chi^2$ is already below one without extinction, meaning we are in the data-starved regime of model fitting, and in the following we choose to consider the simpler models that provide good fits to our data. We discuss the equally good solutions provided by more complex -- and more underconstrained -- models that include additional extinction in the next section.

\begin{figure}
\begin{center}
\includegraphics[width=8.4cm]{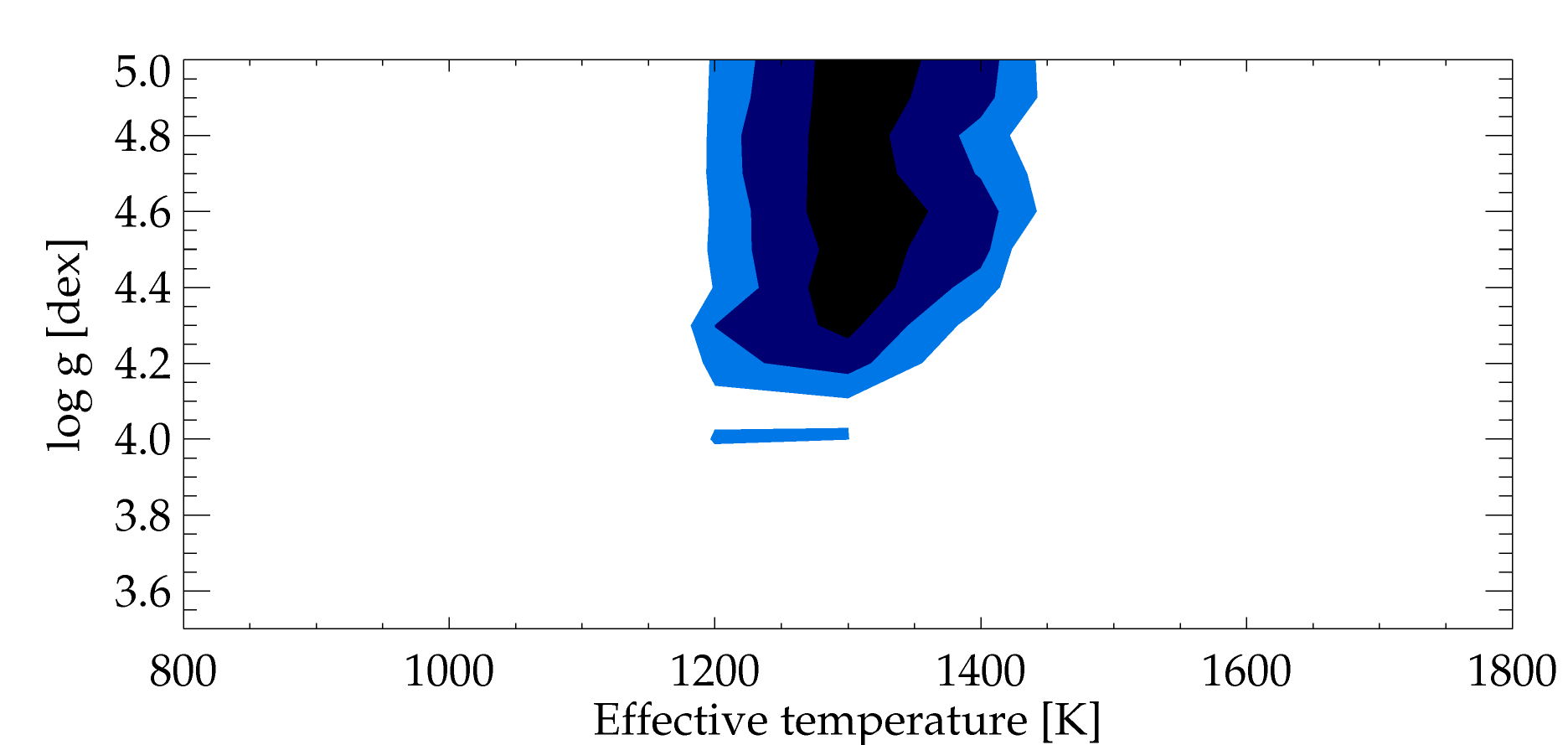}
\caption{  $\chi ^2$ contour map of ExoRem atmosphere model fits at solar metallicity. Deep blue is the 1$\sigma$ contour, blue is 2$\sigma$, and light blue is 3$\sigma$. }
\label{exochi2}
\end{center}
\end{figure}

The $\chi^2$ map is given in
Fig.~\ref{exochi2} and strongly points to 1300 K object at $\log{g}$ being higher than 4.2 dex. We locally enhanced the sampling of the temperature grid by computing models at 1250~K and 1350~K, but the 1300~K atmosphere still provided the best fit.

 We also forced high metallicity (three times solar) and get a good fit to the observations with a reduced  $\chi^2$ of 1.05, corresponding to an effective temperature of 1500~K and a gravity of \logg =4.5. However, the high-metallicity solutions are significantly degenerate, and can provide good fit to the data for temperatures ranging for 1300~K (with \logg =3.1) to 1500~K (with \logg =4.9).

This atmosphere model analysis does bring a constraint on the effective temperature of the object, which, in the absence of external extinction, appears to be in the 1250-1500~K range, but does not provide any robust constraint on the gravity, which could vary by almost two decades if all metallicities are considered. Our analysis provides a slightly narrower constraint (\logg $>$4.2) if we make the hypothesis of solar metallicity, but it is clear that atmosphere models by themselves cannot constrain the mass of \obj.

\subsubsection{Breaking the degeneracy with evolutionary models and absolute flux}
\begin{figure*}
\begin{center}
\begin{tabular}{cc}
\includegraphics[width=8.4cm]{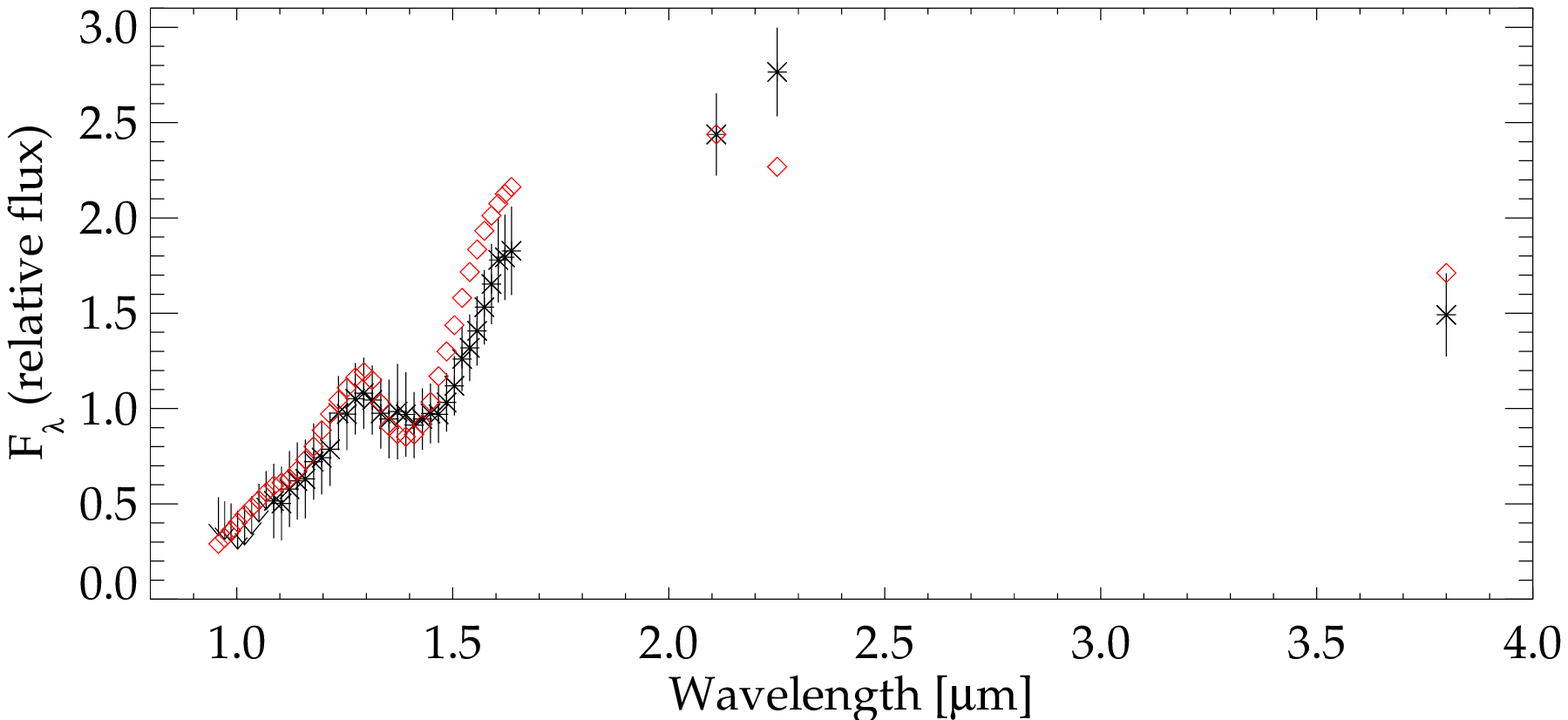} &
\includegraphics[width=8.4cm]{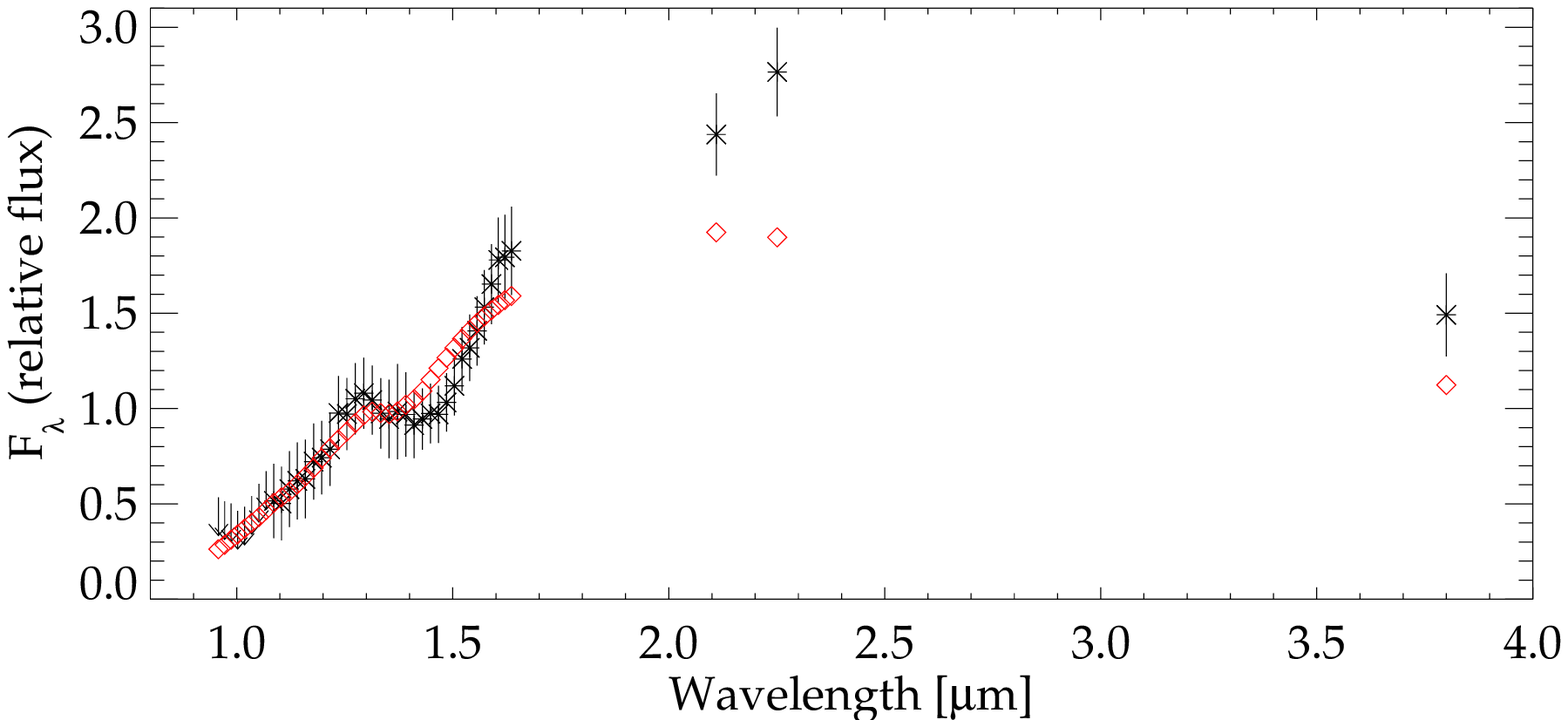} 
\end{tabular}
\caption{{\bf Left:} Observed spectra of \obj (red) compared with the lowest gravity atmosphere model compatible with evolutionary models, \teff=1200~K, \logg =4.2. The associated $\chi  ^2$ is 1.58. {\bf Right:} Observed spectra of \obj (red) compared with the highest gravity atmosphere model compatible with evolutionary models, \teff=1400~K, \logg =5.0.  The associated $\chi ^2$ is 1.27.}
\label{extremefit}
\end{center}
\end{figure*}

  As discussed in Sect. \ref{absflux}, we can use information on the absolute flux and evolutionary models to select only the atmosphere models that can be related to a physically plausible object, with a mass and radius that account for its gravity and are at least roughly consistent with evolutionary model predictions. Figure \ref{evol} clearly shows that constraints arising from evolutionary models break most of the degeneracy of Exorem atmosphere models and that the physically consistent solutions populate a relatively narrow diagonal of the parameter space. The lowest plausible temperature (1200~K) must be associated with low gravity (\logg=4.2) and a large radius and  is only marginally consistent with what would be a 50~Myr old planetary mass (10--13\Mjup) companion. Though not excluded by our analysis, we  stress that it is unlikely for \obj to reside in the planetary mass range, because it would  not only require  an age that is close to the youngest possible according to the independent age constraint from the host star, but also a combination of effective temperature and gravity that is only marginally consistent with evolutionary models predictions. In addition, the match to the data is not very good for this extreme solution with a reduced $\chi ^2$ of 1.58 (see Fig. \ref{extremefit}).  \\
 For the atmosphere model best fit temperature of 1300~K, the range of gravity that is compatible with evolutionary models is restricted to \logg between 4.3 and 4.6, corresponding to 12-20\Mjup objects aged 100--200~Myr, in good agreement with our estimation of \host age. For a 1400~K atmosphere, the range of physically consistent gravity is restricted to \logg between 4.8 and 5.0. Since gravity between  \logg = 4.6 and 4.8 would be compatible with an intermediate temperature of 1350~K, we consider in detail only the most extreme compatible case, that of a 1400~K, \logg =5.0 atmosphere, which corresponds to a $\sim$40\Mjup companion aged 600--700~Myr. As can be seen in Fig. \ref{extremefit}, with an associated reduced $\chi ^2$ of 1.27 this highest temperature, highest gravity solution compatible with evolutionary models provides a relatively poor fit to the data. It is striking to note that our independent upper age estimate for the host star (700 Myr) again coincides rather well with the constraints arising from the atmospheric and evolutionary analysis of the companion.\\
 If we follow the opposite approach and explore the solutions that are associated with our best estimate of the age of the host star (200-300~Myr), Fig. \ref{evol} shows that the effective temperature of the \obj should be between 1300~K and 1400~K, with a gravity restricted to \logg between 4.6 and 4.8, corresponding to a  20-30\Mjup brown dwarf companion. We highlight that even though this approach leads to solutions that happen to be close to our best fitting atmosphere model, it is almost independent of atmosphere modelling. Indeed, in Fig. \ref{evol} atmosphere models are only used to derive the scaling factor to match the observed absolute flux. However, these atmosphere models at 1350~K and \logg=4.6 to 4.8 do provide a very good fit to the observed spectrum (see Fig.\ref{exochi2}). This makes the hypothesis of a 200-300~Myr, 20-30\Mjup brown dwarf companion especially interesting.

\begin{table*}
\caption{Physical parameters associated with solar metallicity best fitting models explored in this article that are consistent with absolute flux and substellar evolution models constraints.\label{recapmodels} }
\begin{tabular}{|l|c c c c c c c|} \hline
 Model type & Teff(K) & log $g$ & $\log (Lbol/ L\odot )$  & Extinction(A$_K$) &M(\Mjup)$^1$& Compatible age (Myr)$^1$ & Reduced $\chi ^2$ \\ \hline \hline
 ExoREM & 1200 & 4.2 & -4.4 & None &10-13 & 50-100 & 1.58  \\ \hline
 ExoREM & 1300 & 4.4 & -4.3 & None &12-20 & 100-200 & 0.98  \\ \hline
 ExoREM & 1400 & 5.0 & -4.5 & None &35-40 & 600-700 & 1.27  \\ \hline \hline
 Extincted BT-Settl & 1650 & 4.5 & -4.0$^2$ & 0.66 & 20-25 & $\sim$100 & 0.41  \\ \hline 
 Extincted BT-Settl & 1700 & 5.0 & -4.1$^2$ & 0.59 & 35-40 & 300-400 & 0.51  \\ \hline 
\end{tabular}
\tablefoot{$^1$ Mass and age are estimated using \citet{Baraffe.2003} evolutionary models.
$^2$ This is the bolometric luminosity before extinction: the actual observed bolometric luminosity is  fainter.}
\end{table*}

\subsection{Atmosphere models artificially reddened}
  Since  our object appears extremely red, and because the only way to match its observed spectra with that of known L dwarfs is to add extra-photospheric absorption (see Section. \ref{sec:refobj}), we add a similar reddening to the atmosphere model fitting. We caution here that we add one free parameter to a fit that is already underconstrained,  with reduced $\chi^2$ values below 1 even without extinction, so we obviously find very low reduced $\chi^2$, actually as low as 0.28, with extinction. We therefore do not claim that the reddened fits are better in a statistical sense, but our aim is simply to explore whether these reddened atmospheres can also be associated with a physically consistent set of masses, radii, and gravities. With reddening we obtain good fits with all model sets, and consistently get higher effective temperatures (1600--1700~K). We choose in the following to describe the results obtained with  reddened  \textit{BT-Settl} models. 

\begin{figure*}
\begin{center}
\begin{tabular}{cc}
\includegraphics[width=8.4cm]{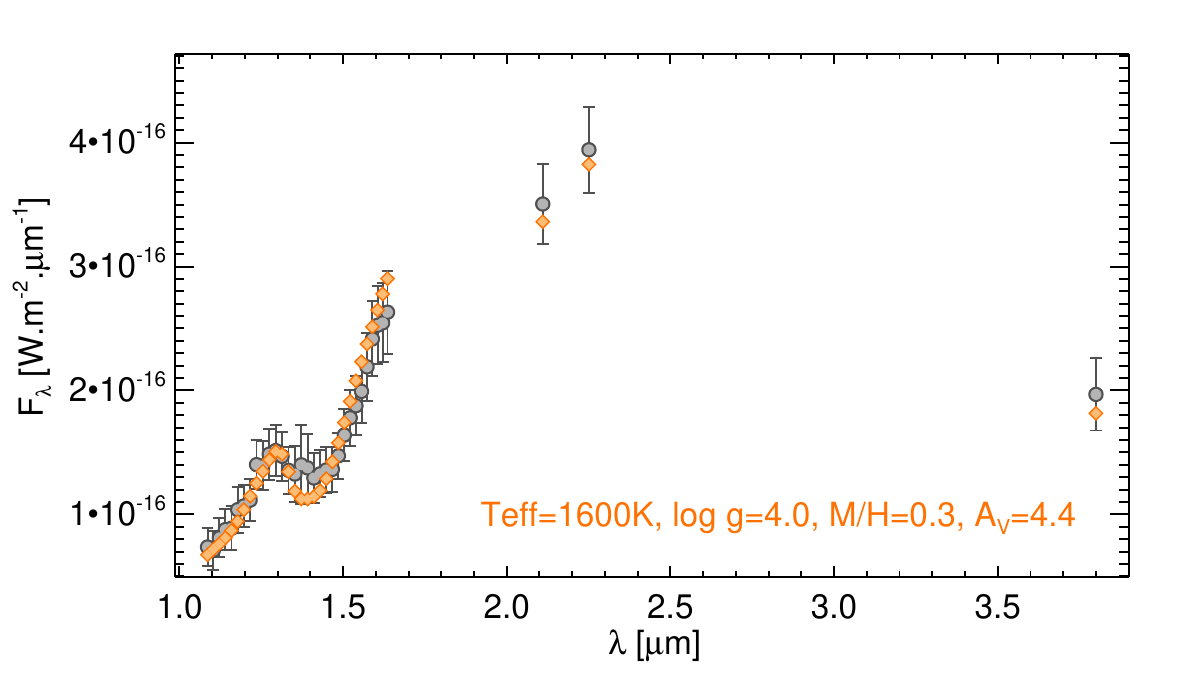} &
\includegraphics[width=8.4cm]{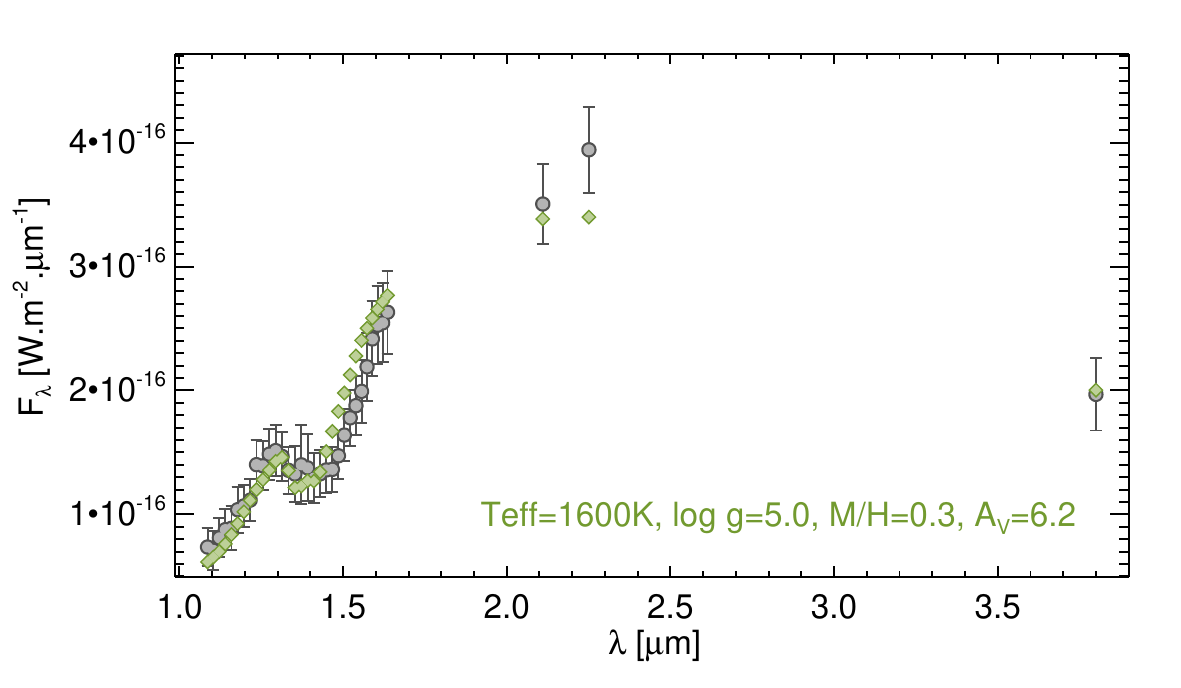} 
\end{tabular}
\caption{{\bf Left:} Best fit of observation (in black) with reddened BT-Settl atmosphere (in orange). {\bf Right:}  Best fit ($\chi ^2$) with reddened BT-Settl atmosphere when forcing higher gravity.}
\label{redsettl}
\end{center}
\end{figure*}

The best fit is obtained for an atmosphere with an effective temperature of 1600~K, a gravity of \logg =4.0, a slightly super-solar metallicity of [M/H]=0.3~dex, and a reddening of $A_V$=3.8mag ($A_K$=0.42), see Fig. \ref{redsettl}. If the reddening is assumed to come from within the object, the flux is not lost, but redistributed, and the resulting radii and masses are very similar to those found for the \textit{BT-Settl} models without reddening, also at 1600~K, and are associated with an object of a few \Mjup with a radius larger than 1.5~\Rjup, which is much smaller than the radius of the object that is necessary to match the absolute flux (<1.0\Rjup), and therefore physically inconsistent. The reddened solutions therefore can be physically plausible only if the reddening actually causes a flux loss and is linked to an extinction  outside of the object, as in the case of interstellar extinction or absorption by a circumstellar or circum-companion disc. Another plausible explanation could be dust released by a recent giant impact in the system. Even if this hypothesis is quite speculative, it is worth noting that the age of the host star is similar to the age of the Sun at the time of the Late Heavy Bombardement, and \citet{Wyatt.2016} claim that observing the effects of giant impacts in systems hosting a debris disc is not unlikely. However, even after accounting for the associated flux loss, the radius associated with the best reddened fit is still too small (<1.3\Rjup) to be consistent with the 1.6\Rjup that the \citet{Baraffe.2003} evolutionary models associate with a 1600~K, \logg =4.0, atmosphere of 9-10\Mjup object aged 10~Myr. Even if at such young ages evolutionary models might be affected by significant systematics \citep[see e.g.][]{Marley.2007}, we also note that this combination of parameters is not compatible with our independent age constraints on the host star. 

Finally, there is a degeneracy between reddening and the gravity of the fits, meaning that a family of models which provide fits almost as good as the best fit exist at higher gravity and higher reddening, and is associated with physical parameters that are consistent with the observed absolute flux of the object. This family of higher gravity best fits does not exist without reddening;  in this case very low gravity would be necessary to obtain  very red spectra and to match our observations. For a gravity of \logg =4.5 and an effective temperature of 1650~K, we get a physically consistent best fit corresponding to an object of 20--25\Mjup aged approximately 100~Myr. For \logg=5.0, the best fit corresponds to an effective temperature of 1700~K, which is consistent with evolutionary models for a 35--40\Mjup brown dwarf aged 300-400~Myr. In both case, these higher gravity solutions are in good agreement with our independent age constraints.

\subsection{Conclusions from atmosphere model synthesis}
All the atmosphere model fits that lead to solutions at least moderately consistent with absolute flux and substellar evolution model constraints are presented in Table \ref{recapmodels}. Our best fit solutions with the ExoREM atmosphere model favour an extremely dusty atmosphere with an effective temperature around 1300~K, solar metallicity, and an intermediate gravity from \logg =4.4 to \logg =4.8. The \citet{Baraffe.2003} evolution models show that such atmospheric parameters are associated with a moderately young (100--300~Myr) 15--30\Mjup low-mass brown dwarf. This age estimate is in very good agreement with the independent age constraint on the host star.  

 We also studied the alternative hypothesis of extra-photospheric absorption and showed that reddened atmosphere model spectra can also match both the SED of \obj and its absolute flux. The solutions that are compatible with our observation correspond to  1600--1700~K, 20--40\Mjup, extincted L dwarf, aged 100--400~Myr. The source of the absorption has to be located outside the companion, and cause an actual loss of flux rather than just a flux redistribution toward the red to match the absolute flux.

The mass estimates with or without additional reddening are also in good agreement with the upper limit on mass that can be derived from the Gaia-Tycho2 astrometric solution (see Appendix \ref{app:astrom}).

\section{Orbital constraints and planet-disc interactions}
\subsection{Constraints on the orbit of the companion}
\label{s:astrom}
\begin{table}
\caption{Available astrometry (position angle and separation) of \obj  \label{astrom}}
\begin{tabular}{l|cc} \hline 
Date& P.A.($^o$)  & separation(mas)\\ \hline \hline
 2015-Oct-05$^1$ &  69.95$\pm$0.55   & 270$\pm$2.6  \\
 2016-Aug-08 $^2$ & 61.6$\pm$1.9   & 269$\pm$10.4  \\
 2016-Sep-16 $^3$ &  62.25$\pm$0.11   & 265$\pm$2  \\\hline \hline
\end{tabular}
\tablefoot{ $^1$ SPHERE $H$ band \citet{Milli.2017}
 $^2$ NACO $L'$ band \citet{Milli.2017}
$^3$ SPHERE $K$ band this article}
\end{table}

The on-sky astrometric calibration of the IRDIS and IFS  used the tools described by \citet{Maire.2016astrom}, with a resulting uncertainty of  0.048$^\circ$ on the true north position. With this new SPHERE astrometric measurement and the two existing data points by \cite{Milli.2017}, we were able to constrain the orbit of the companion. 

\begin{figure*}
\includegraphics[width=18cm]{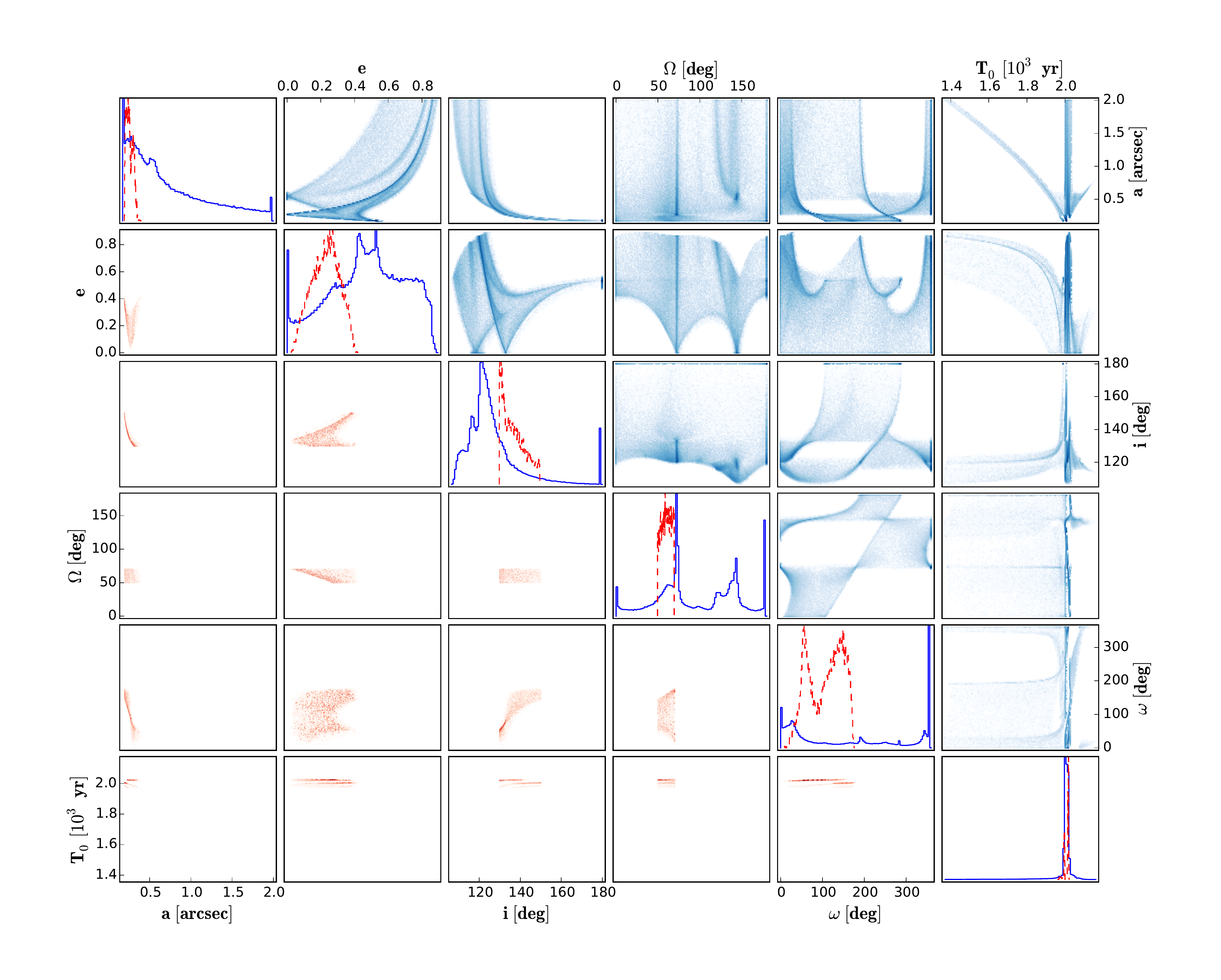}
\caption{
Results of our LSMC orbit fit for all fitting parameters. On the diagonal we show the distributions of each individual parameter (normalised to the distribution peak), while the other tiles show correlation plots between pairs of orbital parameters. In the upper right corner in blue hues (solid blue lines on the diagonal), we show all orbits that fit the available astrometry with an $\chi^2_{red} < 2$. In the lower left corner in red hues (red dashed lines on the diagonal), we show all the well fitting orbit solutions that are co-planar with the disk. 
%{\it lower panel should be updated if decision to include}}
 \label{main-orbit-results} }
\end{figure*}

For this purpose we used the least-squares Monte Carlo (LSMC) approach, as described in \cite{2013MNRAS.434..671G}. The concept of this method is to generate several million random sets of orbit solutions within some global boundaries and then use these as starting points of a least-squares minimisation. The main advantage of the approach is that we can confidently exclude certain parts of the parameter space. The approach was successfully used to constrain the orbits of several brown dwarf and planetary mass companions, e.g.  \cite{2014MNRAS.444.2280G} and \cite{2016A&A...587A..56M}. A detailed comparison to the Markov chain Monte Carlo approach is given in \cite{Vigan.2016}. \\
For the \host system we run an LSMC fit with $ 5\times 10^6$ individual orbits. We used flat input distributions of the orbital elements in order to constrain the possible orbits purely based on the available astrometry. We arbitrarily restricted the semi-major axes of the orbit solutions to values smaller than 2\, arcsec to somewhat constrain the parameter space. This only excludes extremely eccentric long-period solutions for which it would be very unlikely to find the companion in such close proximity to the host star as we observe it. We assumed a total system mass of 1.35\,M$_\odot$ (1.32\,M$_\odot$ for the primary, see Section \ref{s:age}  and 0.03\,M$_\odot$ for the companion) and used a distance of 40.7\,pc as measured by Gaia (\citealt{Gaia.2016a}, \citealt{Gaia.2016b}, \citealt{gaia_dr1}). \\ 
We find approximately $2.7 \times 10^5$ orbits with a reduced $\chi^2$ smaller than 2. In Fig.~\ref{main-orbit-results} we show all these solutions in the upper right part of the figure (displayed in blue). Since we do not have precise radial velocity measurements of the companion we restrict the displayed solutions to longitudes of the ascending node $\Omega$ between 0\,deg and 180\,deg. We note that the distributions of the other orbital elements look identical for both possible ranges of $\Omega$. In addition, in the lower left part of Fig.~\ref{main-orbit-results} (displayed in red) we show the orbit solutions that have been restricted in inclination $i$ and $\Omega$ in order to be coplanar with the observed debris disc. Taking the uncertainties of the disc inclination and position angle into account, we used a parameter range of 130\,deg to 150\,deg for the inclination (values larger than 90\,deg are associated with retrograde motion of the companion) and 50\,deg to 70\,deg for the longitude of the ascending node. We find approximately $1.2 \times 10^4$ such solutions with reduced $\chi^2$ smaller than 2. We caution that the best fitting orbits may change when new data points become available. Given the low number of astrometric data points, we are still very sensitive to small deviations in astrometric calibration or to calibration offsets between NACO and SPHERE.

%We further show the 3 best fitting unconstrained orbits and the 3 best fitting coplanar orbits in Fig.~\ref{unconstrained-best-fit} with the associated orbital elements in Tab.~\ref{orbit-elements}.

\subsection{Can the companion shape the debris disc without  additional planets?}
\begin{figure}
\centering
\includegraphics[width=8.4cm]{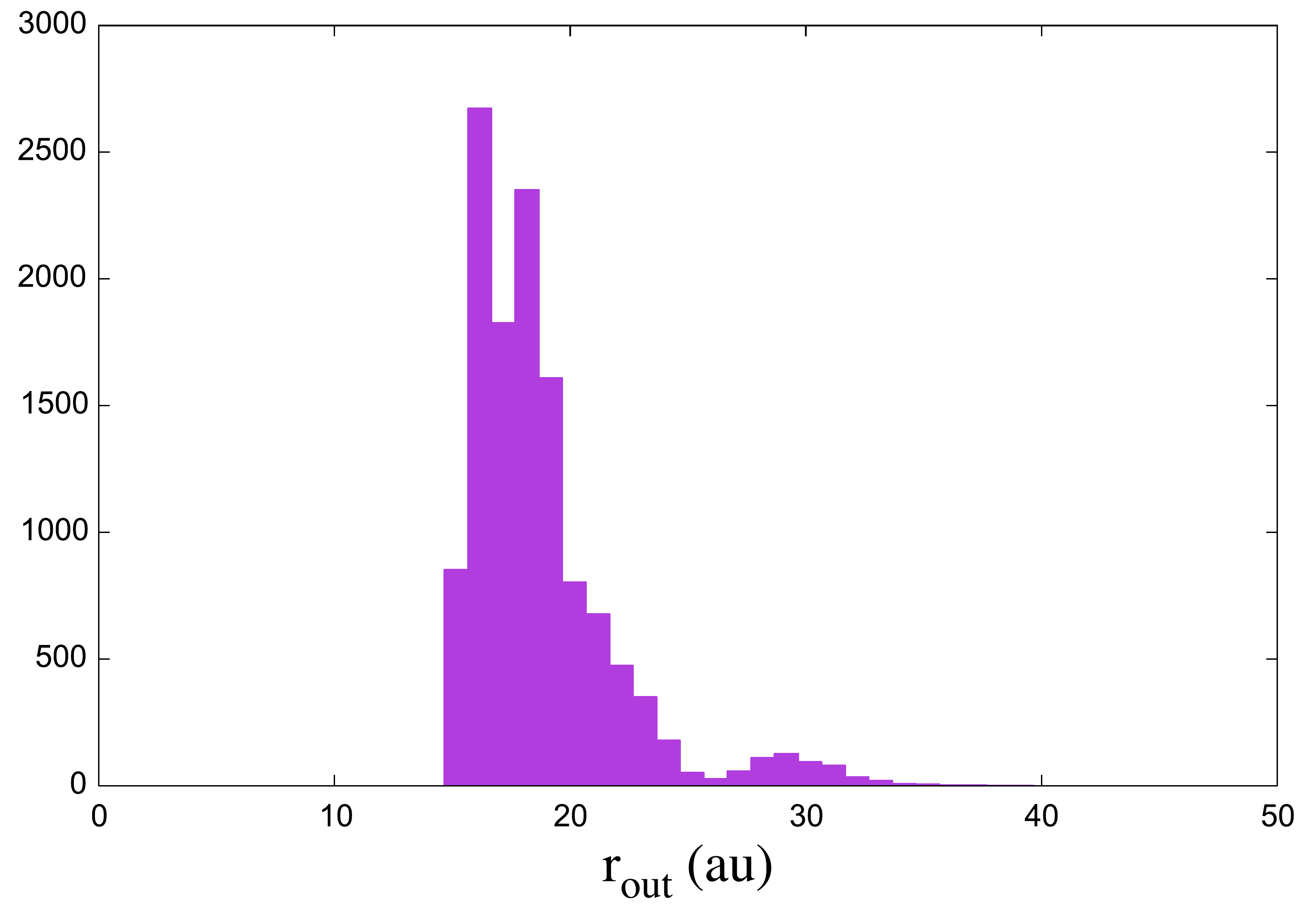}
\caption{Distribution of the outer extent of the chaotic zone for each orbit}
\label{chaos}
\end{figure}

In order to establish if the companion interacts with the disc, possibly shaping its inner edge, we have to determine the extension of chaotic zone of the former. More specifically, since the planet is interior to the disc, we are interested in the outer semi-amplitude of the chaotic zone, $\Delta a_{ext}$. It depends on a certain power law of the ratio between the mass of the planet and the mass of the star, $\mu$, on the semi-major axis of the planet, $a_p$, and on its eccentricity, $e_p$ \citep{Lazzoni.2017sub}. For each of the $12426$ coplanar orbits found to be compatible with the astrometric measurements, we used the equation in Wisdom (1980) modified for the eccentric case of $e_p\le0.3$ 
\begin{equation}
\Delta a_{ext}= 1.3 \mu^{2/7} a_p(1+e_p)
\label{eq1}
,\end{equation}
whereas for $e_p>0.3$ we applied the equation obtained by Mustill $\&$ Wyatt (2012), also modified for the eccentric case 
\begin{equation}
\Delta a_{ext}= 1.8 \mu^{1/5} e_p^{1/5} a_p(1+e_p).
\label{eq2}
\end{equation}
We did a further approximation in this last equation, substituting in the original formula the mean value of the eccentricities of the particles in the disc, $e$, with the eccentricity of the planet. This simplification is acceptable since the planet induces  an eccentricity on dust grains that is proportional to its own (Mustill $\&$ Wyatt 2009).\\
In Fig. \ref{chaos} we show the distribution of the outer extent of the chaotic zone obtained by all coplanar orbits. Due to the high uncertainty on the age of the star, the mass of the companion is not as accurately  determined, ranging between $12$ $M_J$ and $60$ $M_J$. We used as mean mass for the planet $30$ $M_J$;  however, we note that the clearing zone is not strongly influenced by $M_p$ since in equation \ref{eq1} and \ref{eq2} it is divided by the mass of the star $M_*$. Placing the edge at $50AU\pm5AU$ \citep[][]{Milli.2017}, it is clear that the planet never gets near enough to the disc  to shape its inner edge.\\

\begin{figure*}
\centering
\includegraphics[width=18cm]{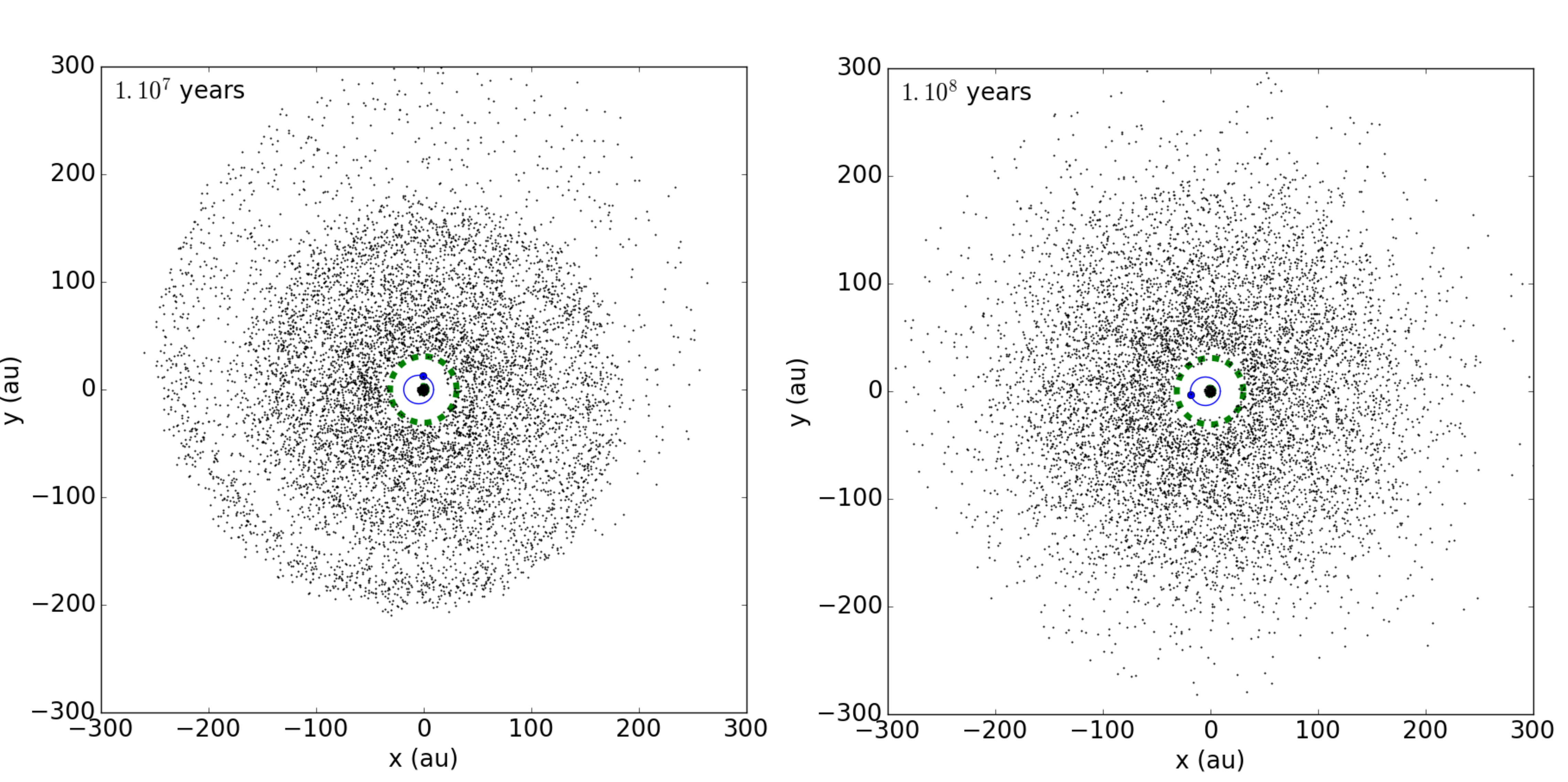}
\caption{Representation of an N-body simulation of 10,000 test particles initially ranging from 1 to 200 AU after (a) 10 million years (b) 100 million years. A companion has been injected at t=0 with semi-major axis 14.2 AU and eccentricity 0.31. Its orbit is depicted in blue, while the gap outer extent given by Eq. \ref{eq2} is in green.}
\label{MCMCdisc}
\end{figure*}

A companion can open a gap in a disc, but it could also deplete it and change its overall shape. We ran an N-body simulation to study the impact of the best fitting orbits on the disc and to confirm the results obtained through the approximate analytical approach. We used the code SWIFT\_RMVS \citep{swift}, and let 10,000 massless test particles evolve over 100 million years. The initial disc was set with a uniform radial distribution and extended from 1 to 200 AU. The evolution after 10 and 100 million years are plotted in Fig.  \ref{MCMCdisc}, for a companion with the best fitting orbit, with semi-major axis 14.2 au and eccentricity 0.31. We see that the companion creates a one-arm spiral that propagates through the disc during the first ten million years. Then the spiral dissipates, leading to a blurry outer extent between 150 and 200 AU. On the other hand, the inner edges of the disc are sharp, and closely match  the analytical prediction of a gap in the 3--31 AU range. This computation confirms that the orbit of the observed companion could not be responsible for a gap in the disc at $50 \pm 5$ AU. We also performed an N-body simulation for other best fitting orbits and found very similar outcomes, in good agreement with the analytical approach.\\

\section{Conclusions}
  We carried out a detailed study of \obj and its host star. We show that the host star is most probably aged between 50~Myr and 700~Myr, and that it has nearly solar metallicity. Our detailed analysis of the atmospheric properties of the companion show that many atmosphere models can provide a decent fit to our low-resolution observations over a range of effective temperature spanning 1300--1600~K and gravity between \logg =3.5 and \logg =5.0  for both solar metallicity and super-solar metallicity. The only parameter that appears absolutely necessary in order to fit the very red and relatively smooth spectra of \obj is dust. When dust content is a parameter of the tested model atmosphere, the best fit is always obtained for the highest photospheric dust content available in the grid.
 However, our knowledge of the distance of the object, and hence of its absolute flux, allows us to associate a mass and a radius for a given effective temperature and gravity, and to significantly break the degeneracy of the plausible atmospheric parameters through a comparison with substellar evolution models. The only solutions for which the atmosphere is associated with a set of effective temperatures and gravity that is physically consistent with the  substellar evolution models are those which are close to our best fit with the ExoREM atmosphere model, favouring an extremely dusty atmosphere with an effective temperature around 1300~K, solar metallicity, and a gravity of \logg =4.4. Comparison with the \citet{Baraffe.2003} evolution models show that these atmospheric parameters are consistent with a moderately young (100--300~Myr) 15--30\Mjup low-mass brown dwarf. These physical parameters are also in good agreement with the independent constraint on age derived from the study of the host star. The resulting mass-ratio between \obj and \host would be around 2\%, which together with a semi-major axis of 10~AU could be compatible with a formation along core-accretion scenario \citep{Pollack.1996}. Detailed core-accretion modelling by \citet{Mordasini.2012} show that it is very unlikely that core accretion could form a 15--30\Mjup companion around a solar metallicity star. However, in the unlikely case the system  resided near the lower end of our age estimate range (50~Myr),  \obj would be in the planetary mass range, and in this specific case, it would be compatible with a formation via core-accretion.
 Though not necessary to match our observed spectra, we note that super-solar metallicity atmosphere models also provide good fits to the data in a range of effective temperatures and gravities that are close to those derived for solar metallicity models. Though our current data cannot rule out or favour super-solar  metallicity, future observation at higher spectral resolution, perhaps with ELT {\textit Harmoni}---the separation between \obj and its host-star is probably too small and the contrast too high for JWST to provide useful constraints---should be able to resolve the atomic and molecular lines that are crucial in order to define both the gravity and the metallicity of L dwarfs \citep{Allers.2013,Faherty.2014}. \\
 We note that though this companion is extremely red, our atmosphere fitting as well as the independent constraints on the age of the system are much more consistent with an intermediate gravity (\logg 4.4--5.0) than with the very low gravity that is sometimes associated with very red L dwarfs \citep[e.g.][]{Liu.2013}. This suggests that other parameters in addition to gravity,  such as enrichment at formation \citep[e.g.][]{2016arXiv161003216L}, inhomogeneous dust cloud coverage \citep[][]{2016ApJ...829L..32L}, or even a circumplanetary disc \citep{2017MNRAS.464.1108Z} could also explain the peculiar properties of some of the very red L dwarfs. Our empirical comparison with known L dwarfs shows that the spectrum of \obj is best matched by the spectra of atypically red L3-L5 dwarfs, which are themselves artificially reddened by a modelled extra-photospheric absorption by fosterite dust. We also explored whether similarly reddened atmosphere model spectra could match both the SED of \obj and its absolute flux. We show that the source of the absorption has to be located outside the companion, hence causing a real loss of flux (and not only a flux redistribution toward the red), so that the reddened atmosphere models can match the absolute flux. In this case a 1600--1700~K, 20--40\Mjup, extincted L dwarf would provide a very good match to our observations and our age constraint on the system.\\
 However, we recall that atmosphere models can also match the extremely red SED of \obj without requiring the addition of complexity and free parameters associated with an external source of absorption. We conclude that \obj might be either the most extreme known case of extremely dusty and red, 1300--1400~K, 15-30\Mjup L dwarfs or a very red, slightly warmer, and more massive L dwarf embedded in some kind of external light-absorbing structure, such as a circum-companion disc or dust cloud released by a recent giant impact in the system.
 Finally, our dynamical analysis of the companion orbit and of the dusty disc detected around its host star shows that the companion cannot be responsible for the shape of the inner edge of the disc. Together with our direct imaging detection limits that reach well within the planetary mass range for separations larger that that of the observed companion, this means that the disc is probably shaped by one or more additional, as yet undetected, exoplanets with a mass of a few \Mjup at most.

\begin{acknowledgements}
This work has made use of the SPHERE Data Centre, jointly operated by OSUG/IPAG (Grenoble), PYTHEAS/LAM/CeSAM (Marseille), OCA/Lagrange (Nice), and Observatoire de Paris/LESIA (Paris). SPHERE is an instrument designed
and built by a Consortium consisting of IPAG (Grenoble, France), MPIA (Heidelberg,
Germany), LAM (Marseille, France), LESIA (Paris, France), Laboratoire
Lagrange (Nice, France), INAF–Osservatorio di Padova (Italy), Observatoire astronomique de l’Université de Genève (Switzerland), ETH Zurich (Switzerland), NOVA (Netherlands),
ONERA (France), and ASTRON (Netherlands) in collaboration with
ESO. SPHERE was funded by ESO, with additional contributions from CNRS
(France), MPIA (Germany), INAF (Italy), FINES (Switzerland), and NOVA
(Netherlands). SPHERE also received funding from the European Commission
Sixth and Seventh Framework Programmes as part of the Optical Infrared Coordination Network for Astronomy (OPTICON) under grant number RII3-Ct-
2004-001566 for FP6 (2004–2008), grant number 226604 for FP7 (2009–2012),
and grant number 312430 for FP7 (2013–2016)
This publication makes use of VOSA, developed under the Spanish Virtual Observatory project supported from the Spanish MICINN through grant AyA2011-24052.
This research has made use of the SIMBAD database,
operated at CDS, Strasbourg, France.
We acknowledge financial support from the French ANR GIPSE, ANR-14-CE33-0018.We acknowledge financial support from the Programme
National de Planétologie (PNP) and the Programme National de Physique Stellaire
(PNPS) of CNRS-INSU. This work has also been supported by a grant from
the French Labex OSUG@2020 (Investissements d’avenir – ANR10 LABX56).

O.A.\ acknowledges funding from the European Research Council under the European Union's Seventh Framework Programme (ERC Grant Agreement n.~337569), and from the French Community of Belgium through an ARC grant for Concerted Research Action.

C.M. acknowledges the support of the Swiss National Science Foundation via grant BSSGI0\_155816 “PlanetsInTime”.

We acknowledge financial support from the “Progetti Premiali” funding 
scheme of the Italian Ministry of Education, University, and Research.

S.D. warmly thanks Ennio Poretti for the useful discussions.

This work has been in particular  carried out in the frame of the National Centre for Competence in Research ‘PlanetS’ supported by the Swiss National Science Foundation (SNSF).
V.C. acknowledges support from CONICYT through CONICYT-PCHA/Doctorado Nacional/2016-21161112, and from the Millennium Science Initiative (Chilean Ministry of Economy), through grant RC130007. V.C. and O.A. acknowledge support from the European Research Council under the European Union's Seventh Framework Program (ERC Grant Agreement n. 337569).
\end{acknowledgements}

\bibliographystyle{aa}
\bibliography{biball}

\begin{appendix}
\section{Rotation period}
\label{app:rot} 
We can reasonably infer that the P=0.996\,d signal has a stellar origin and is most likely linked to the rotation of the star for the following reasons:

{\it a)} We collected three average magnitudes per night, which  are enough to show
   the significant magnitude variation that the star undergoes in a time of about 5 hr. The
   same pattern of variation repeats  each night and can be well fitted by a sinusoid
   with  P = 0.996d, but sampling a slightly progressing rotation phase. This implies
   that the observed magnitudes progressively decrease toward the phase minimum.

{\it b)} The $c-ck$ time series, which has same sampling as the $v-c$
    time series, does not show that intra-night pattern of variation exhibited by the variable and its  periodogram 
did not detect any significant periodicity at confidence levels greater than 50\%.

{\it c)} When a time series has a periodicity different from 1 day, but the most significant
   power peak at about 1 day (owing to the spectral window),
   the filtering of the about 1d periodicity removes it and shows the
  true periodicity as the most significant. In our case, even filtering the periodicity corresponding to the maximum power peak on the spectral window,
  the most significant power peak still remains at P = 0.996\,d.

{\it d)} The upper limit on rotation period as derived from stellar radius and projected
rotational velocity is about 2 days, thus allowing only a limited range of values
around 1 day.

Another possibility is that the observed periodicity is due to $\gamma$ Dor-like pulsations. Indeed, the observed photometric amplitude is well above those observed in Pleiades F stars by K2 \citep{Stauffer16} and more consistent with $\gamma$ Dor variables \citep{balona11}.
Such spot-dominated photometric variability is observed in mid-F stars only at ages significantly younger than the Pleiades \citep[see e.g.][for the case of $\beta$ Pic MG]{messina16}. When observed with the time sampling
characteristic of ground-based observations, $\gamma$ Dor variability
appears to be dominated by one or a few periods and for this reason the classification as spot dominated or $\gamma$ Dor variability is ambiguous in several cases \citep{rebull16}.
HD~206893 has an effective temperature slightly cooler than bona fide $\gamma$ Dor stars,
but compatible with several $\gamma$ Dor candidates \citep{alicavus16}.
The pulsation frequency of $\gamma$ Dor variables is correlated with
the projected rotation velocity \citep{alicavus16} and there are
indications  that these stars oscillate at or close to the rotation
period \citep{balona11}.

Our data are insufficient to draw firm conclusions on the origin of the photometric
variations, but considering this correlation, 
in this article we consider the detected periodicity as the true rotation period of the star.

\section{Independent data reductions}\label{app:otherred}

    We also analysed our data using totally independent pipelines, namely VIP Optimal-PCA \citep{Wertz.2017}, based on Principal Component Analysis \citep[PCA; e.g.][]{Amara.2012,Soummer.2012} and ANDROMEDA \citep{Cantalloube.2015}, based on \textit{smart}-ADI \citep{Lagrange.2010}.

\begin{figure}
\includegraphics[width=8.4cm]{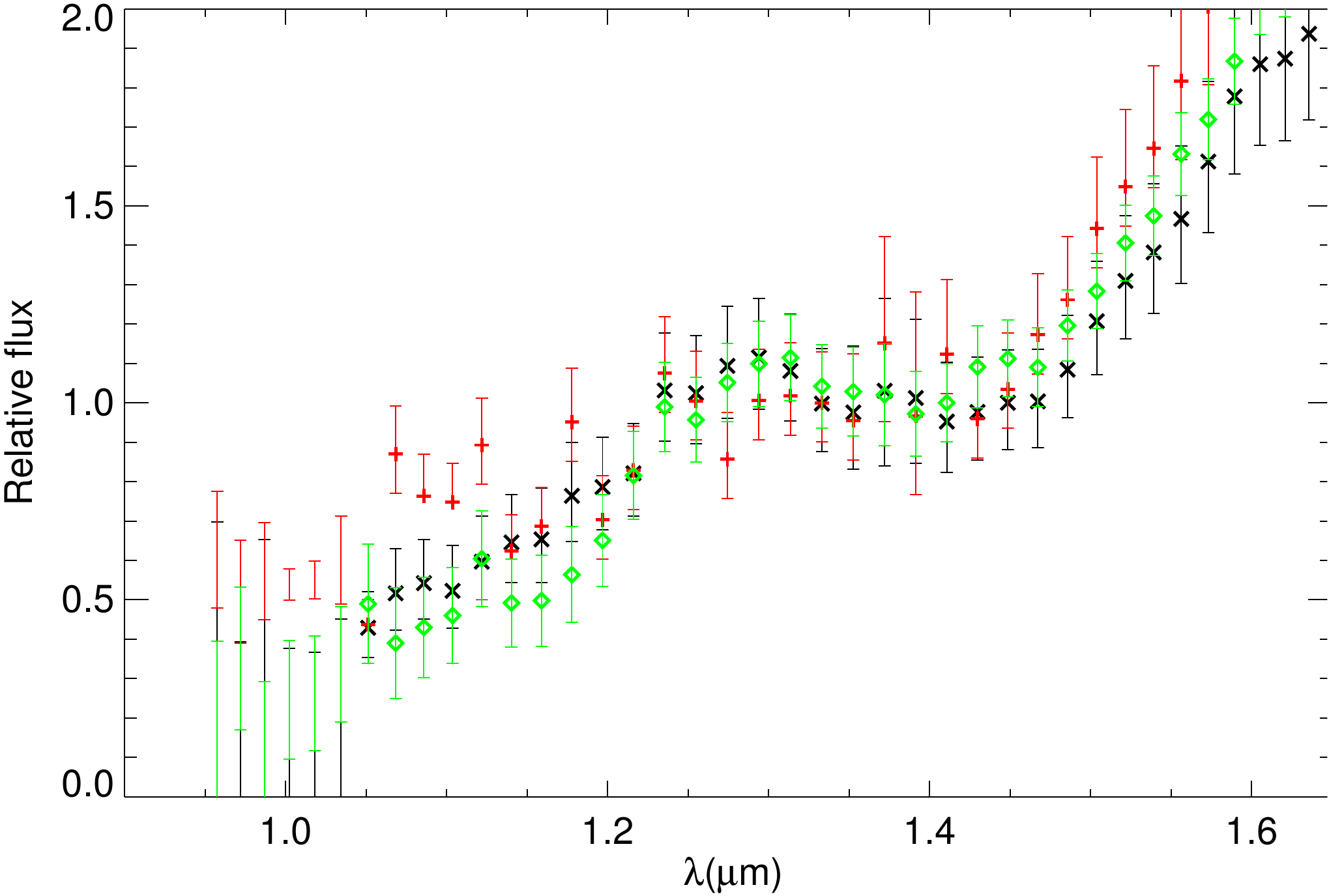}
 \caption{Extracted spectrum of \obj through TLOCI-ADI (black), VIP Optimal-PCA  (red), and ANDROMEDA (green)\label{app:spectra}}
\end{figure}

 The comparison between the spectrum of \obj extracted by these three algorithms is presented in Fig. \ref{app:spectra}. All three reductions show a very red spectra, with non-detection or marginal detections in the blue part of the spectrum and high signal-to-noise detections in the red channels. According to the estimated error bars, the difference between the extracted fluxes are mostly consistent with Gaussian noise, but a difference can be noticed in the measured signal of the $J$-band peak. Though the difference is moderate in $\sigma$ unit, it is systematic;  the estimated flux in the $J$-band peak is systematically lower through the VIP Optimal-PCA analysis than through the TLOCI-ADI and the ANDROMEDA analyses. Since we tested more carefully the spectral fidelity (ability to retrieve  the signal of injected fake planets, within the error bars) of the TLOCI-ADI analysis when setting up our pipeline, we primarily consider the TLOCI-ADI extracted spectrum in this article. This choice is backed up by the similar SED obtained when using the ANDROMEDA algorithm.
It is important to note that since speckle noise is correlated from channel to channel and that for a separation of approximately 8 lambda/D, it takes 150-200~nm, or 8-12 channels for a speckle residual to move one FWHM. It is therefore normal that extractions using different speckle subtraction algorithms show non-Gaussian differences over several contiguous channels.

 We nevertheless checked the impact of this systematic difference by fitting the spectra extracted through the VIP-optimal-PCA algorithm, and in this case, the best fitting \textit{BT-Settl} atmosphere model was that of a 1550~K, \logg=3.5 atmosphere, almost identical to the best fit obtained with TLOCI-ADI (1600~K, \logg=3.5). This confirms that our study of the atmospheric properties of \obj is not significantly affected by the systematic differences that arise when using different algorithms to extract the spectrum of the companion. 

% Over 41 data points, we have one channel for which the two independent reductions have a 3$\sigma$ difference and 3 channels for which the difference is between 2 and 3 $\sigma$, which is actually only slightly above Gaussian statistics. The dispersion , in sigma units, of the difference between both extracted spectrum is 1.01, which is in very good agreement with Gaussian statistics, and hints that the noise that we measure for each spectra dominates over unaccounted systematic noise. 

 \section{Characterisation of the spectral correlation of noise in IFS data} \label{app:correl}
\begin{center}
\begin{figure}
\includegraphics[width=8.4cm]{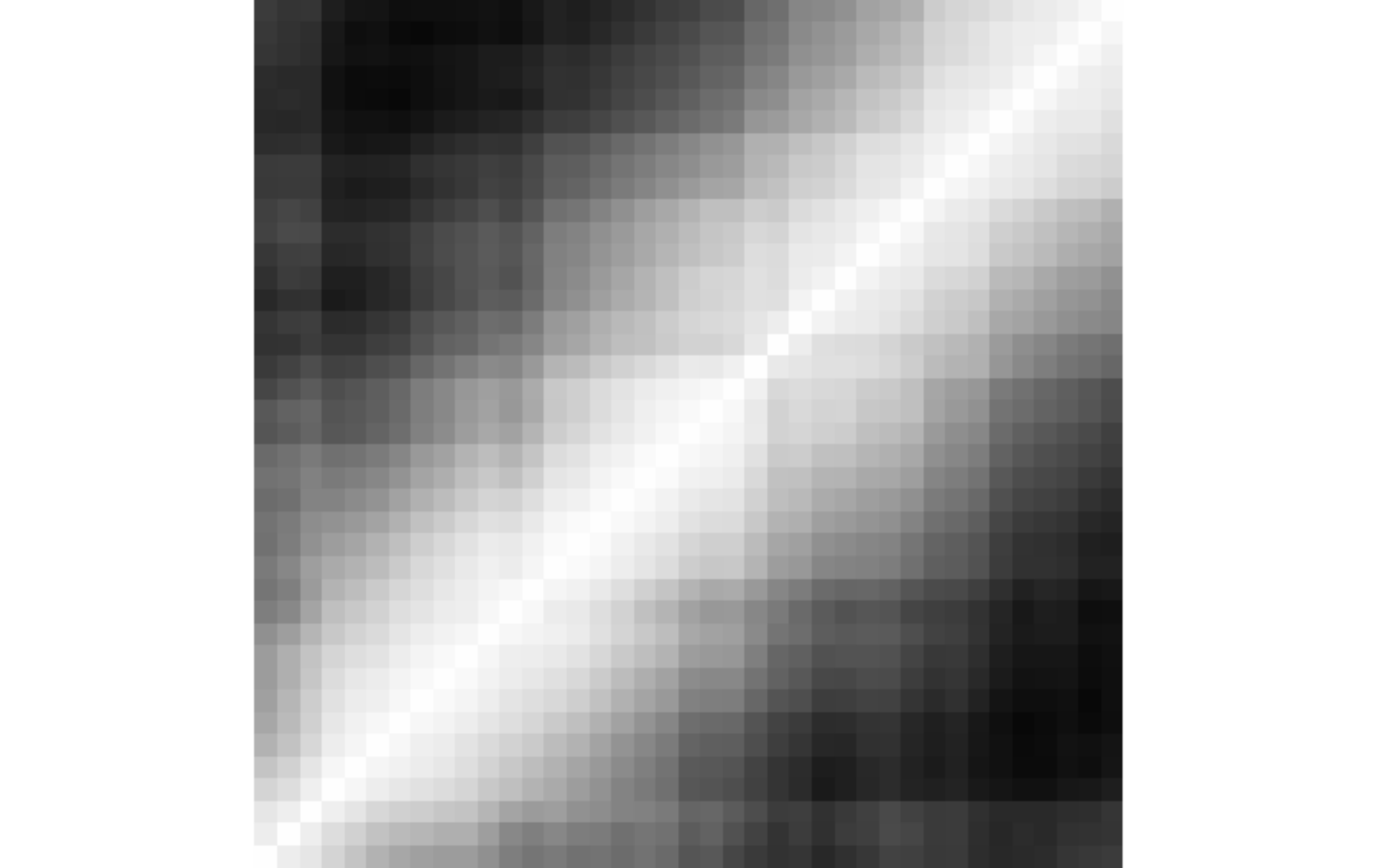}
 \caption{ Correlation matrix of the noise in the residual image TLOCI-ADI analysis. The x- and y-axes have 39 values associated with the 39 IFS channel. The pixel ${i,j}$ therefore identifies the correlation of noise between channels $i$ and $j$. We use a linear scale for which a perfect correlation of 1 is white and a slight anticorrelation of -0.1 is black.  \label{correl}}
\end{figure}
\end{center}

  We derived the correlation of the noise in IFS spectral channels in the same way as \citet{Samland.2017}. We therefore followed the prescription of \cite{Greco.2016} to derive the correlation of IFS spectral channels at the separation of the detected companion. Since the noise is dominated by speckle noise, the spectral correlation of noise is highly dependent upon the analysis technique that is used, and we therefore directly measure it from the final residual wavelength cube after TLOCI-ADI has been applied. For the resulting correlation matrix, shown in Fig. \ref{correl}, the matrix component $\phi_{ij}$ measures the correlation between channels $i$ and $j$ at the separation of the companion.  The value of the correlation varies between -1 (pure anticorrelation) and +1 (pure correlation). Zero is adopted for spectral channels that are uncorrelated. Though correlation of noise in data is not in itself a problem, it needs to be taken into account to avoid biasing the fit of models or empirical template spectra. In the following this is taken into account by following the prescription of \citet{Derosa.2016} by introducing the covariance matrix $C$ into the calculation of the $\chi ^2$ of the fit and $C$ is directly derived from the correlation matrix $\phi$, with  $C=\Sigma \times \phi \times \Sigma $, where $\Sigma$ is a diagonal matrix for which $\Sigma_{ii}$ is the value of noise of the extracted spectrum on channel $i$. The correlation matrix for the TLOCI-ADI reduction of our IFS dataset is shown in Fig. \ref{correl}.
% As expected, it is lower in the optimal-PCA approach, which uses a different PCA-Klip parameter for each spectral channel.
 It is also important to remember than the spectral resolution of IFS in $Y-H$ mode is 30, which means that three adjacent channels fall within the same spectral resolution element and are expected to have their signals correlated regardless of any reduction-induced correlation of noise.\\

\section{Goodness of fit}
\label{app:goodness}
The goodness-of-fit indicator  takes into account the filter width $w$ when comparing each of the $k$ template spectra $\textsl{F}_{k}$ to the $n$ photometric data points of HD~206893b, each of which have  a given error bar $\sigma _{i}$
\begin{equation}
\label{eq:reg}
G_{k}=\sum_{i=1}^{n} w_{i} \left ( \frac{f_{i} - \alpha_{k}\textsl{F}_{k,i}}{\sigma _{i}} \right )^{2}
,\end{equation}  
where $\alpha_{k}$ is a multiplicative factor between the companion spectrophotometry and that of the template which minimises $G_{k}$ and is given by

\begin{equation}
\label{eq:alpha}
\alpha_{k} = \frac{\sum_{i=1}^{n}w_{i}f_{i}\textsl{F}_{k,i}/\sigma _{i}^{2}}{\sum_{i=1}^{n}w_{i}\textsl{F}_{k,i}/\sigma _{i}^{2}}
\end{equation}The correlation of noise was then taken into account in the spectral fitting by following the method described in \cite{Derosa.2016}.

\begin{comment}

\begin{equation}
\label{eq:cn}
G_{k}=R_{k}^{T}C^{-1}R_{k} + \sum_{i=1}^{n} w_{i} \left ( \frac{f_{i} - \alpha_{k}\textsl{F}_{k,i}}{\sigma _{i}} \right )^{2}
\end{equation}  

$C$ is the correlation matrix  determined following \cite{Greco.2016} and $R_{k} = f  - \alpha_{k}\textsl{F}_{k}$.
% The resulting correlation matrixes for two of the independant data analysis methods presented in this article are shown on Fig.\ref{correl}.

$\alpha_{k}$ values do not necessarily corresponds to the values found in equation \ref{eq:alpha} anymore.  We therefore found it via a two-steps process. We first adopted the value found in  equation \ref{eq:alpha}. We then considered an exploration of  10$^{4}$ values of $\alpha_{k}$ plus or minus 100 times the value determined at the previous step and determined the one which was minimizing $G_{k}$. This exploration leads to stable minima of $G_{k}$ for the considered template spectra.  
\end{comment}

\section{Astrometric upper limits on companion mass}
\label{app:astrom}
\begin{center}
\begin{figure}
\includegraphics[width=8.4cm]{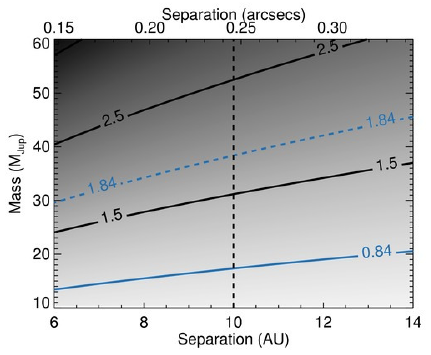}
\caption{Mass limit for the observed $\Delta \mu$ adopting $R_0=1$ in the
\cite{makarov05} equation. (corresponding to pole-on circular orbit)
The solid blue line represents the observed  $\Delta \mu$ value, the dashed
line the 1$\sigma$ limit.
%Lower panel: results for the OLD orbits 1 and 2 by C. Ginski.
%{\it lower panel should be updated if decision to include}}
 \label{f:deltamu}}
\end{figure}
\end{center}

The presence of a companion may be revealed as a difference on short-term proper motions
taken at different epochs or between short-term and long-term proper motions
(Bonavita et al. 2017, submitted).
%\citep{bonavita17}.
The proper motion difference depends on companion mass, orbital separation, and
orbital parameters \citep{makarov05}. 
The astrometric data of \host do not reveal any significant differences in  proper motion. The Gaia-Tycho2 proper motion difference is $\Delta \mu_{\alpha} = -0.16\pm90 $ mas/yr
and $\Delta \mu_{\delta} = 0.86\pm1.0$ mas/yr.
We checked the constraints on the mass of the substellar companions that can be derived
from these measurements, using the procedure by Fontanive et al. in prep. (Fig. ~\ref{f:deltamu}).
The constraints are rather loose, but still make the hypothesis that \obj would be a massive brown dwarf somewhat unlikely. These dynamical limits are in general agreement with the mass derived from theoretical models for most of the adopted age range, but are marginally  in tension (1$\sigma$) with the companion mass associated with the older ages, such as $\sim$50\Mjup for an age of 700~Myr.

\section{Benchmark dusty L dwarfs}
\label{App:A}
\begin{table*}
\caption{Properties of  dusty, peculiar, and possibly young dwarfs later than L4 considered in our study}
\label{tab:AppA}
\begin{center}
\begin{tabular}{lllll}
\hline\hline
Source name                                                             &               Spectral type            &                               Cand. Member                            &       Mass   &     Ref \\
                                                                                                &                                                               &                                                                               &       ($\mathrm{M_{Jup}}$)            & \\
\hline
2MASS J11193254-1137466         &               L7                      &                 TWA                             &      4.3-7.6         &       1, 2    \\
2MASS J17081563+2557474 &       L5              &       \dots                           &               \dots           &       1 \\
WISEP J004701.06+680352.1 & L7  &       AB Dor   &      $\sim$18  &     3, 4 \\
PSO J318.5338-22.8603   &       L7      &       $\beta$ Pic             &       $8.3\pm0.5$     &       5, 6 \\
ULAS J222711-004547 & L7        &       \dots & \dots & 7 \\
WISE J174102.78-464225.5 & L7 & $\beta$ Pic or AB Dor & 4-21 & 8 \\
WISE J020625.27+264023.6 & L8 & \dots & \dots & 9, 22 \\
WISE J164715.57+563208.3 & L9 & Argus  & 4-5 & 9, 10 \\
2MASS J00011217+1535355 &       L4      & AB Dor        &       $25.3 \pm 1.0$    &        11, 12, 13 \\
2MASS J21543454-1055308 & L4    & Argus?        &       \dots & 14, 22 \\2MASS J22064498-4217208 &       L4      &       AB Dor  &       $23.1\pm6.4$ & 15 \\ 
2MASS J23433470-3646021 & L3-L6 & AB Dor        &       \dots & 12 \\
2MASS J03552337+113343 & L5     &       AB Dor & 13-30 & 16, 17, 18 \\
2MASS J21481628+4003593  & L6   &       Argus?  & \dots &               10, 15, 19, 22\\
2MASSW J2244316+204343 & L6     &       AB Dor  &       11-12   &       15, 20 \\
2MASSJ05012406–0010452 & L4   & Columba       & $10.2^{+0.8}_{-1.0}$ & 12, 21 \\
\hline
\end{tabular}
\end{center}
\tablefoot{References: 1-\cite{2015AJ....150..182K}, 2-\cite{2016ApJ...821L..15K}, 3-\cite{2012AJ....144...94G}, 4-\cite{2015ApJ...799..203G}, 5-\cite{Liu.2013}, 6-\cite{2016ApJ...819..133A}, 7-\cite{2014MNRAS.439..372M}, 8-\cite{2014AJ....147...34S}, 9-\cite{2011ApJS..197...19K}, 10-\cite{Gagne.2014a}, 11-\cite{2004AJ....127.3553K}, 12-\cite{2015ApJS..219...33G}, 13-\cite{2015ApJ...798...73G}, 14-\cite{2014ApJ...792L..17G}, 15-\cite{Faherty.2016}, 16-\cite{2006AJ....132..891R}, 17-\cite{2009AJ....137.3345C}, 18-\cite{2013AJ....145....2F}, 19-\cite{2008ApJ...686..528L}, 20-\cite{2003ApJ...596..561M}, 21-\cite{2008MNRAS.390.1517C}, 22-\cite{Liu.2016}}\\
\tablefoot{Distances: $^{k}=$kinematic distance,  $^{\pi}$=parallax,  $^{p}$=photometric.}\\
\end{table*}

\end{appendix}

\end{document}